\RequirePackage{rotating}
\documentclass[a4paper,fleqn,usenatbib]{mnras}

\usepackage{newtxtext,newtxmath}

\usepackage[T1]{fontenc}
\usepackage{ae,aecompl}
\usepackage{hhline}

\usepackage{graphicx}	
\usepackage{amsmath}	
\usepackage{caption}

\newcommand\msun{\mbox{M$_{\odot}$}}
\newcommand\lsun{\mbox{L$_{\odot}$}}
\newcommand\kms{\mbox{km~s$^{-1}$}}
\newcommand\HI{H\,{\sc i}}

\newcommand\askapsoft{{\sc askapsoft}}

\newcommand\clean{{\sc clean}}

\newcommand\sofia{{\sc sofia}}
\newcommand\profound{{\sc profound}}

\title[WALLABY: \HI\ content of the Eridanus Supergroup]{WALLABY Pre-Pilot Survey: \HI\ Content of the Eridanus Supergroup}

\author[B.-Q. For et al.]{
  B.-Q. For,$^{1,2}$\thanks{E-mail: biqing.for@uwa.edu.au}
  J. Wang,$^{3}$
  T. Westmeier,$^{1,2}$
  O.I. Wong,$^{4,1,2}$
  C. Murugeshan,$^{4,2}$
  \newauthor
  L. Staveley-Smith,$^{1,2}$
  H.M. Courtois,$^{5}$
  D. Pomar$\rm{\grave{e}}$de,$^{6}$
  K. Spekkens,$^{7}$
  B. Catinella,$^{1,2}$
  \newauthor
  K.B.W. McQuinn,$^{8}$
  A. Elagali,$^{9}$
  B.S. Koribalski,$^{10,11,2}$
  K. Lee-Waddell,$^{1,4}$
  J.P. Madrid,$^{12}$
  \newauthor
  A. Popping,$^{1}$
  T.N. Reynolds,$^{1,2}$ 
  J. Rhee,$^{1,2}$
  K. Bekki,$^{1}$
  H. D$\rm{\grave{e}}$nes,$^{13}$
  P. Kamphuis,$^{14}$ \&
  \newauthor
  L. Verdes-Montenegro$^{15}$\\
  $^{1}$International Centre for Radio Astronomy Research, University of Western Australia, 35 Stirling Hwy, Crawley, WA 6009, Australia\\
  $^{2}$ARC Centre of Excellence for All Sky Astrophysics in 3 Dimensions (ASTRO 3D)\\
  $^{3}$Kavli Institute for Astronomy and Astrophysics, Peking University, Beijing 100871, China\\
  $^{4}$CSIRO Space $\&$ Astronomy, PO Box 1130, Bentley, WA 6102, Australia\\
  $^{5}$Univ Claude Bernard Lyon 1, IP2I Lyon, IUF, F-69622, Villeurbanne, France\\
  $^{6}$Institut de Recherche sur les Lois Fondamentales de l’Univers, CEA Universit$\'e$ Paris-Saclay, France\\
  $^{7}$Royal Military College of Canada, PO Box 17000, Station Forces, Kingston, ON K7K7B4, Canada\\
  $^{8}$Rutgers University, Department of Physics and Astronomy, 136 Frelinghuysen Road, Piscataway, NJ 08854, USA\\
  $^{9}$Telethon Kids Institute, Perth Children’s Hospital, Perth, Australia\\
  $^{10}$CSIRO Space $\&$ Astronomy, PO Box 76, Epping, NSW 1710, Australia\\
  $^{11}$School of Science, Western Sydney University, Locked Bag 1797, Penrith, NSW 2751, Australia\\
  $^{12}$The University of Texas Rio Grande Valley, One West University Blvd, Brownsville, TX 78520, USA\\
  $^{13}$ASTRON, The Netherlands Institute for Radio Astronomy, Oude Hoogeveensedijk 4, 7991 PD, Dwingeloo, the Netherlands\\
  $^{14}$Ruhr University Bochum, Faculty of Physics and Astronomy, Astronomical Institute, 44780 Bochum, Germany\\
  $^{15}$Instituto de Astrof$\acute{i}$sica de Andaluc$\acute{i}$a, CSIC, Glorieta de la Astronom$\acute{i}$a, E-18080, Granada, Spain
}

\date{Accepted 2021 July 30. Received 2021 July 29; in original form 2021 July 5}

\pubyear{2021}

\begin{document}
\label{firstpage}
\pagerange{\pageref{firstpage}--\pageref{lastpage}}
\maketitle

\begin{abstract}
  We present observations of the Eridanus supergroup obtained with the Australian Square Kilometre Array Pathfinder (ASKAP)
  as part of the pre-pilot survey for the Widefield ASKAP $L$-band Legacy All-sky Blind Survey (WALLABY).
  The total number of detected \HI\ sources is 55, of which 12 are background galaxies not associated with the Eridanus supergroup.  
  Two massive \HI\ clouds are identified and
  large \HI\ debris fields are seen in the NGC~1359 interacting galaxy pair, and the face-on spiral galaxy NGC~1385.
  We describe the data products from the source finding algorithm
  and present the basic parameters. The presence of distorted \HI\ morphology in all detected galaxies suggests
  ongoing tidal interactions within the subgroups. The Eridanus group has a large fraction of \HI\ deficient galaxies
  as compared to previously studied galaxy groups. These \HI\ deficient galaxies are not found at the centre of the group.
  We find that galaxies in the Eridanus supergroup do not follow the general trend of the atomic gas fraction versus
  stellar mass scaling relation, which indicates that the scaling relation changes with environmental density.
  In general, the majority of these galaxies are actively forming stars.  
\end{abstract}




\section{Introduction}

The star formation-density and morphology-density relations highlight the dependence of galaxy evolution on
environment.
A galaxy cluster is the most striking place to look for
environmental effects on galaxy evolution. Observations show that
the star formation is suppressed in cluster environments \citep{Gomez03, Cortese19} and the
fraction of early-type (elliptical and lenticular)
galaxies increases in denser cluster environments as compared to the fraction of
late-type (spiral and irregular) galaxies \citep{Dressler80, Postman05}. 
Recent studies of clusters in the local Universe have shown conflicting results.  
While a morphology-density relation remains within each halo,
the total fraction of early-type galaxies is nearly constant across three orders of magnitude in
cluster halo mass (13 $\leq \log M_{\rm h}/(h^{-1}$ \msun) $\leq$ 15.8) \citep{Hoyle12}. The study of \citet{Simard09}
also shows no clear trend of the early-type fractions in clusters as a function of cluster velocity dispersion.
These results suggest that 
most morphological transformation is happening in groups prior to merging
into massive clusters. The small velocity dispersion of galaxy groups also results in more
galaxy mergers than in clusters (e.g., \citealp{Hickson97}). A combination of nature and nurture must be
at play to drive such differences. The evolutionary effects imposed by different environments are
important to understand. 

In order to gain a full picture of galaxy evolution, it is crucial to investigate the
gas and stellar content of galaxies. Hydrogen gas makes up the baryonic matter
that is channeling through large-scale structure filaments to haloes, cooling to neutral form in
galaxy disks and then to 
molecular gas to form stars. Observations of \HI\ are a direct way to trace
recent and ongoing interaction between galaxies and their environment 
due to its large extent beyond the stellar disk \citep{Hibbard96}. 
Interactions can occur via a number of physical processes, such as cold gas accretion with fresh gas
channeling from the intergalactic medium (IGM) directly into the galaxies, hot mode gas
accretion with gas being shock-heated to high halo virial temperatures and falling back to the galaxy centre as
it cools \citep{Katz03, Keres05}, tidal
stripping due to gravitational interaction \citep{Chung09, For14} and ram-pressure stripping when galaxies
pass through a dense IGM \citep{Kenney04, Abramson11}. These processes largely contribute to 
gas removal and accretion in galaxies, which in turn, affect their star formation efficiency \citep{CCS21}.

There is evidence that galaxies have undergone ``pre-processing'' in group environments before
falling into clusters \citep{ZM98, Mahajan13}. Thus, galaxy groups are the ideal place for following the
galaxy evolutionary path. In addition, galaxy groups are more common than clusters
\citep{Tully15}. 
Examples include the studies of the 
NGC~3783 group \citep{Kilborn06}, the Sculptor group \citep{Westmeier17}, NGC~2997 and NGC~6946 \citep{Pisano14}
using single-dish telescopes as well as the Hickson compact groups using the interferometer \citep{VM01}.  
Systematic surveys of a large sample of galaxy groups are not an easy task in part due to the
trade-off between angular resolution and sensitivity. Single-dish telescopes provide
excellent sensitivity but at arcminutes angular resolution. In order to
resolve gas disks and \HI\ debris, high angular resolution imaging with the use of interferometers is required. 
However, the required observing time is significantly more for an interferometer to reach the equivalent low \HI\ 
column density sensitivity of a single-dish telescope. In addition,
a traditional interferometer has a small field of view (FOV), which makes targeting clusters a better option 
for providing large samples than galaxy groups for the equivalent amount of observing time.

The Widefield ASKAP $L$-Band Legacy All-sky Blind SurveY (WALLABY, \citealp{Koribalski20}) makes use of the
large FOV capability of the Australian Square Kilometre Array Pathfinder (ASKAP; \citealp{Johnston07})
to image \HI\ galaxies out to a redshift $z\sim$0.26 and across 3$\pi$ steradian of the sky ($-$90\degr < $\delta$ < +30\degr).
The estimated root-mean-square (RMS) noise level is anticipated to be 1.6 mJy per beam per 18.5~kHz channel. The survey is expected to
detect 500,000 galaxies \citep{Koribalski20}. 
The main advantage of ASKAP over traditional radio interferometers is the use of a state-of-the-art phased array feed (PAF),
which allows the formation of multiple beams on the sky simutaneously to achieve a fast survey speed. 
For WALLABY, the targeted angular and spectral resolution of 30\arcsec\ and 4~\kms\ at $z=0$ are  
two to thirty times better than some of the single-dish surveys, most notably the \HI\ Parkes
All-Sky Survey (HIPASS; \citealp{Barnes01}) and the Arecibo Legacy Fast ALFA survey (ALFALFA; \citealp{G05}). 
WALLABY will revolutionize the \HI\ extragalactic field by providing a large statistical sample of galaxies at high resolution
across a wide range of environments from voids to groups and clusters.

To verify the observational feasibility of ASKAP and to improve the efficiency of the ASKAP data reduction pipeline, a WALLABY
early science program was carried out in 2018 using 12 to 16 ASKAP antennas. 
Studies of the NGC~7162 group \citep{Reynolds19}
and the LGG~351 group of Galaxies 351 \citep{For19} show that the majority of galaxies in these groups are
actively forming stars indicating inefficiency of gas removal processes in the loose group environment. 
\HI\ debris due to tidal interaction has also been identified in the NGC~7232 galaxy group, with one gas cloud 
postulated to be the progenitor of a long-lived tidal dwarf galaxy \citep{LW19}. Detailed studies of individual
galaxies, NGC~1566 \citep{Elagali19} and IC~5201 \citep{Kleiner19}, 
have also highlighted the importance of high resolution imaging. The resolved \HI\ kinematics allowed
them to recover the dark matter fractions by modelling the observed rotational curves and
to investigate the asymmetries of \HI\ morphology
of these galaxies.

\subsection{Eridanus Supergroup}

\begin{figure}
  \centering
  \includegraphics[width=\columnwidth]{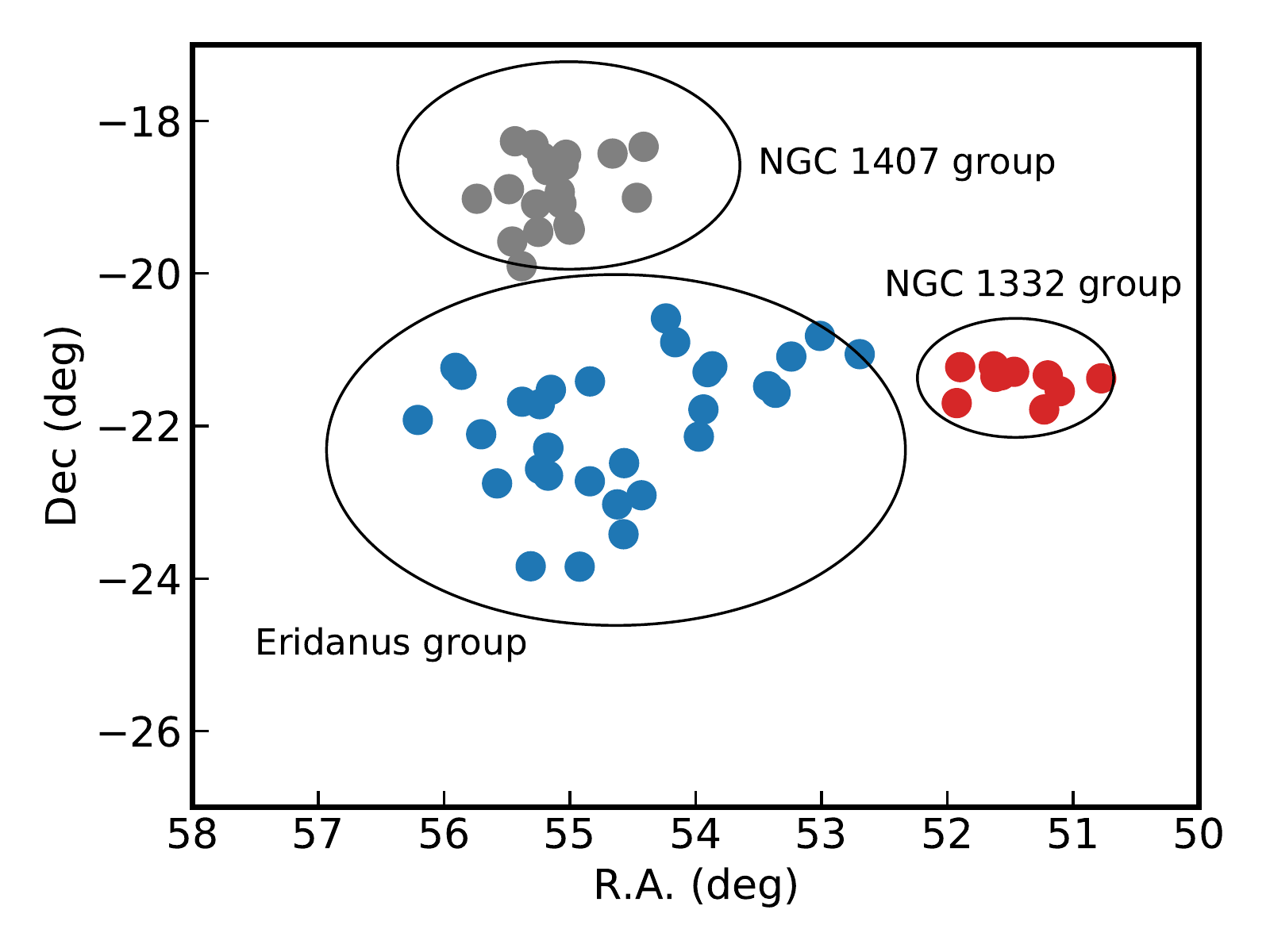}
  \caption{Distribution of galaxies in the Eridanus supergroup. Group members of NGC~1407, NGC~1332 and Eridanus
    groups as identified in \citet{B06} are represented by grey, red and blue dots, respectively.
    The ellipses indicate that maximum radial extent of the groups. 
    \label{groups}}
\end{figure}

Studies of the Eridanus concentration/region can be traced back to 1930s (see \citealp{Baker33}).
A later in-depth study by \citet{dv75} found that Group 31 along with galaxies associated with NGC~1332 and
NGC~1209 formed the ``Eridanus Cloud''. This cloud lies on the Eridanus-Fornax-Dorado filamentary structure
and is connected to the ``Great Wall'' feature at $\sim$4000~\kms\ in the background \citep{costa88, Willmer89}.
Several studies have shown the complex nature of the region. \citet{Willmer89} showed that
the region is made up of three or four subclumps, which are dynamically bound to each other and condensing to
form a cluster. On the other hand, \citet{OD05a} considered that the galaxies in the region as loose groups and
in an intermediate evolutionary stage between the Ursa-Major group and the Fornax cluster.
A re-analysis of this region
using the 6dF Galaxy Survey (6dFGS; \citealp{Jones04}) suggests that there are 3 distinct groups,
namely the NGC~1407, NGC~1332 and Eridanus groups (\citealp{B06}, hereafter B06).  
These groups also form part of the supergroup, which is defined as a group of groups that may eventually
merge to form a cluster. Figure~\ref{groups} shows the group members as identified in \citet{B06}. 

The Eridanus supergroup is chosen as a pre-pilot field
because it presents a different environment than the early
science fields, which allows us to probe the subcluster merging/pre-processing scenario.
It is thought to be an intermediate step of evolutionary path for clusters assembly,
similar to the Ursa Major Supergroup \citep{Wolfinger16}.
Our goals of this study are to investigate (1) the \HI\ content of galaxies
in the Eridanus field; (2) if the galaxies follow the typical \HI\ scaling relation; (3) if their star formation rates reveal any difference
in the merging/pre-processing environment, (4) if they are deficient in \HI\ and (5) if the \HI\ deficiency parameter correlates
with the projected distance from the group centre. 

This paper is structured as follows. 
In Section 2, we describe the previous \HI\ observations,
ASKAP observations, its data reduction and quality assessment. In Section 3, we describe the
source finding methodology, present the \HI\ spectra, moment maps, mask and S/N maps, as well as
compare the WALLABY integrated fluxes with
previous \HI\ studies. In Section 4, we re-define the Eridanus supergroup membership.
In Section 5, we derive physical parameters: recession velocity, distance, \HI\ linewidth, stellar mass,
\HI\ mass and \HI\ deficiency parameter. In Sections 6 and 7, we compare the derived quantities
with the \HI\ scaling relation and star forming main-sequence. In Sections 8 and 9, we show a 3-dimensional 
model of the environment the Eridanus supergroup resides in and a summary and conclusions. 
We also describe the \HI\ and optical
morphologies in Appendix A. 

Throughout the paper, we adopt a $\Lambda$ cold dark matter cosmology model ($\Lambda$CDM) with
$\Omega_{\rm M} = 0.27$, $\Omega_{\rm K} = 0$, $\Omega_{\rm \Lambda} = 0.73$ and $H_{\rm 0} = 73$ \kms\ Mpc$^{-1}$.
These are the default parameters for distances and cosmological corrected quantities in the 
NASA/IPAC Extragalactic (NED) interface \citep{Spergel07}.

\section{Data}

\subsection{Previous \HI\ Observations}

The Eridanus field has previously been observed as part of
the single-dish Parkes \HI\ blind basketweave (BW) survey, with a 
follow-up study using the Australia Telescope Compact Array (ATCA; \citealp{Waugh05})
and one using the Giant Metrewave Radio Telescope (GMRT; \citealp{OD05a, OD05b}).
The BW survey of the Eridanus field covered $\sim$100 deg$^{2}$ and 
is centred on $\alpha = 3^{\rm h}22^{\rm m}00^{\rm s}$ (J2000),
$\delta =-22$\degr00\arcmin00\arcsec.
The BW scanning technique was aimed at
achieving higher sensitivity than HIPASS, with RMS noise level
a factor of 2 better (i.e., $\sim$7~mJy). The Hanning-smoothed velocity resolution is 26.4~\kms.
The follow-up ATCA observations targeted 24 \HI\ sources with a median 
RMS of 6.5~mJy and a sythesized beam of $\sim$113\arcsec $\times$ 296\arcsec. Some of the \HI\ sources
remained unresolved.  
The GMRT also targeted the Eridanus supergroup with 46 scientifically usable observations out of the 57 observed targets.
The primary beam of the
GMRT observation is $\sim$24\arcmin. The GMRT final image cubes are
either low ($\sim$50\arcsec) or high resolution (25\arcsec\ or 30\arcsec) with a velocity resolution of
$\sim$13.4~\kms. The 3$\sigma$ \HI\ column density detection limit is $\sim1\times10^{20}$~cm$^{-2}$
for the GMRT high resolution images.

\subsection{ASKAP Observations}\label{obs}

ASKAP is a radio interferometer located in the remote outback of Western Australia
and is part of the Murchison Radio-astronomy Observatory (MRO). It consists of 36 $\times$ 12-m antennas,
with each antenna equipped with a second-generation MK II PAF.
ASKAP is designed to provide an instananeous large field of view (5.5\degr$\times$5.5\degr) with
36 dual-polarisation beams,
a bandwidth of 288~MHz as well as 
high angular and spectral resolutions \citep{McConnell16, Hotan21}. 
During the interim period between the ASKAP early science program and the pilot survey,
WALLABY carried out a pre-pilot survey targeting the Eridanus field in March 2019.
The survey attempted to utilise the full 36 antennas for the two interleaving footprints
(footprints A and B)
in contrast to the 12 to 16 antennas and the multiple observations per footprint for the early science program.

\begin{figure}
  \includegraphics[width=\columnwidth]{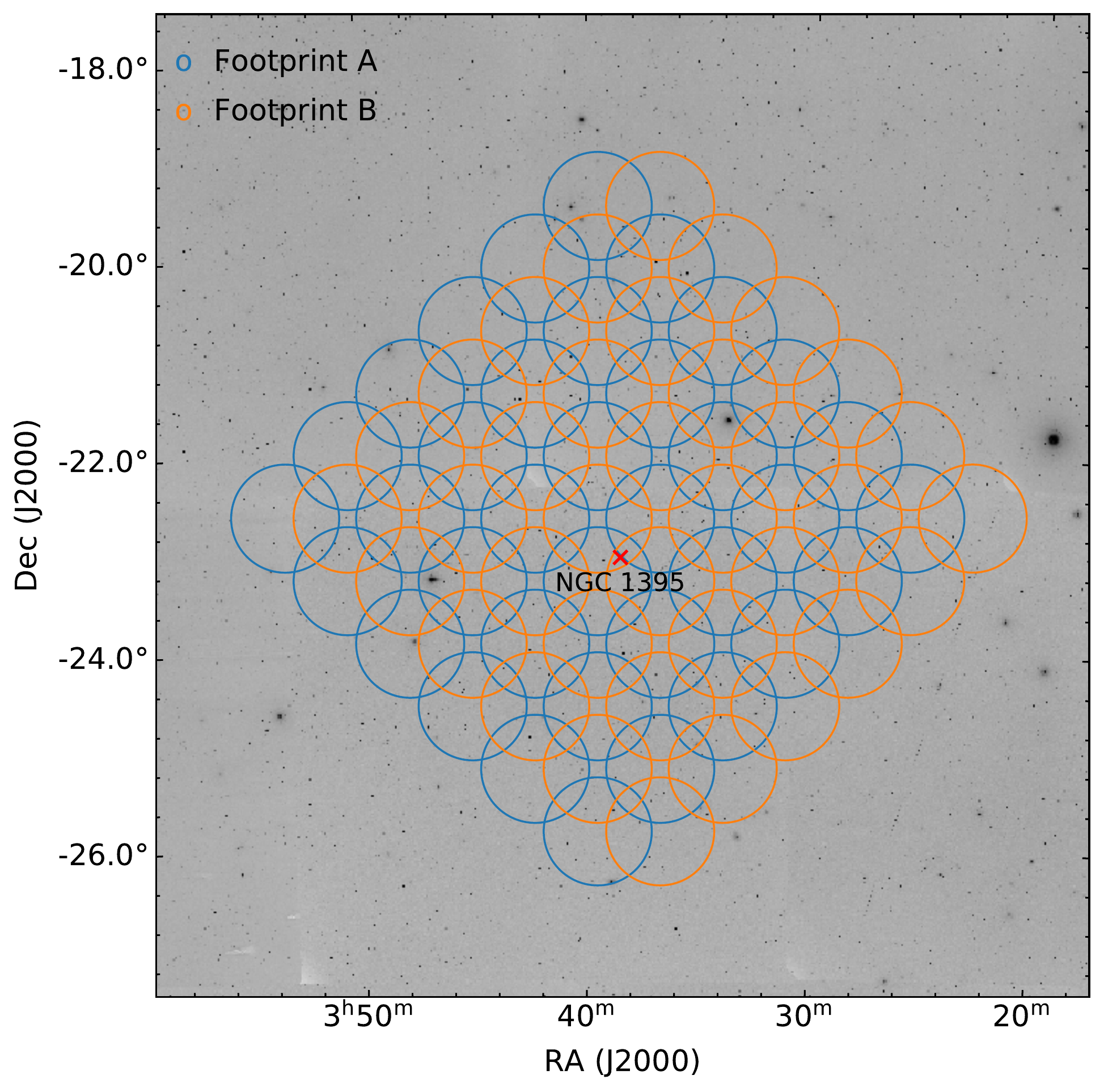}
  \caption{ASKAP footprints for the observations as listed in Table~\ref{obslog} overlaid on to the optical DSS-2 red image.
    The observed footprints
    have been rotated by position angle of 45\degr\ relative to a standard ASKAP footprint. Each beam has an FWHM of 1\degr.
    The blue and orange circles represent footprint A and B, respectively. 
    The brightest elliptical galaxy, NGC 1395, in the Eridanus supergroup is marked with cross.
    \label{obs_footprints}}
\end{figure}

Each footprint has a 6$\times$6 beam pattern and was rotated by 45\degr\ on the sky. 
These observations were carried out during the day for
footprint A and mostly at night time for footprint B. The primary calibrator, PKS~1934$-$638, was
observed for about 2~h at the beginning of each observation. 
Each observation is given a scheduling block identification number (SBID\footnote{The SBID can be used
to search for the corresponding data set in CSIRO ASKAP Science Data Archive (CASDA).}) and
the total integration time for both footprints is 10.8 h. 
We present the observing log in Table~\ref{obslog}
and the on-sky positions of interleaved footprints in Figure~\ref{obs_footprints}. 

\begin{table*}
  \centering
  \begin{minipage}{200mm}
\caption{ASKAP Observations log of the Eridanus Field.}
\label{obslog}
\begin{tabular}{ccccccccc} %
  \hline
    UT date     &  Footprint &   R.A. & Decl.  & Calibrator SBID & Science SBID & Integration time &       Bandwidth     &       Central frequency      \\
  (yyyy-mm-dd)  & & (h:m:s) & (\degr~\arcmin~\arcsec)   &     &         & (h)  &       (MHz)   &       (MHz)    \\
    \hline
    2019-03-13 & A & 3:39:30 & $-$22:30:00 & 8169 & 8168 & 5.8 & 288 & 1295.5 \\
    2019-03-13 & B & 3:36:44.52 & $-22$:37:54.69 & 8169 & 8170 & 5.0 & 288 & 1295.5 \\
    \hline
\end{tabular}
\end{minipage}
\end{table*}

\subsection{Data Reduction and Quality Assessment}\label{datared}

The data are processed automatically using \askapsoft\ version 0.24.7 \citep{Whiting20,WRO20}
installed on the ``Galaxy'' supercomputer at the Pawsey Supercomputing Centre.
Only baselines shorter than 2~km and half of the full 288~MHz bandwidth (i.e., 1295.5--1439.5~MHz) are processed.
This is a huge improvement in data processing capability as compared to the early science program,
in which only a small frequency range and few beams were processed at a given time (see e.g., \citealp{For19}). 

All 36 beams are used for bandpass calibration, and the autocorrelation of each beam measurement set
is flagged out completely. Radio frequency interference (RFI) and antenna flagging are performed on a beam-by-beam basis.
The overall flagged visibility fraction ranges from 10 to 30 percent across all beams and
the mininum number of utilised antennas per beam is 31.
Subsequently, the derived bandpass solution is applied to each beam. 
Flux density calibration in each frequency channel is performed using PKS~B1934$-$638
and gain calibration is performed via self-calibration. 
Continuum images of each beam are used as models to subtract continuum sources in the $uv$-domain.
For the final spectral line imaging, a pixel size of 6\arcsec\ is adopted. 
We use Wiener filtering with a robustness of 0.5 in addition to Gaussian tapering to a resolution of $\sim$30\arcsec.
Multi-scale \clean\ is used to deconvolve the image \citep{Cornwell08}. 
Once the spectral line image cube per beam is created,
image based continuum subtraction of the residual continuum emission is performed by
fitting a low-order polynomial to each spectrum. Finally, 
we create a primary beam corrected beam-by-beam mosaic  
image cube. 
Channels in beams that are severely affected by artefacts have been masked out in the image cube.
A particular type of artefact
in the form of large-scale stripes across the beam is found to have been 
caused by deconvolution failing during the imaging step.
The origin of the divergence comes from over-flagging of channels that contain Galactic emission. 

We evaluate the data quality of each footprint image cube based on a set of metrics.
These metrics are established based on the data in the WALLABY early science field of M~83 \citep{For19},
and include RMS, minimum and maximum flux densities,
1 percentile noise level and median absolute deviation of median flux (MADMF).  
Two sets of these values are calculated and recorded during the pipeline processing.
Each set consists of values for three types of image cubes, i.e., before and after continuum subtraction image cubes as well as a residual image cube. 
The first and second sets give values for each channel of the mosaicked image cubes and for each channel of each beam, respectively. 
The RMS values of the footprint A and B mosaics increase as a function of frequency (5--6~mJy), which is consistent with the theoretical RMS trend.
The RMS values of combined mosaic across all beams vary from 2.4 to 4.4~mJy with central beams having the lowest RMS noise level. 

\begin{figure}
  \includegraphics[width=\columnwidth]{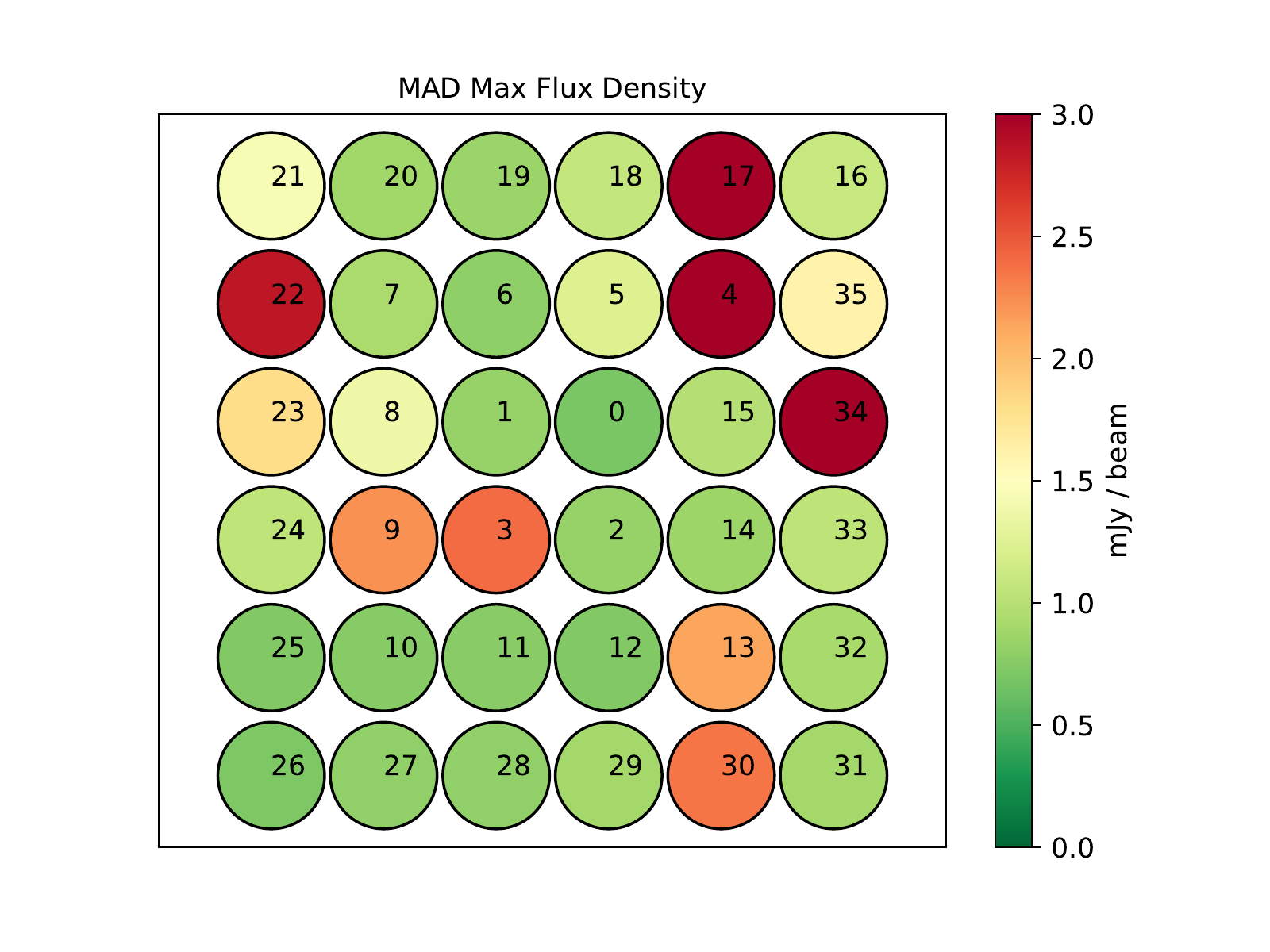}
  \caption{Median absolute deviation of maximum flux density (MADMFD) for each beam over frequency
    range 1295.5--1439.48~MHz for footprint A.
    Beam numbers for the ASKAP 6\degr$ \times $6\degr\ footprint are labeled. 
    \label{madmfd}}
\end{figure}

To evaluate the effect of broadband RFI/artefacts, we examine the
median absolute deviation of maximum flux density (MADMFD) of each beam. This metric is sensitive
to strong artefacts. In Figure~\ref{madmfd}, we show 
the variation of MADMFD, and the largest values correspond to the beams with artefacts. We also examine
the distribution of flux density values for all voxels in each beam at the 1 percentile level. Excessive negative
flux density is shown to be related to bandpass calibration and/or sidelobe issues. We fit each distribution with a Gaussian and
use the variance as a metric to determine if beams have been affected by those issues. The metric shows a similar
result to that of MADMFD for this data set. 

A stand-alone script\footnote{Available at \url{https://github.com/askap-qc/validation/blob/master/wallaby\textunderscore hi\textunderscore val.py}}   
is used to generate an HTML style summary report for each SBID. 
This report provides basic observation information, statistical plots generated from the pipeline,
plots for visibilities and antenna flagging statistics etc. 
The report of each footprint is included along with the data release on CASDA.

\section{Source Finding and Cataloging}\label{sf}

We use the Source Finding Application (\sofia\footnote{Available at \url{https://github.com/SoFiA-Admin/SoFiA}}; \citealp{Serra15})
version 2.0 \citep{Westmeier21} to search for \HI\ sources in the mosaicked cube from $cz$ = 500 to 8500~\kms.
Channels that contain the Milky Way emission were excluded.  
\sofia\ was set to automatically flag some known artefacts, such as
continuum residuals or RFI, prior to the search.
We scale the data by the local RMS noise level and apply a threshold of 3.5$\sigma$
for the smooth + clip source detection algorithm. A radius of 2
pixels and 2 spectral channels are used to merge the detected voxels into objects. 
\sofia\ generates a cubelet, an associated mask cube,
a signal map with the total number of \HI\ detected channels for each pixel, a spectrum, integrated \HI\ intensity (0th moment),
velocity field (1st moment) and velocity dispersion (2nd moment) maps for each detected source.
To filter out false positive sources, \sofia\ has implemented a reliability filter.
The filter works by comparing the density of detections with positive
  and negative flux in parameter space to estimate the reliability of all positive detections
  under the principal assumption that astronomical sources must have positive flux,
  while the stochastic noise and any artefacts present in the data are symmetric about zero.
  A user-defined reliability threshold can then be applied to discard unreliable detections.
  We refer the reader to \citet{Serra12} for a detailed description of the reliability filter.

\begin{figure*}
  \includegraphics[scale=1.3]{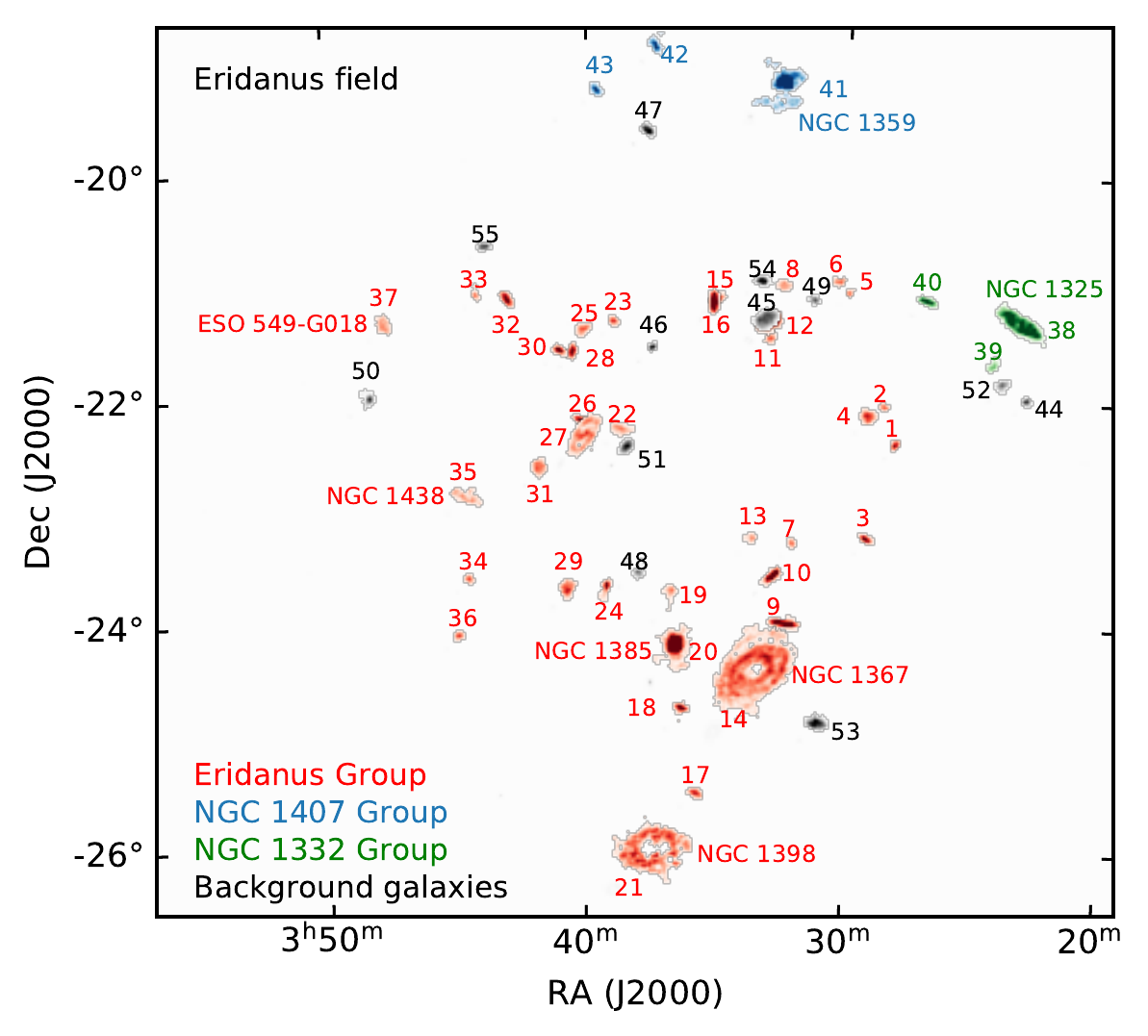}
  \caption{Integrated \HI\
    intensity map of individual sources in the Eridanus Supergroup and in the background. 
    Sources of the Eridanus group,
    the NGC 1407 group, the NGC 1332 group and background galaxies are in red, blue, green and black, respectively.
    All sources are enlarged by a factor of 4.5 for clarity.
    Designated ID number for all 55 \HI\ detected sources are labelled.}
    \label{onsky}
\end{figure*}

The final catalogue consists of 55 \HI\ sources including two new \HI\ sources 
(WALLABY~J033911-222322 and WALLABY J033723-235753; see \citealp{Wong21})
that do not have optical counterparts in DR8 DESI Legacy Imaging Survey.
The membership of each source is discussed in Section~\ref{member}. 
In Figure~\ref{onsky}, we present the integrated \HI\ intensity map of
individual detected sources enlarged by a factor of 4.5 in the Eridanus supergroup and 
the full mosaic with all sources labelled by
their designated
catalogue identification number (ID) (refer to Table~\ref{catalogue1}). Within the WALLABY Eridanus FOV,
38 galaxies were also observed by the GMRT. Among those, 22 are detected by both WALLABY and GMRT.
The remaining 16 non-detections are detected by GMRT but not by ASKAP. 
As a blind survey, WALLABY recovers an additional 19 \HI\ sources that were not targeted in the GMRT study.
The properties and parameters of each source are given in Tables~\ref{catalogue1} and \ref{catalogue2}. 

\subsection{\HI\ intensity, velocity field, S/N maps and \HI\ spectra}

\begin{figure*}
  \includegraphics[scale=0.22]{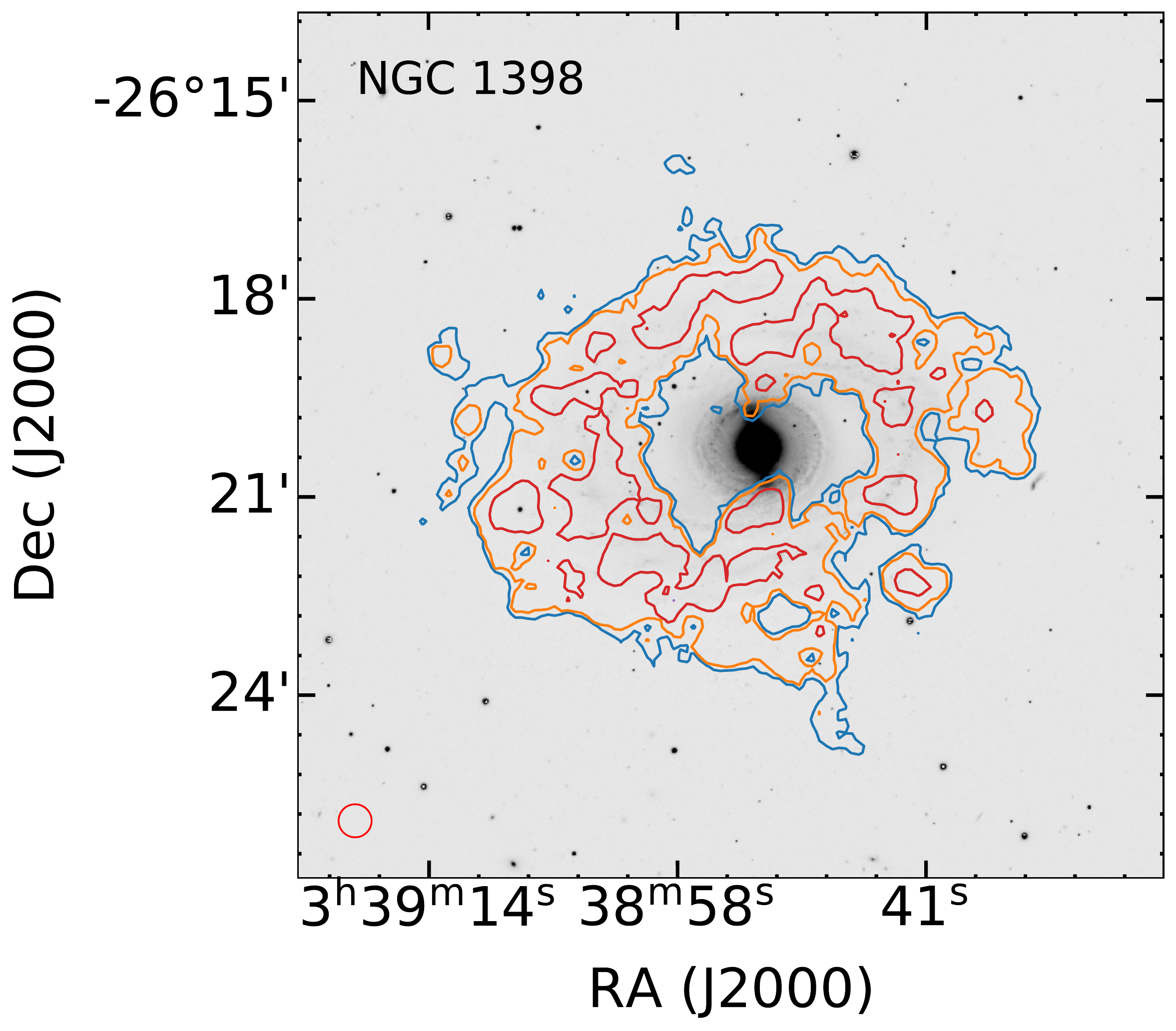}
  \includegraphics[scale=0.22]{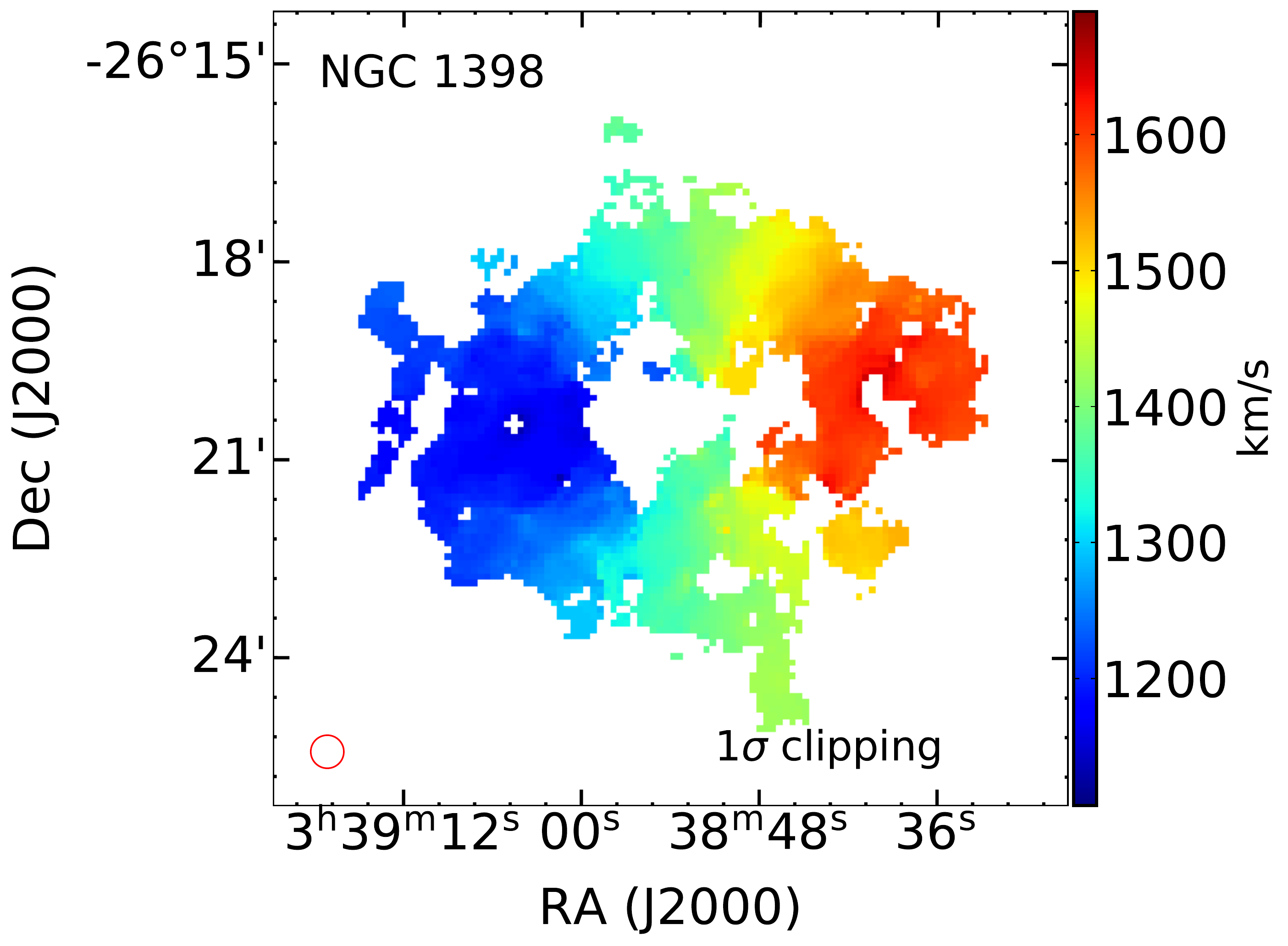}
  \raisebox{.1\height}{\includegraphics[scale=0.39]{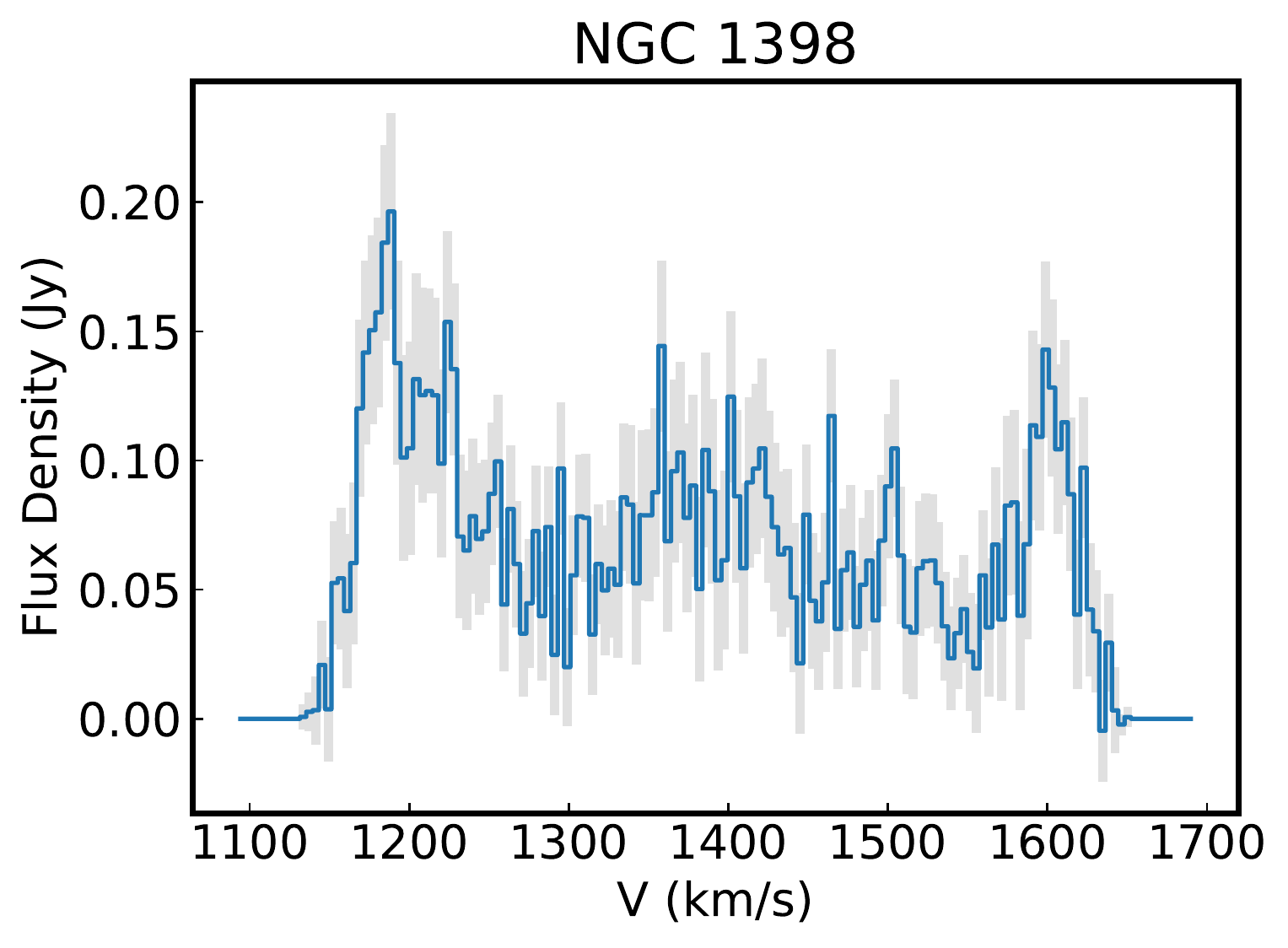}}
  \includegraphics[scale=0.22]{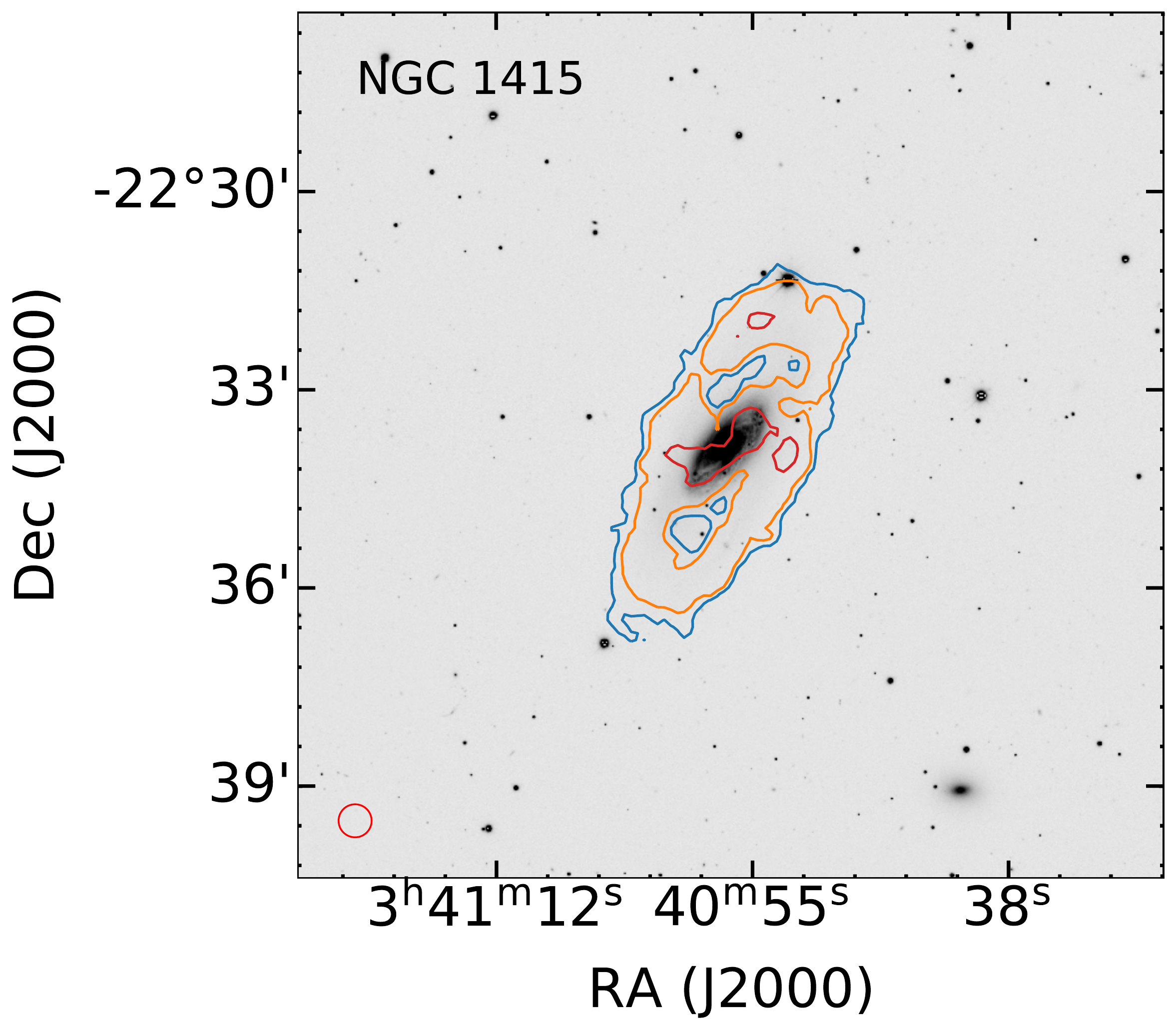}
  \includegraphics[scale=0.22]{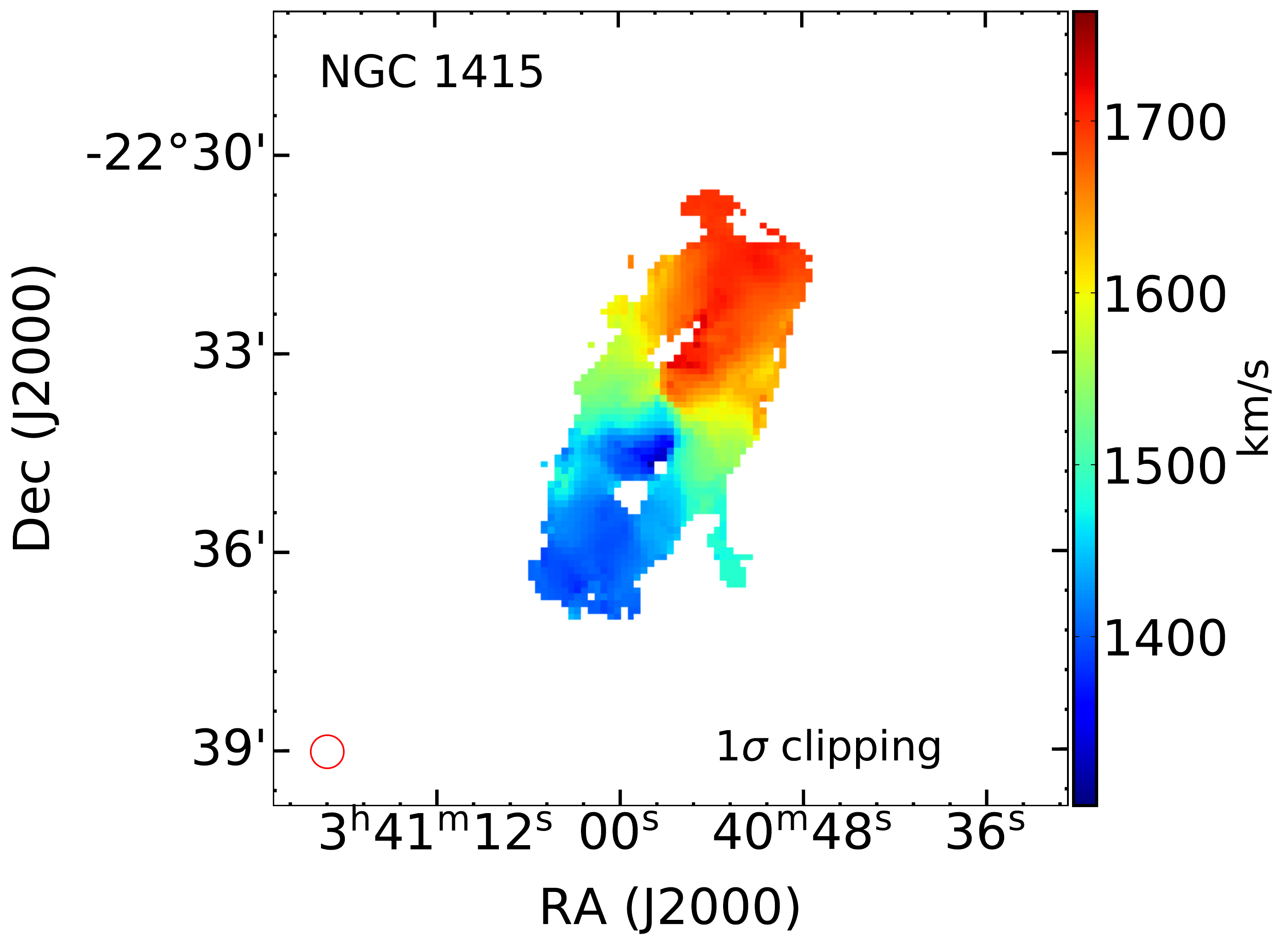}
  \raisebox{.1\height}{\includegraphics[scale=0.38]{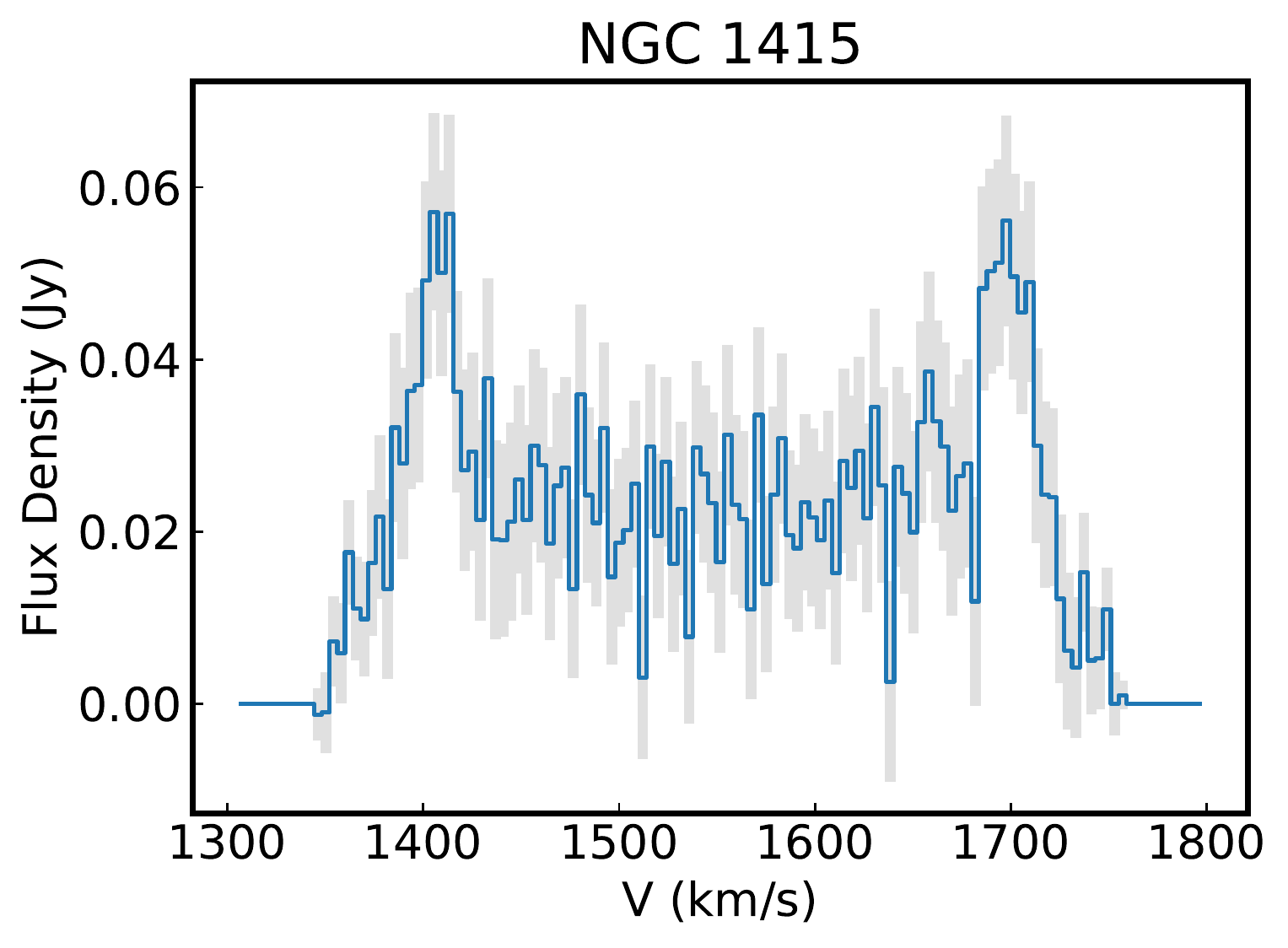}}
  \caption{Examples of WALLABY \HI\ sources. 
    A mean correction factor of 20$\%$ has been applied to
    integrated \HI\ column density maps, flux density and local RMS of spectra.
    The synthesised beam of 30\arcsec$\times$30\arcsec\
    is plotted at the left bottom corner of each sub-plot as a reference.
    \textit{Left:} Integrated \HI\ column density maps of individual sources overlaid onto the DR8 DESI Legacy Imaging Survey $g$-band stacked images.
    The colour scale of blue, orange, red, purple, and olive green represent the contour levels of
    1, 2, 5, 10, 20$\times10^{20}$ cm$^{-2}$, respectively. The 3$\sigma$ of \HI\ column density sensitivity level for
    a typical RMS of 1.7~mJy beam$^{-1}$ over 20~\kms\ is 1.2$\times10^{20}$ cm$^{-2}$.
    \textit{Middle:} 1st moment velocity field maps of individual sources.
    Pixels above $n-\sigma$ are plotted and the value for $n$ is given at bottom of each sub-plot.   
    \textit{Right:} Spectra of individual sources derived by \sofia. Grey area represents the upper and lower limit of uncertainties, 
    $\sigma_{\rm RMS}\times\sqrt{N_{\rm pix}/A_{\rm beam}}$, where the Gaussian beam area, $A_{\rm beam} = \pi\theta_{\rm bmaj}\theta_{\rm bmin}/(4\ln2)$.
    {\it Note}: This figure is published in its entirety as Supporting Information with the electronic version of the paper. A portion is shown here. 
}
    \label{combine}
\end{figure*}

We convert the \HI\ intensity maps (Jy Hz) from \sofia\ to integrated \HI\ column density ($N_{\rm HI}$) maps 
by using Eq. 76 in \citet{Meyer17}. We also create 1$\sigma$ flux density sensitivity maps by using the channel cubes from \sofia. 
Subsequently, we divide each integrated \HI\ column density map by its corresponding
1$\sigma$ $N_{\rm HI}$ sensitivity map to obtain a signal-to-noise ratio (S/N) map.
The S/N ratio maps are used to clip the velocity field maps, in which pixels below a certain $\sigma$ are masked out.
The spectrum of each source is created by integrating the flux densities of all spatial pixels in each spectral channel (provided
by \sofia). 
Figure~\ref{combine} shows the $N_{\rm HI}$ contours overlaid onto the DR8 DESI Legacy Imaging Survey
$g$-band co-added image\footnote{Co-added cutout images are available at \url{https://datalab.noao.edu/ls/dataAccess.php}}
(refer to section~\ref{smass} for detail),  
clipped velocity field maps and spectra in $V$ = c$z$, where c is the speed of light and $z$ is the \HI\ derived redshift.
Most of the
detected galaxies are marginally resolved. Using these maps and images, we describe the \HI\ and optical morphologies
of individual galaxies in Appendix A. 

\subsection{\HI\ integrated flux} \label{intflux}

\begin{figure}
  \includegraphics[width=\columnwidth]{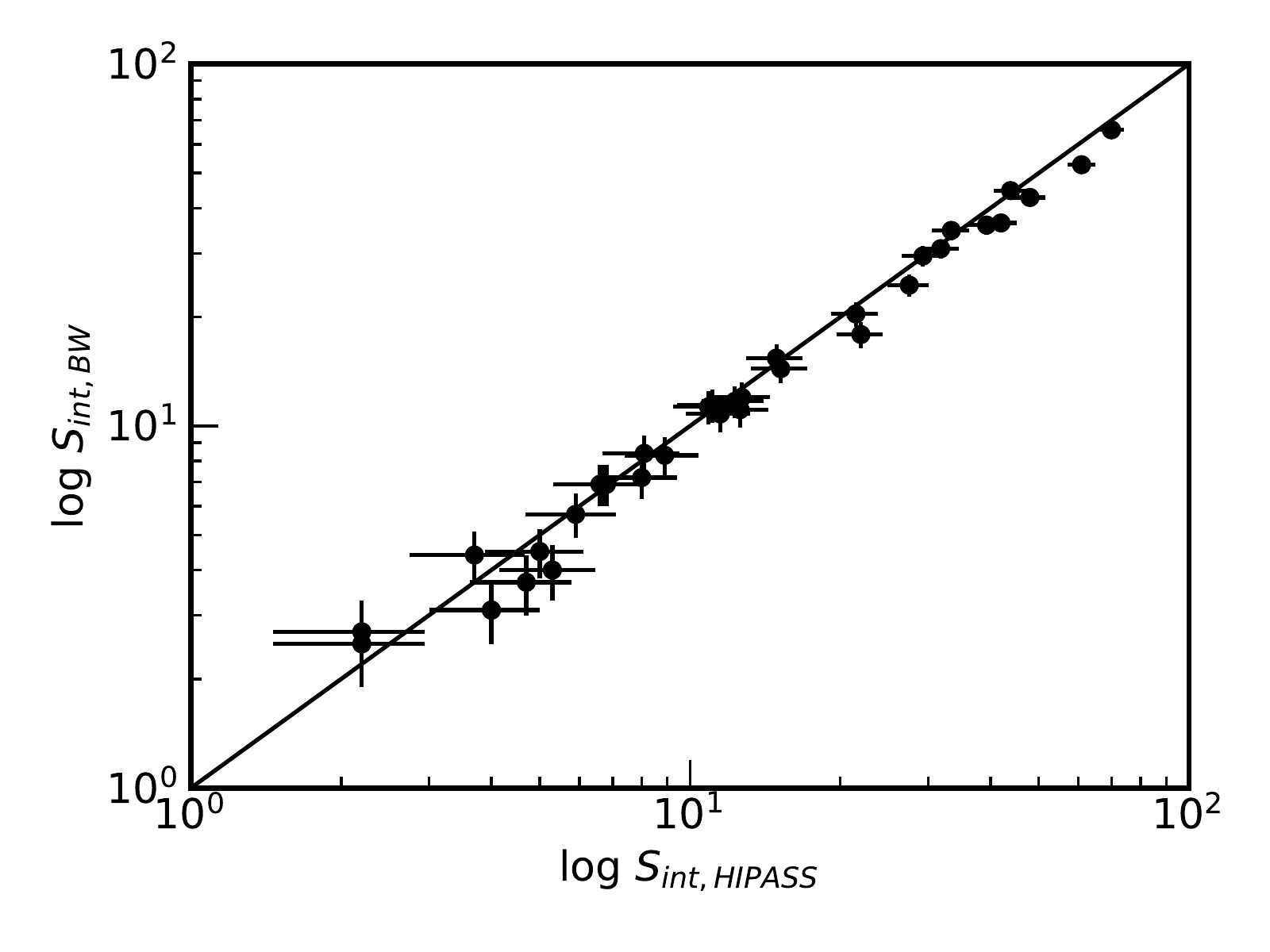}
  \includegraphics[width=\columnwidth]{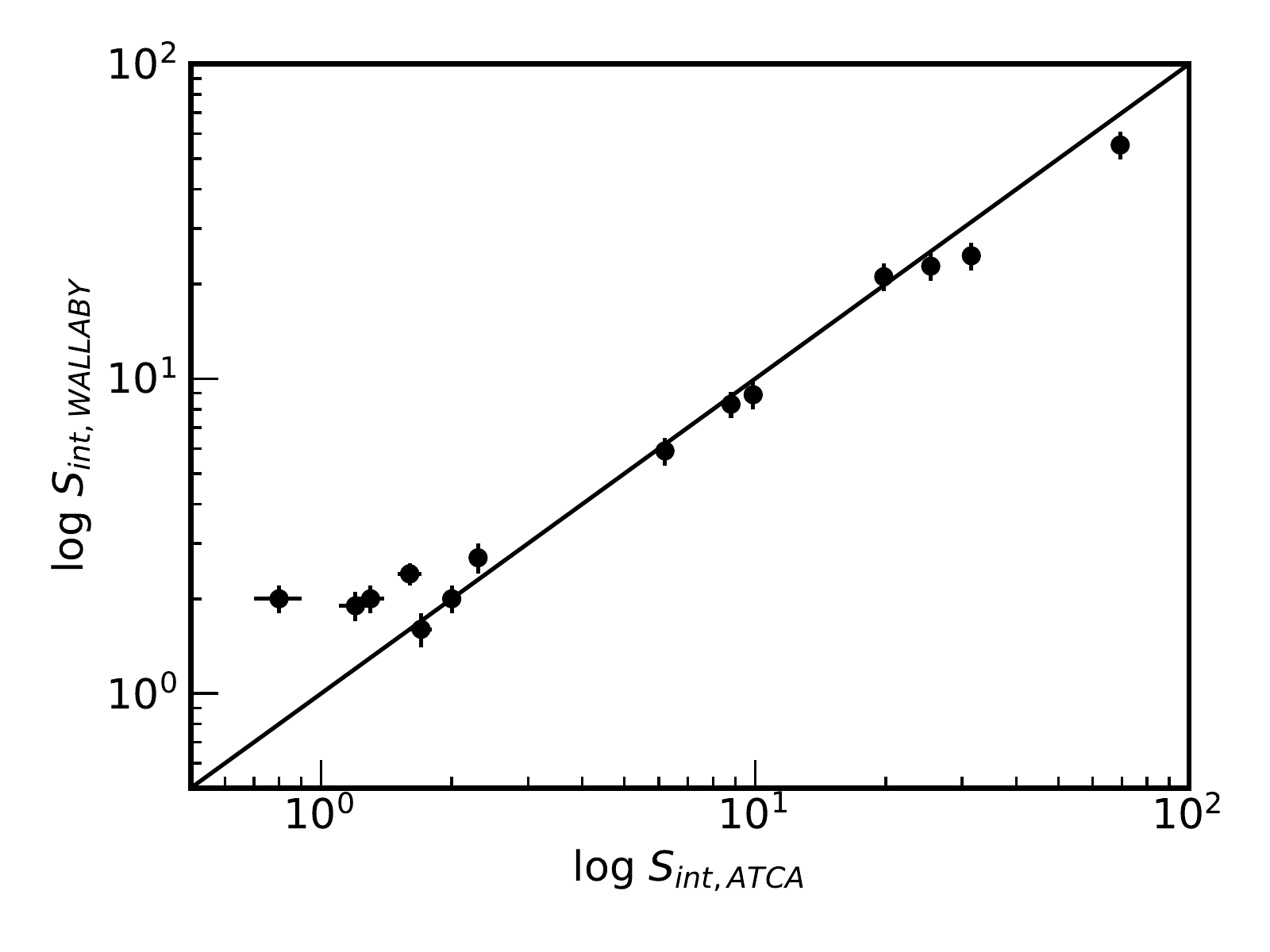}
  \includegraphics[width=\columnwidth]{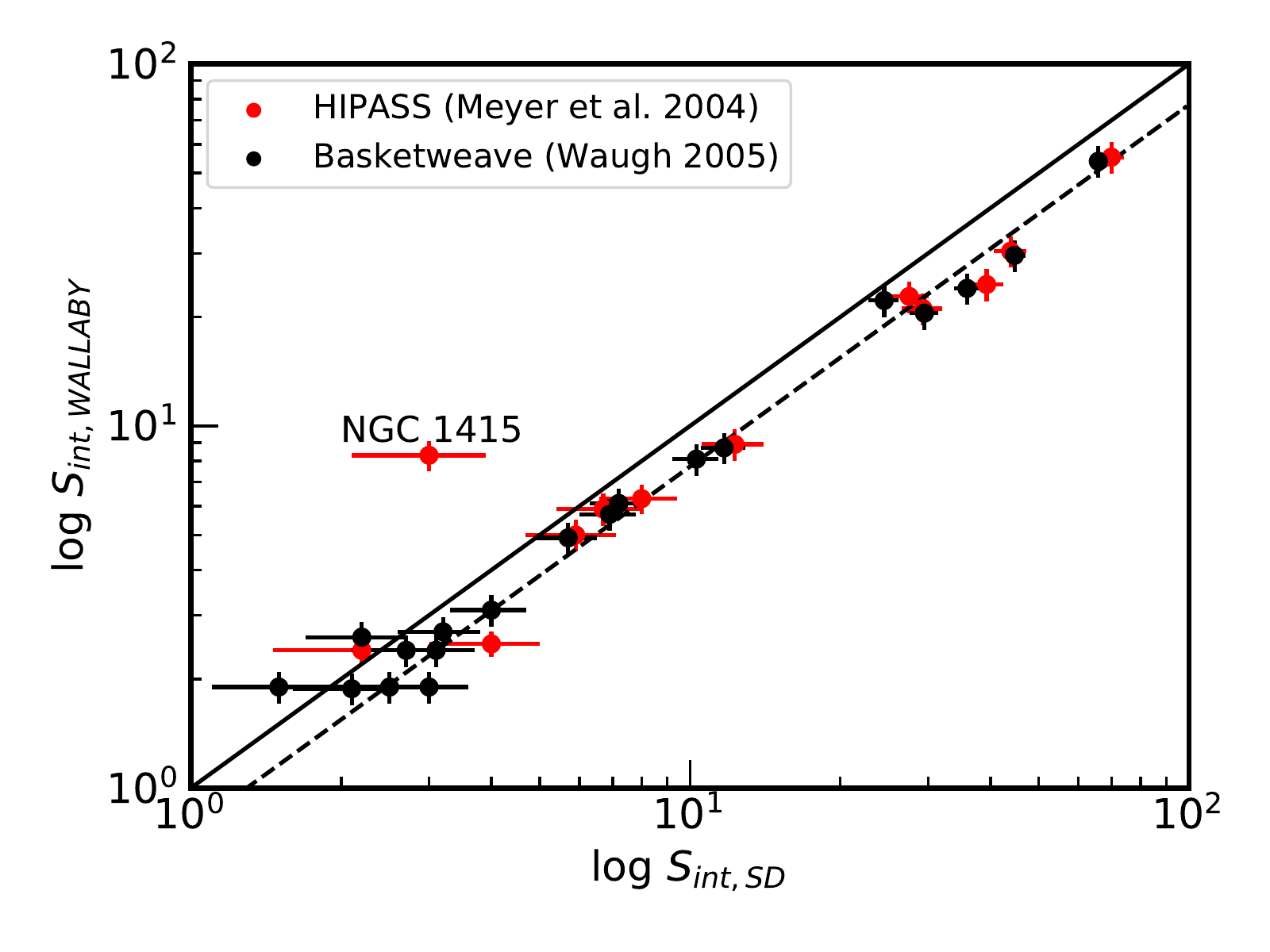}
  \caption{Comparison of $S_{\rm int}$ between various studies. The solid line is the 1:1 unity line.
    We adopt $10\%$ of $S_{\rm int}$ as errors for WALLABY data points as the statistical uncertainties as listed in Table~\ref{catalogue1} do not take into account calibration uncertainties. 
    {\it Top}: Comparison of $S_{\rm int}$ between BW \citep{Waugh05} and HIPASS \citep{Meyer04}.
    {\it Middle}: Comparison of $S_{\rm int}$ between WALLABY pre-pilot and ATCA \citep{Waugh05}.
    {\it Bottom}: Comparison of $S_{\rm int}$ between WALLABY pre-pilot and single-dish data. The black and red dots represent the $S_{\rm int}$ from BW and HIPASS,
    respectively. The dashed line is the fitted line, indicating a $\sim$20\% systematic offset between the WALLABY and single-dish data.
  }
  \label{flux_comp}
\end{figure}

To obtain the integrated flux ($S_{\rm int}$) in Jy Hz, we sum the flux densities across all channels 
within the mask. 
To verify the measured $S_{\rm int}$ from WALLABY and for direct comparison with previous studies, 
we convert the values from Jy Hz to Jy~\kms. 
In Figure~\ref{flux_comp}, we show 
the $S_{\rm int}$ comparison between BW and HIPASS \citep{Meyer04} catalogues (top panel),
the $S_{\rm int}$ comparison between WALLABY and ATCA (middle panel), 
and the $S_{\rm int}$ comparison between WALLABY and the single-dish studies (bottom panel).
The solid line represents the unity line.
Overall, the $S_{\rm int}$ from two Parkes surveys are consistent with each other.
For the interferometers comparison, there are outliers at the fainter end (< 10~Jy~\kms).
The $S_{\rm int}$ values of fainter sources from ATCA are about a factor of two smaller than WALLABY's.
We find that the $S_{\rm int}$ values of these outliers are also a factor of two lower than BW, which suggests
that the issue is not related to WALLABY. However, there does appear to be some missing
flux in the ASKAP/ATCA interferometer results relative to both Parkes studies. 

For the outlier, NGC~1415, a clear systematic offset is seen between the WALLABY and single-dish data.
For NGC~1415, we verify that HIPASS has only detected half of the galaxy. The missing fluxes for
some of the bright WALLABY sources can be explained. For example, NGC~1359 is an interaction galaxy pair, which 
WALLABY has resolved into tidal debris (see Figure~\ref{combine}).
The diffuse emission is likely resolved out in this case. 
Both NGC~1398 and NGC~1359 are located in edge beams,
where the noise level is significantly higher than in central beams. 
NGC~1367/1371 is a large galaxy 
with an 
\HI\ diameter ($d_{\rm HI}$) as observed by the Green Bank Telescope (GBT) of 22.3\arcmin \citep{S19}.
Our 0th-moment map shows a
$d_{\rm HI}$ of 9.9\arcmin, which indicates that our observation does
not recover the extended \HI\ disc of this galaxy. We cannot make a comparison with the GMRT study
because they also lost flux for large galaxies due to inadequate sampling of short spacing data.

We further investigate the missing flux issue by comparing
the non-confused continuum sources with the NRAO VLA Sky Survey (NVSS; \citealp{Condon98}).
We find that our sources are systematically 6$\%$ lower in flux density than NVSS.
However, the above issues (flux density calibration, sensitivity and zero spacing) 
cannot fully 
account for the offset between WALLABY and single-dish data.
Results from the Rapid ASKAP Continuum Survey (RACS; \citealp{McConnell20}) have shown position-dependent
variations of flux density due to inappropriate primary beam corrections. 
To circumvent this issue,
holography beam measurements will be used as part of the instrumental flux calibration
for future observations. 

To estimate the systematic offset in \HI\ fluxes, we compare the WALLABY data with all the single-dish data 
and fit 
an orthogonal distance regression (ODR) model.
Both WALLABY and single-dish data errors are weighted for the ODR model (dashed line). 
We find a $\sim$20$\%$ mean difference. 
For statistical purposes, we use this mean correction factor for the derived parameters for all sources 
for the rest of the paper. We also list the uncorrected $S_{\rm int}$. 

\section{Group Membership}\label{member}

By examining the older (\citealp{Garcia93a, Garcia93b}; hereafter G93 for both)
and newer (\citealp{Tully15,Tully16}; hereafter T15 and TCS16, respectively)
galaxy group catalogues, we find that several galaxies are not included as part of the groups in B06.
This is in part due to different methods and constraints being adopted 
for sample selection. G93 used a combination of friends-of-friends and Materne-Tully \citep{Materne78, Tully80} methods 
to identify galaxy groups based on a galaxy sample with a $B$-band magnitude cutoff of
14.0 in the Lyon-Meudon Extragalactic Database (LEDA; \citealp{Paturel88}).
T15 defined the galaxy group differently based on
Two Micron All Sky Survey (2MASS) data and used scaling relations as additional selection parameters.
There are caveats for all of these methods and selection criteria.

\begin{figure}
  \includegraphics[width=\columnwidth]{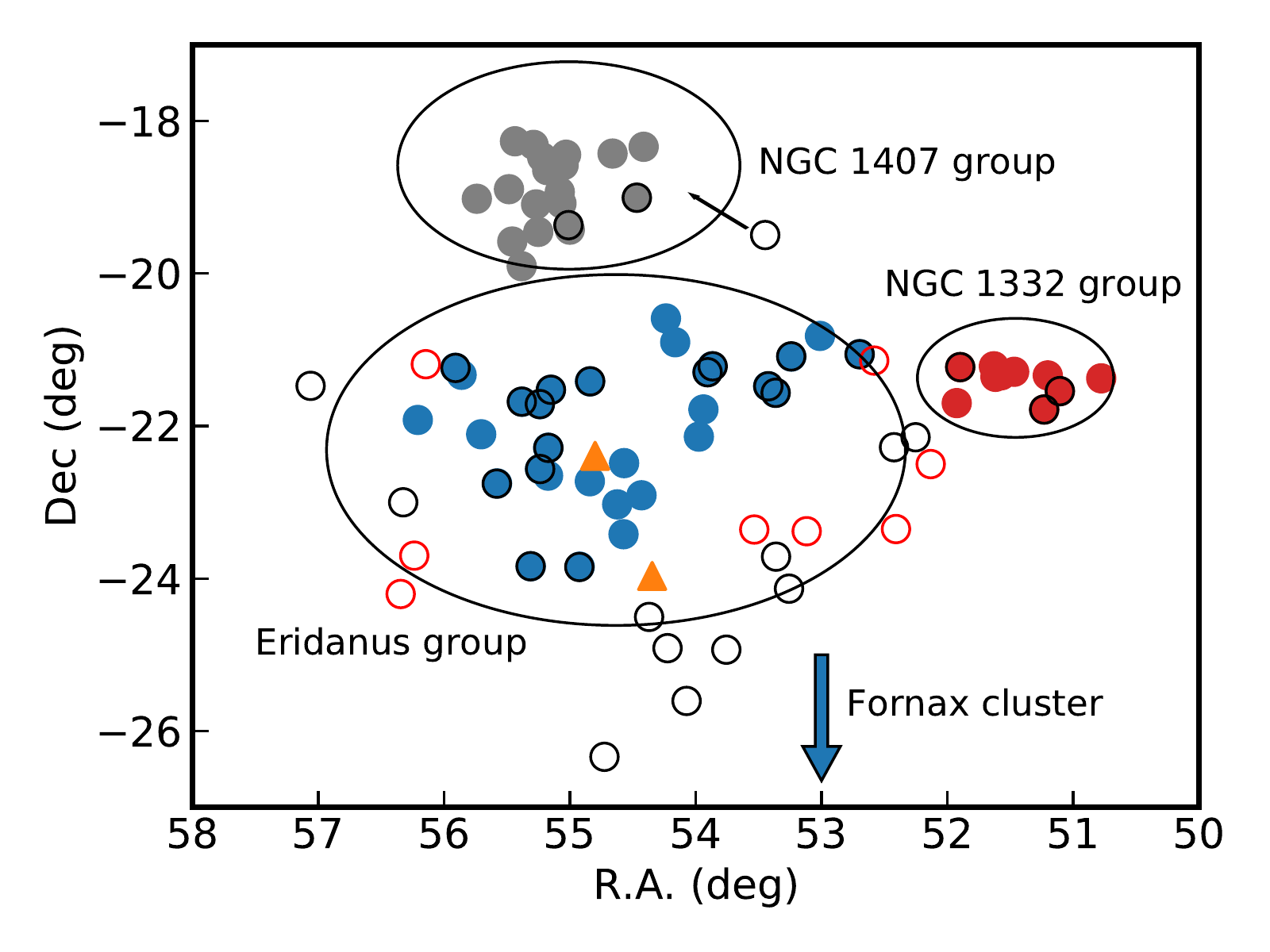}
  \caption{Same as Figure~\ref{groups} except adding detected \HI\ sources in this study. 
    Both red and black edge colour
    circles (including those filled) represent the detected \HI\ galaxies. 
    Two \HI\ clouds are shown in orange triangles.}
    \label{groups_dist}
\end{figure}

In Figure~\ref{groups_dist}, we show the galaxy members of the NGC~1407 (grey), Eridanus (blue) and NGC~1332 (red) groups as identified in B06. 
The ellipses represent the maximum radial extent of the groups based on 
$r_{\rm 500}$ (radius that encompasses overdensity of 500 times the critical density) in Table~3 of B06.
The encompassed virial masses
are 7.9, 2.1 and 1.4 $\times 10^{13}$ \msun\ for NGC~1407, Eridanus and NGC~1332 groups, respectively. 
The locations of two detected \HI\ clouds are shown as orange triangles.  
The black and red edgecolor circles represent the \HI\ detected galaxies from this study.
The former have been identified as group members
in B06, G93 and/or T15 catalogues and the latter do not have a membership confirmation. 
These eight (open red circles) galaxies are relatively faint ($r > 15$~mag, see Table~\ref{catalogue2}) and
do not have 2MASS photometry. 
Thus, it is likely that these eight galaxies were missed in the T15 catalogue.
To assess if these eight galaxies are associated with the Eridanus group,
we calculate the velocity difference between each galaxy and the mean central velocity of the Eridanus group
(1492~\kms) and then compare it to 3 times the velocity dispersion of 228~\kms (refer to T15 and TCS16).
With this criterion, these eight galaxies are considered as part of the Eridanus group.
We also consider detected \HI\ galaxies in either B06, G93 and/or T15 as part of their associated groups. 
Table~\ref{membership} summarises the group membership identified in different catalogues and this study. 

\begin{table*}
\centering
\begin{minipage}{130mm}
  \caption{Group membership.}
\label{membership}
\begin{tabular}{lccccc}
  \hline
Designation	&	$V$	&	B06	&	G93	&	T15/TCS16 & This Stutdy	\\
	        &	(\kms)	        &		&		&          	&	\\
(1)       	&	(2)	&	(3)	&	(4)	&	(5)       & (6)	\\
\hline
\multicolumn{2}{c}{}&
\multicolumn{4}{c}{Group Identification} \\
\hline
WALLABY J034002$-$192200	&       1216	&       NGC 1407	& \ldots        & \ldots	& NGC 1407 \\
WALLABY	J033752$-$190024	&	1218	&	NGC 1407	& \ldots	& \ldots	& NGC 1407 \\
WALLABY	J033019$-$210832	&	1226	&	\ldots	&	\ldots	&	\ldots	& Eridanus \\
WALLABY	J033408$-$232125	&	1262	&	\ldots	&	\ldots	&	\ldots	& Eridanus \\
WALLABY	J033047$-$210333	&	1292	&	Eridanus 	&	\ldots	&	\ldots	& Eridanus \\
WALLABY	J033854$-$262013	&	1373	&	X 6dF	&	\ldots	&	TSK 849, Nest 200100	& Eridanus \\
WALLABY	J032455$-$214701	&	1456	&	NGC 1332	&	\ldots	&	\ldots	& NGC 1332 \\
WALLABY	J033501$-$245556	&	1458	&	X 6dF	&	\ldots	&	TSK 849, Nest 200100	& Eridanus\\
WALLABY	J033723$-$235753$^{*}$	&	1469	&	\ldots	&	\ldots	&	\ldots	& Eridanus \\
WALLABY	J033728$-$243010	&	1497	&	X       	&	LGG 97	&	TSK 849, Nest 200100	& Eridanus \\
WALLABY	J033327$-$213352	&	1509	&	Eridanus 	&	\ldots	&	\ldots	& Eridanus \\
WALLABY	J033527$-$211302	&	1518	&	Eridanus 	&	\ldots	&	\ldots	& Eridanus \\
WALLABY	J034517$-$230001	&	1546	&	X       	&	LGG 97	&	TSK 849, Nest 200100	& Eridanus \\
WALLABY	J034056$-$223350	&	1552	&	Eridanus 	&	\ldots	&	TSK 849, Nest 200100	& Eridanus \\
WALLABY	J034219$-$224520	&	1569	&	Eridanus 	&	\ldots	&	\ldots	& Eridanus \\
WALLABY	J034434$-$211123	&	1578	&	\ldots	&	\ldots	&	\ldots	& Eridanus \\
WALLABY	J034814$-$212824	&	1586	&	X 6dF	&	LGG 97	&	TSK 849	& Eridanus \\
WALLABY	J032425$-$213233	&	1588	&	NGC 1332	&	\ldots	&	TSK 849	& NGC 1322\\
WALLABY	J033617$-$253615	&	1590	&	X 6dF	&	\ldots	&	Nest 200100	& Eridanus \\
WALLABY	J034337$-$211418	&	1612	&	Eridanus 	&	\ldots	&	\ldots	& Eridanus \\
WALLABY	J033941$-$235054	&	1622	&	Eridanus 	&	\ldots	&	\ldots	& Eridanus \\
WALLABY	J033921$-$212450	&	1622	&	Eridanus 	&	\ldots	&	\ldots	& Eridanus \\
WALLABY	J032900$-$220851	&	1627	&	X	&	LGG 97	&	\ldots	& Eridanus \\
WALLABY	J034131$-$214051	&	1644	&	Eridanus 	&	\ldots	&	\ldots	& Eridanus \\
WALLABY	J034036$-$213129	&	1644	&	Eridanus 	&	\ldots	&	\ldots	& Eridanus \\
WALLABY	J032937$-$232103	&	1657	&	X 6dF	&	\ldots	&	\ldots	& Eridanus \\
WALLABY	J033257$-$210513	&	1665	&	Eridanus 	&	\ldots	&	\ldots	& Eridanus \\
WALLABY	J032735$-$211339	&	1686	&	NGC 1332	&	\ldots	&	\ldots & NGC 1332	\\
WALLABY	J034057$-$214245	&	1695	&	Eridanus 	&	\ldots	&	\ldots	& Eridanus \\
WALLABY	J034522$-$241208	&	1733	&	\ldots	&	\ldots	&	\ldots	& Eridanus \\
WALLABY	J032941$-$221642	&	1755	&	X 6dF	&	LGG 97	&	\ldots	& Eridanus \\
WALLABY	J033228$-$232245	&	1755	&	X	&	\ldots	&	\ldots	& Eridanus \\
WALLABY	J034040$-$221711	&	1774	&	Eridanus 	&	\ldots	&	\ldots	& Eridanus \\
WALLABY	J032831$-$222957	&	1774	&	\ldots	&	\ldots	&	\ldots	& Eridanus \\
WALLABY	J033537$-$211742	&	1802	&	Eridanus 	&	\ldots	&	\ldots	& Eridanus \\
WALLABY	J033326$-$234246	&	1810	&	X	&	LGG 97	&	Nest 200100	& Eridanus \\
WALLABY	J034456$-$234158	&	1819	&	X	&	\ldots	&	\ldots	& Eridanus \\
WALLABY	J033653$-$245445	&	1842	&	X	&	\ldots	&	Nest 200100 & Eridanus	\\
WALLABY	J033341$-$212844	&	1859	&	Eridanus 	&	\ldots	&	\ldots	& Eridanus \\
WALLABY	J033911$-$222322$^{*}$	&	1879	&	\ldots	&	\ldots	&	\ldots	& Eridanus \\
WALLABY	J034114$-$235017	&	1885	&	Eridanus 	&	\ldots	&	Nest 200100 & Eridanus	\\
WALLABY	J033302$-$240756	&	1915	&	X 6dF	&	\ldots	&	Nest 200100	& Eridanus \\
WALLABY	J033347$-$192946	&	1964	&	X 6dF	&	LGG 100	&	TSK 863	& NGC 1407\\
\hline
\end{tabular}
Cols (1)-(2): Designation and \HI\ spectral line derived velocity - $cz$.
Col (3): Identification in B06 - member of Eridanus, NGC 1332 or NGC 1407 group;
X represents galaxies that are not identified as any member of the groups but with 6dF identification.
X 6dF indicates galaxies identified in NED but not observed in the 6dF survey. 
Col (4): Identification in G93 -- member of LGG 97 or LGG 100.
Col (5): Identification in T15 or TCS16 -- member of TSK 849, TSK 863 and/or Nest 200100.
Col (6): Identification in this study. 
$*$: \HI\ clouds.
\end{minipage}
\end{table*}

\section{Physical Parameters}

The parameters derived in this section are given in Tables~\ref{catalogue1} and \ref{catalogue2}. 

\subsection{Distance and Recession Velocity}\label{recession}

\begin{figure}
  \includegraphics[width=\columnwidth]{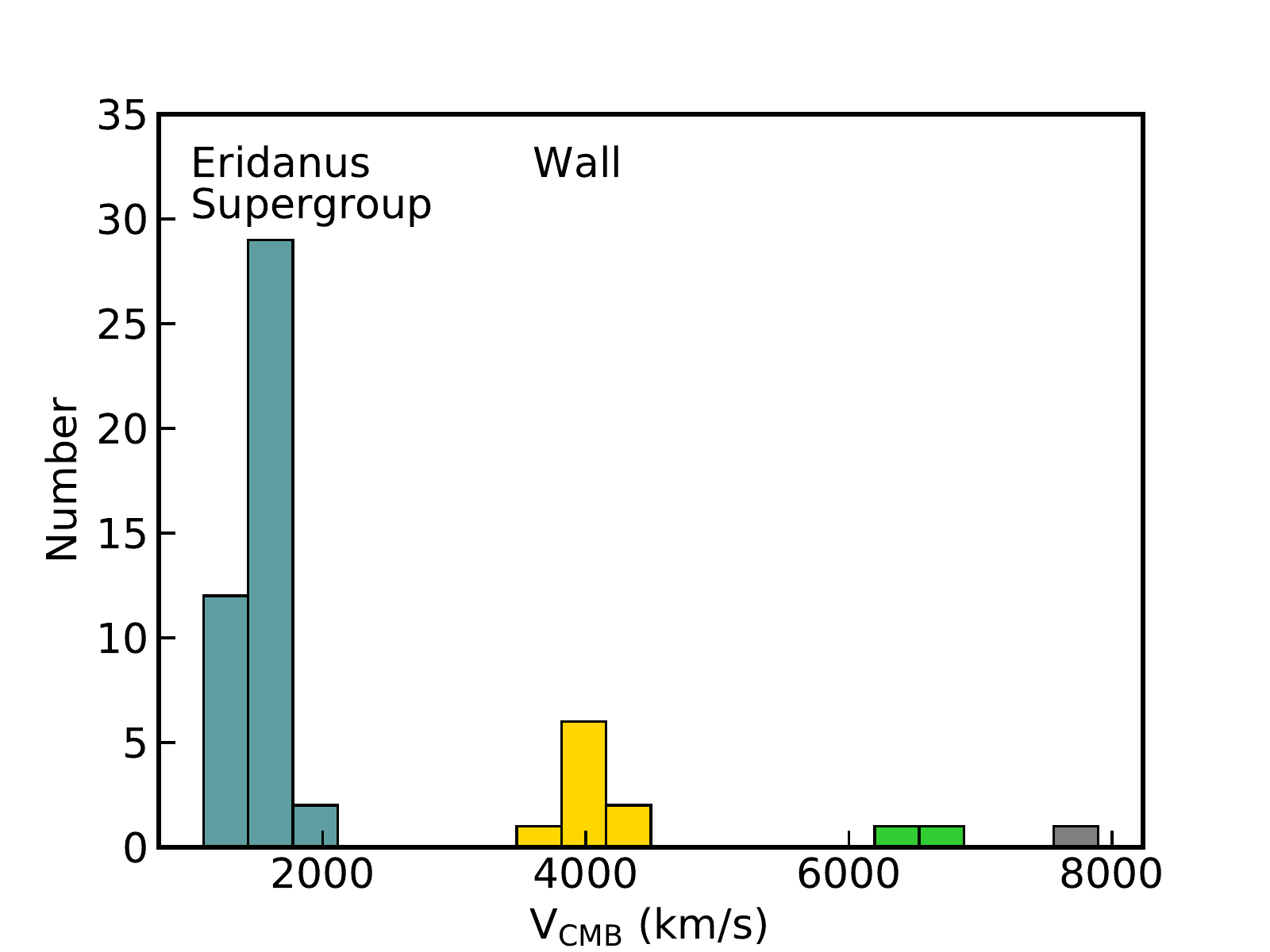}
  \caption{Histogram of recession velocities ($V_{\rm CMB}$).
    Different colours represent \HI\ detections in different groups of velocity.
    \label{hist_vcos}}
\end{figure}

The recession velocity in the
cosmic microwave background (CMB) reference frame, $V_{\rm CMB}$, is given by

\begin{equation}
  V_{\rm CMB} = V_{\rm opt} + V_{\rm apex}[\sin b \sin b_{\rm apex} + \cos b \cos b_{\rm apex} \cos (l-l_{\rm apex})],
\end{equation}
where $l$ and $b$ are Galactic coordinates, $l_{\rm apex}$ = 264.14\degr, $b_{\rm apex}$ = 48.26\degr\ and
$V_{\rm apex} = 371.0$~\kms\ \citep{Fixsen96}.
In Figure~\ref{hist_vcos}, we show the $V_{\rm CMB}$ distribution of \HI\ detected galaxies in the Eridanus field. 
The mean $V_{\rm CMB}$ for the Eridanus supergroup is $\sim1500$~\kms. 
The ``wall'' behind the Eridanus supergroup at 4000~\kms\ is also shown.
There are fewer galaxies beyond 4000~\kms\ as a result of decreasing 
sensitivity of the WALLABY data toward higher redshift. 

For galaxies behind the Eridanus supergroup, we use $D_{\rm H} = V_{\rm CMB}/H_{\rm 0}$
and $D_{\rm L}$ = (1+$z$)$D_{\rm c}$ to calculate the Hubble and luminosity distances, respectively. 
At low redshift ($z$ < 0.05), the co-moving distance, $D_{\rm c} \approx D_{\rm H}$, hence $D_{\rm L} \approx$ (1+$z$)$D_{\rm H}$.

For the Eridanus subgroups, we consider  
redshift-independent measurements for their distances to avoid the
effects of peculiar velocities, which are significant for galaxies with recession
velocities less than 2000~\kms\ \citep{Marinoni98}. 
There are several redshift-independent methods for determining distances, including 
the use of the globular cluster luminosity function (GCLF; \citealp{Richtler03} and references therein).
The GCLF is derived from the total observed population of GCs of a galaxy. 
The distance modulus ($m-M$) derived from the GCLF of NGC~1407, NGC~1332 and NGC~1395 are
$31.6\pm0.1$~mag \citep{Forbes06}, $31.7\pm0.24$~mag \citep{Kundu01} and $31.79\pm0.16$~mag \citep{Escudero18}, respectively.
These correspond to distances of 20.9~Mpc, 21.9~Mpc and 22.8~Mpc.
We note that the distance to the Eridanus group cannot solely rely on the measurement of NGC~1395 because
it is not at the dynamical/virial centre of the group. Eridanus is dynamically young and is still evolving.
Considering the distances and their associated errors of the more mature subgroups (NGC~1407 and NGC~1332), we
adopt a distance of 21~Mpc for all subgroups. 

\subsection{\HI\ Linewidth} \label{linewidth}

The \HI\ linewidth measured at 20 percent of the level of each peak ($w_{\rm 20}$) from \sofia\ needs to
be corrected for instrumental broadening. 
The correction is given by

\begin{equation}
  w^{c}_{\rm 20} = \frac{\sqrt{w^{2}_{\rm 20}-(\Delta s)^{2}}}{\Delta\nu} \times \Delta V_{\rm rest}. 
\end{equation}
$w_{\rm 20}$ has a default unit of Hz from \sofia.
The instrumental broadening, $\Delta s$ = $2\Delta\nu\lambda$, where $\lambda$ is the broadening parameter and 
$\Delta\nu$ is assumed to be equal to the observed channel width in Hz \citep{S05}. We adopt $\lambda = 0.5$ for a
typical profile shape. The channel width in the source rest frame ($\Delta V_{\rm rest}$) in units of \kms\
can be calculated using the following equation: 

\begin{equation}
  \Delta V_{\rm rest} \simeq \Delta\nu \frac{c(1+z)}{\nu_{\rm HI}},
\end{equation}
where $c$ is the speed of light, $z$ is redshift and $\nu_{\rm HI}$ is rest frequency of the \HI\ emission line \citep{Meyer17}.
Subsequently, equation~2 can be simplified as follows,

\begin{equation}
  w^{c}_{\rm 20} = \sqrt{w^{2}_{\rm 20}-(\Delta s)^{2}} \times \frac{c(1+z)}{\nu_{\rm HI}}. 
\end{equation}

\begin{sidewaystable*}
  \centering
    \scriptsize
  \begin{minipage}{210mm}
  \caption{Source catalogue and derived parameters.}\label{catalogue1}
  \begin{tabular}{lllccccccccrrrrr}
  \hline
ID	&	Designation		&	Other ID	&	$\alpha$ (J2000)	&	$\delta$ (J2000)	&	$\nu_{\rm obs}$	&	$z$	&	$V$	&	$V_{\rm CMB}$	&	$D_{\rm H}$	&	$D_{\rm L}$ 	&	$S_{\rm int}$ 	&	$S_{\rm int}^{c}$ 	&	err$_{S_{\rm int}^{c}}$	&	S/N  & $w^{c}_{\rm 20}$	\\
	&			&		&	(\degr)	&	(\degr)	&	(MHz)	&		&	(\kms)	&	(\kms)	&	(Mpc)	&	(Mpc)	&	(Jy Hz)	&	(Jy Hz)	&	(Jy Hz) &        & (\kms)	\\
(1)	&	(2)		&	(3)	&	(4)	&	(5)	&	(6)	&	(7)	&	(8)	&	(9)	&	(10)	&	(11)	&	(12)	&	(13)	&	(14)	& (15)   & (16)\\
\hline
\multicolumn{16}{c}{Eridanus Group} \\
\hline
1	&	WALLABY	J032831$-$222957	&	ESO 481$-$G028	&	52.132460	&	$-$22.499216	&	1412.05	&	0.0059	&	1774	&	1643	&	21.0	&	21.0	& 5956	&	7147   &  629	&	11.4  	    &	91.4	\\
2	&	WALLABY	J032900$-$220851	&	ESO 548$-$G025	&	52.252679	&	$-$22.147546	&	1412.74	&	0.0054	&	1627	&	1495	&	21.0	&	21.0	& 3009	&	3611   &  467	&	7.7             &	135.5	\\
3	&	WALLABY	J032937$-$232103	&	ESO 481$-$G030	&	52.408316	&	$-$23.350973	&	1412.60	&	0.0055	&	1657	&	1528	&	21.0	&	21.0	& 9160	&	10991  &  756	&	14.5            &	134.3	\\
4	&	WALLABY	J032941$-$221642	&	NGC 1347	&	52.424026	&	$-$22.278345	&	1412.14	&	0.0059	&	1755	&	1625	&	21.0	&	21.0	& 12727	&	15272  &  999	&	15.3            &	84.8	\\
5	&	WALLABY	J033019$-$210832	&	LEDA 832131	&	52.581366	&	$-$21.142435	&	1414.62	&	0.0041	&	1226	&	1095	&	21.0	&	21.0	& 2244	&	2692   &  458	&	5.9             &	49.7	\\
6	&	WALLABY	J033047$-$210333	&	ESO 548$-$G029	&	52.697580	&	$-$21.059272	&	1414.31	&	0.0043	&	1292	&	1161	&	21.0	&	21.0	& 4076	&	4892   &  619	&	7.9             &	138.5	\\
7	&	WALLABY	J033228$-$232245	&	ESO 482$-$G003	&	53.117925	&	$-$23.379351	&	1412.14	&	0.0059	&	1755	&	1630	&	21.0	&	21.0	& 3347	&	4017   &  465	&	8.6             &	69.6	\\
8	&	WALLABY	J033257$-$210513	&	ESO 548$-$G034	&	53.240898	&	$-$21.087152	&	1412.56	&	0.0056	&	1665	&	1537	&	21.0	&	21.0	& 5408	&	6490   &  740	&	8.8             &	84.5	\\
9	&	WALLABY	J033302$-$240756	&	ESO 482$-$G005	&	53.258442	&	$-$24.132440	&	1411.39	&	0.0064	&	1915	&	1793	&	21.0	&	21.0	& 29797	&	35756  &  1192	&	30.0            &	172.7	\\
10	&	WALLABY	J033326$-$234246	&	IC 1952	        & 	53.361541	&	$-$23.712911	&	1411.88	&	0.0060	&	1810	&	1688	&	21.0	&	21.0	& 23915	&	28698  &  1119	&	25.6            &	268.1	\\
11	&	WALLABY	J033327$-$213352	&	ESO 548$-$G036	&	53.364655	&	$-$21.564527	&	1413.29	&	0.0050	&	1509	&	1383	&	21.0	&	21.0	& 4659	&	5591   &  642	&	8.7             &	142.2	\\
12	&	WALLABY	J033341$-$212844	&	IC 1953	        & 	53.422847	&	$-$21.478898	&	1411.65	&	0.0062	&	1859	&	1733	&	21.0	&	21.0	& 42479	&	50974  &  1514	&	33.7            &	214.7	\\
13	&	WALLABY	J033408$-$232125	&	GALEXASC J033408.06$-$232130.1	&	53.534437	&	$-$23.357173 & 1414.45	& 0.0042&	1262	&	1140	&	21.0	&	21.0	& 3553	&	4263   &  473	&	9.0             &	50.6	\\
14	&	WALLABY	J033501$-$245556	&	NGC 1367/NGC1371&	53.756598	&	-24.932413	&	1413.53	&	0.0049	&	1458	&	1340	&	21.0	&	21.0	& 262608&	315130 &  4405	&	71.5            &	398.2	\\
15	&	WALLABY	J033527$-$211302	&	ESO 548$-$G049	&	53.865093	&	$-$21.217356	&	1413.25	&	0.0051	&	1518	&	1394	&	21.0	&	21.0	& 9332	&	11199  &  731	&	15.3            &	108.8	\\
16	&	WALLABY	J033537$-$211742	&	IC 1962	        &	53.905289	&	$-$21.295060	&	1411.92	&	0.0060	&	1802	&	1678	&	21.0	&	21.0	& 27984	&	33580  &  939	&	35.8            &	164.1	\\
17	&	WALLABY	J033617$-$253615	&	ESO 482$-$G011	&	54.072398	&	$-$25.604380	&	1412.91	&	0.0053	&	1590	&	1475	&	21.0	&	21.0	& 7755	&	9306   &  777	&	12.0            &	143.6	\\
18	&	WALLABY	J033653$-$245445	&	ESO 482$-$G013	&	54.223804	&	$-$24.912764	&	1411.73	&	0.0061	&	1842	&	1727	&	21.0	&	21.0	& 11623	&	13948  &  850	&	16.4            &	129.4	\\
19	&	WALLABY	J033723$-$235753	&	\ldots   	&	54.346062	&	$-$23.964901	&	1413.48	&	0.0049	&	1469	&	1352	&	21.0	&	21.0	& 5298	&	6357   &  608	&	10.4            &	44.9	\\
20	&	WALLABY	J033728$-$243010	&	NGC 1385	&	54.370368	&	$-$24.503023	&	1413.35	&	0.0050	&	1497	&	1381	&	21.0	&	21.0	& 99974	&	119969 &  2278	&	52.7            &	210.1	\\
21	&	WALLABY	J033854$-$262013	&	NGC 1398	&	54.725011	&	$-$26.336988	&	1413.93	&	0.0046	&	1373	&	1263	&	21.0	&	21.0	& 144298&	173158 &  6160	&	28.1            &	473.5	\\
22	&	WALLABY	J033911$-$222322	&	\ldots	        &	54.799365	&	$-$22.389573	&	1411.56	&	0.0063	&	1879	&	1762	&	21.0	&	21.0	& 8900	&	10680  &  761	&	14.0            &	49.6	\\
23	&	WALLABY	J033921$-$212450	&	LEDA 13460	&	54.838350	&	$-$21.414127	&	1412.76	&	0.0054	&	1622	&	1504	&	21.0	&	21.0	& 4810	&	5772   &  575	&	10.0            &	126.0	\\
24	&	WALLABY	J033941$-$235054	&	ESO 482$-$G027	&	54.924783	&	$-$23.848401	&	1412.76	&	0.0054	&	1622	&	1509	&	21.0	&	21.0	& 9187	&	11024  &  943	&	11.7            &	82.5	\\
25	&	WALLABY	J034036$-$213129	&	ESO 548$-$G069	&	55.150403	&	$-$21.524984	&	1412.66	&	0.0055	&	1644	&	1527	&	21.0	&	21.0	& 7541	&	9049   &  663	&	13.6            &	65.2	\\
26	&	WALLABY	J034040$-$221711	&	ESO 548$-$G070	&	55.170794	&	$-$22.286666	&	1412.05	&	0.0059	&	1774	&	1659	&	21.0	&	21.0	& 8939	&	10727  &  587	&	18.3            &	163.6	\\
27	&	WALLABY	J034056$-$223350	&	NGC 1415	&	55.235964	&	$-$22.564029	&	1413.09	&	0.0052	&	1552	&	1438	&	21.0	&	21.0	& 39261	&	47113  &  1927	&	24.5            &	376.8	\\
28	&	WALLABY	J034057$-$214245	&	NGC 1414	&	55.238411	&	$-$21.712588	&	1412.42	&	0.0057	&	1695	&	1579	&	21.0	&	21.0	& 11451	&	13741  &  746	&	18.4            &	159.1	\\
29	&	WALLABY	J034114$-$235017	&	ESO 482$-$G035	&	55.311683	&	$-$23.838321	&	1411.53	&	0.0063	&	1885	&	1773	&	21.0	&	21.0	& 15075	&	18090  &  1328	&	13.6            &	215.9	\\
30	&	WALLABY	J034131$-$214051	&	NGC 1422	&	55.381751	&	$-$21.680892	&	1412.66	&	0.0055	&	1644	&	1529	&	21.0	&	21.0	& 9704	&	11645  &  802	&	14.5            &	162.8	\\
31	&	WALLABY	J034219$-$224520	&	ESO 482$-$G036	&	55.580660	&	$-$22.755806	&	1413.01	&	0.0052	&	1569	&	1457	&	21.0	&	21.0	& 11446	&	13735  &  862	&	15.9            &	131.1	\\
32	&	WALLABY	J034337$-$211418	&	ESO 549$-$G006	&	55.908227	&	$-$21.238423	&	1412.81	&	0.0054	&	1612	&	1499	&	21.0	&	21.0	& 13324	&	15989  &  832	&	19.2            &	134.7	\\
33	&	WALLABY	J034434$-$211123	&	LEDA 135119	&	56.144596	&	$-$21.189812	&	1412.97	&	0.0053	&	1578	&	1466	&	21.0	&	21.0	& 3015	&	3618   &  454	&	8.0             &	69.3	\\
34	&	WALLABY	J034456$-$234158	&	LEDA 13743$^{*}$	&	56.236634	&       $-$23.699457	&	1411.84	&	0.0061	&	1819	&	1712&	21.0	&	21.0	& 3959	&	4751   &  681	&	7.0             &	81.3	\\
35	&	WALLABY	J034517$-$230001	&	NGC 1438	&	56.324313	&	$-$23.000331	&	1413.12	&	0.0052	&	1546	&	1438	&	21.0	&	21.0	& 11401	&	13681  &  1027	&	13.3            &	309.0	\\
36	&	WALLABY	J034522$-$241208	&	LEDA 792493	&	56.344383	&	$-$24.202442	&	1412.24	&	0.0058	&	1733	&	1628	&	21.0	&	21.0	& 4405	&	5286   &  561	&	9.4             &	40.1	\\
37	&	WALLABY	J034814$-$212824	&	ESO 549$-$G018	&	57.060158	&	$-$21.473564	&	1412.93	&	0.0053	&	1586	&	1480	&	21.0	&	21.0	& 8818	&	10582  &  1013	&	10.5            &	235.6	\\
\hline                         
  \end{tabular}                
  {\it Note.} This table is available in its entirety as Supporting Information with the electronic version of the paper. A portion is shown here for guidance regarding its form and content. 
$*$: Unresolved \HI\ source. Other possible optical ID is 6dFGS~J034456.8-234200. 
  Cols (1)--(3): Identification number, designation and other identification. 
  Cols (4)--(5): $\alpha$ and $\delta$ (J2000) coordinates are based on \HI\ detection.
  Col (6): $\nu_{\rm obs}$ is the detected central frequency of the \HI\ detection.
  Col (7): $z$ is redshift, defined as $z = (\nu_{\rm rest} - \nu_{\rm obs})/\nu_{\rm obs}$, where $\nu_{\rm rest}$ is \HI\ rest frequency at 1420.405751~MHz and $\nu_{\rm obs}$ is the observed frequency.
  Col (8): $V = cz$.
  Cols (9)--(11): $V_{\rm CMB}$, $D_{\rm H}$ and $D_{\rm L}$ are recession velocity, Hubble distance and luminosity distance (see Section~\ref{recession}).
  Cols (12)--(13): $S_{\rm int}$ and $S_{\rm int}^{c}$ are integrated flux and corrected integrated flux (see Section~\ref{intflux}).   
  Col (14): err$_{S_{\rm int}^{c}}$ is statistical uncertainty of $S_{\rm int}^{c}$, calculated as $\sigma_{\rm RMS}\times\sqrt{N_{\rm pix}/A_{\rm beam}}\times\Delta\nu$,
  where the Gaussian beam area, $A_{\rm beam} = \pi\theta_{\rm bmaj}\theta_{\rm bmin}/(4\ln2)$ and $\Delta\nu$ is the channel width in Hz. 
  Col (15): S/N of $S_{\rm int}^{c}$, calculated as $S_{\rm int}^{c}$ /err$_{S_{\rm int}^{c}}$.
  Col (16): $w^{c}_{\rm 20}$ is \HI\ linewidth measured at 20 percent of the level of each peak corrected for instrumental broadening (see Section~\ref{linewidth}).  
\end{minipage}
\end{sidewaystable*}

\subsection{Stellar mass} \label{smass}

We use the existing colour-stellar mass relations of \citet{Bell03} (hereafter B03) to derive stellar masses of our galaxies.
This relies on the photometry measurements from the co-added images of the DESI Legacy Imaging Surveys DR8 \citep{Dey19},
which are primarily based on the Dark Energy Camera Legacy Survey (DECaLS) southern observations.
The straight-forward way to obtain co-added images is through the cutout tool, in 
which the acquired galaxy is centred and stitched if it falls between two or more CCD tiles. 
However, these cutout images are weighted averages of the individual images and subjected
to varying point-spread-functions (PSF). To verify if the use of such co-added cutout images would result in inaccurate
photometry measurements, we obtain another set of co-added images
by first finding out the identification names of the bricks in which galaxies are located and 
then retrieving the co-added images via the NERSC server\footnote{Available at \url{https://portal.nersc.gov/project/cosmo/data/legacysurvey/dr8/south/coadd/}}.
By comparing five fluxes obtained
from cutout and non-cutout (non-stitched) co-added images, we find that the difference is small and within 0.02~mag.
This implies that the issue with varying PSF is minor and we decide to use the cutout co-added images
for the measurement to bypass the effort of stitching non-cutout co-added images manually.

We perform the $g$ and $r$-band photometry measurement using \profound\ \citep{Robotham18}.
\profound\ is used in rather than the traditional
circular or ellipse based method because it is useful for 
measuring fluxes of low surface brightness (LSB) galaxies, detecting faint signal with high noise level 
background and isolating foreground/background objects. Our sample consists faint dwarfs and LSB galaxies.
To identify background objects, we use the 3-colour
composite images for guidance. The background galaxies tend to be redder in colour. 
We also cross-check the \profound\ photometry measurements of a few bright galaxies with the
DR8 tractor catalogue and find them to be consistent within 0.02~mag.  

To calculate the extinction corrected (intrinsic) magnitude, we adopt 
the wavelength-dependent extinction law that is parametrised by $R_{V} = A_{V}/E(B-V)$ \citep{CCM89}.  
Using the reddening law in \citet{F99} and with $R_{V}$ = 3.1, the tabulated $A_{\lambda}$ values
for Dark Energy Survey (DES) $g$ and $r$-bands are $A_{g}$ = 3.237~$E(B-V)$ and $A_{r}$ = 2.176~$E(B-V)$, respectively
(see Table 6 of \citealp{SF11}).  
We estimate the Galactic dust extinction, $E(B-V)$, by using
the re-calibrated SFD all-sky extinction maps \citep{SFD98, SF11}.
The intrinsic magnitude is then calculated as  
mag$_{\rm 0}$ = mag - $A_{\lambda}$.
Subsequently, we obtain the stellar masses by employing the
mass-to-light ratio ($M/L$) relation in B03 as follow:

\begin{equation}
\log \left(\frac{M_{*}}{L_{r}}\right) = -0.306 +1.097(g-r),
\end{equation}
where $L_{r}$ (luminosity) = $M_{r, \rm abs} - M_{\rm sun, abs}$ in $L_{\odot}$.
The absolute magnitude of the Sun ($M_{\rm sun, abs}$) of different DES wavebands is given in \citet{Willmer18} and the 
$D_{\rm L}$ in Table~\ref{catalogue1} is used to calculate the $M_{r, \rm abs}$.  
This $M/L$ relation adopts the ``diet'' \citet{Salpeter55} initial mass function (IMF)
that contains only 70$\%$ of the mass due to a lower number of faint low-mass stars in
their samples. To use the \citet{Chabrier03} IMF, the $M_{*}$ values
need to be scaled by 0.61 according to \citet{MD14}. This is to put our data points on the same
scale with other surveys for comparison in Section~\ref{scaling}. 

We find that stellar masses for galaxies in the Eridanus
supergroup and background galaxies range from
$1.2\times10^{6}$ to $9.5\times10^{10}$ \msun\ and 
$6.3\times10^{7}$ to $2.1\times10^{10}$ \msun, respectively. 
By comparing the scaled stellar masses with those calculated with the $M/L$-colour relations of
\citet{Zibetti09} (hereafter Z09), we find that the difference is relatively small with an average $\pm0.02$~dex for all galaxies. 
The galaxies with high stellar mass ($M_{*} \gtrsim\ 10^{10}$~\msun) tend to have lower derived $M_{*}$
when using the colour-stellar mass relations of B03.
Using the colour-stellar mass relations of Z09, 
the lowest mass galaxy (WALLABY~J033408$-$232125) in this study is 0.32~dex smaller in stellar mass.
The differences are in part due to different stellar population synthesis models being adopted
in the two studies (see Z09 for a detailed discussion). We adopt the stellar masses derived from B03.
We do not find that the differences alter our conclusions.

\subsection{HI mass and HI-deficiency parameter}\label{himass_def}

\begin{figure}
  \includegraphics[width=\columnwidth]{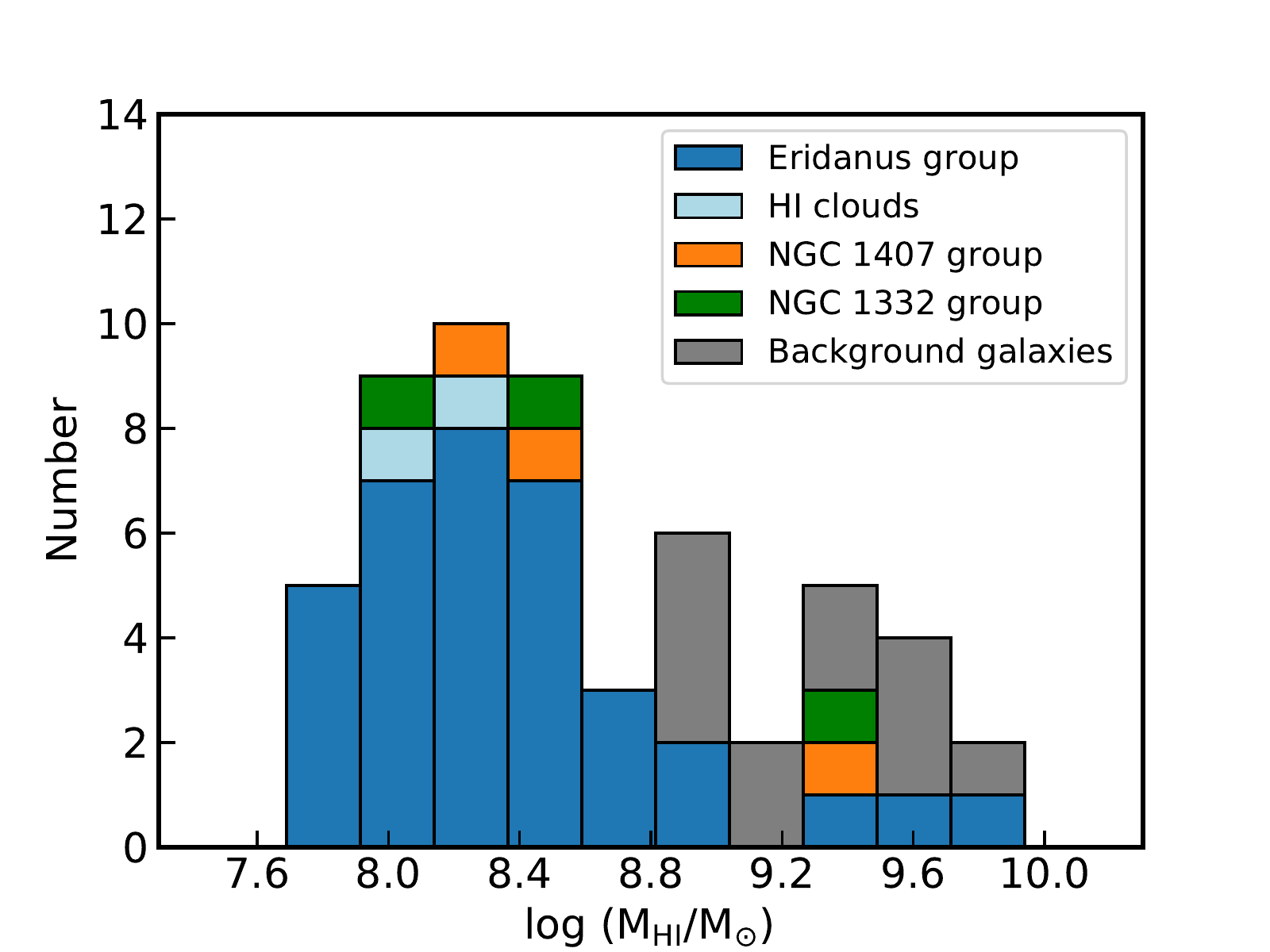}
  \caption{Histogram of \HI\ masses on a logarithmic scale. Galaxies and \HI\ clouds in the Eridanus group are represented in blue and light blue. 
    Galaxies in the NGC~1407 group, the NGC~1332 group and background are also represented in orange, green and grey, respectively. 
    \label{hist_himass}}
\end{figure}

Assuming that the \HI\ sources are optically thin, we can calculate the \HI\ mass using the following equation:

\begin{equation}
  \left(\frac{M_{\rm HI}}{\msun}\right)\simeq49.7\left(\frac{D_{\rm L}}{\rm Mpc}\right)^{2}\left(\frac{S_{\rm int}}{\rm Jy~Hz}\right),
\end{equation}
where $D_{\rm L}$ is the luminosity distance and $S_{\rm int}$ is the integrated flux \citep{Meyer17}. 
In Figure~\ref{hist_himass}, 
we show the histogram of the HI masses of the detected galaxies and HI clouds in the 
Eridanus group (blue and light blue),
galaxies in the NGC~1407 and NGC~1332 groups (orange and green), and background galaxies (grey).
The background galaxies have a \HI\ mass distribution of $\sim10^{8.9-10}$~\msun. 
The \HI\ mass of the Eridanus supergroup covers a range of $10^{7.7-9.8}$~\msun\ and has a median value of $10^{8.3}$~\msun.
We have recovered a significant number of galaxies with \HI\ masses lower than those found in \citet{OD05b},
where their derived \HI\ mass ranges from $\sim10^{8.3}$ to $10^{9.9}$~\msun. 

\begin{figure}
  \includegraphics[width=\columnwidth]{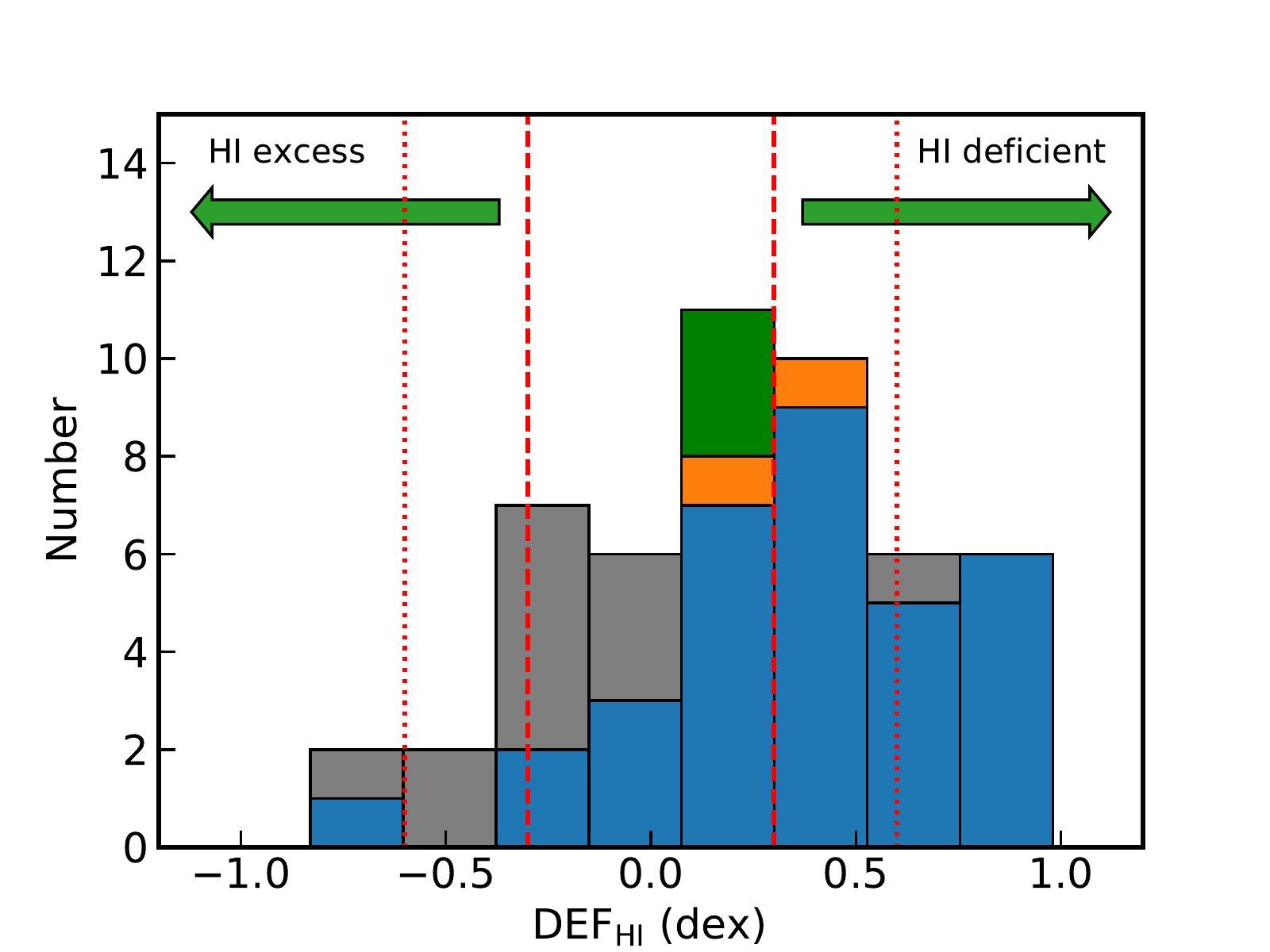}
  \caption{Same as Figure~\ref{hist_himass} except on the distribution of \HI\ deficiency parameter. The red dashed lines mark the boundaries of galaxies with
    normal \HI\ content ($-0.3 <$ DEF$_{\rm HI}$ $< +0.3$) with respect to the expected \HI\ from the scaling relation of \citet{Denes14}.
    Galaxies with DEF$_{\rm HI}$ > +0.3 and < $-0.3$ are considered \HI\ deficient and \HI\ excess, respectively. The red dotted lines indicate the
    conservative definition of \HI\ deficient and \HI\ excess, i.e. DEF$_{\rm HI}$ > $+0.6$, and DEF$_{\rm HI}$ < $-0.6$. 
    \label{hist_DEF}}
\end{figure}

To quantify the relative \HI\ content of galaxies, we use the \HI\ deficiency parameter, DEF$_{\rm HI}$. 
It is defined as the logarithmic difference between the expected and observed \HI\ masses of a galaxy,

\begin{equation}
  {\rm DEF}_{\rm HI} = \log (M_{\rm HI, exp}/\msun) - \log (M_{\rm HI, obs}/\msun)
\end{equation}
\citep{HG84}. 
The expected \HI\ mass of each galaxy can be determined by its morphology and size \citep{HG84, Solanes96, Jones18} or
magnitude based \HI\ scaling relations (\citealp{Denes14}; hereafter D14).
The morphology and size method is subject to inhomogenous
morphological classification and mostly limited to bright galaxies with known classification from previous studies.
We have explored this method and find it unsuitable for our study given that there is a large number of new LSB galaxies and dwarfs 
without a homogenous morphological classification. 
Thus, we adopt the magnitude based \HI\ scaling relation to determine the expected \HI\ mass,
which is given by

\begin{equation}
  \log (M_{\rm HI, exp}/\msun) = 3.43 - 0.3\times M_{r},
\end{equation}
where $M_{r}$ is the absolute magnitude in SDSS $r$-band\footnote{SDSS $r$-band is similar to the $r$-band in DECaLS \citep{Dey19}.}
(see Table~3 of D14). 
We consider a galaxy
to be \HI\ deficient if DEF$_{\rm HI}$ > +0.3 and \HI\ excess if DEF$_{\rm HI}$ < $-0.3$.
These criteria correspond to less than half or more than twice \HI\ mass
on average than expected.
The histogram of DEF$_{\rm HI}$ for all galaxies is shown in Figure~\ref{hist_DEF}. 
20 galaxies are considered to be \HI\ deficient and one \HI\ excess in the Eridanus group. 
There is one \HI\ deficient galaxy within our limited sample of the NGC~1407 group. There are also
four background galaxies considered to be \HI\ excess, and 
with one galaxy considered to be \HI\ deficient.

\begin{figure*}
  \includegraphics[scale=0.4]{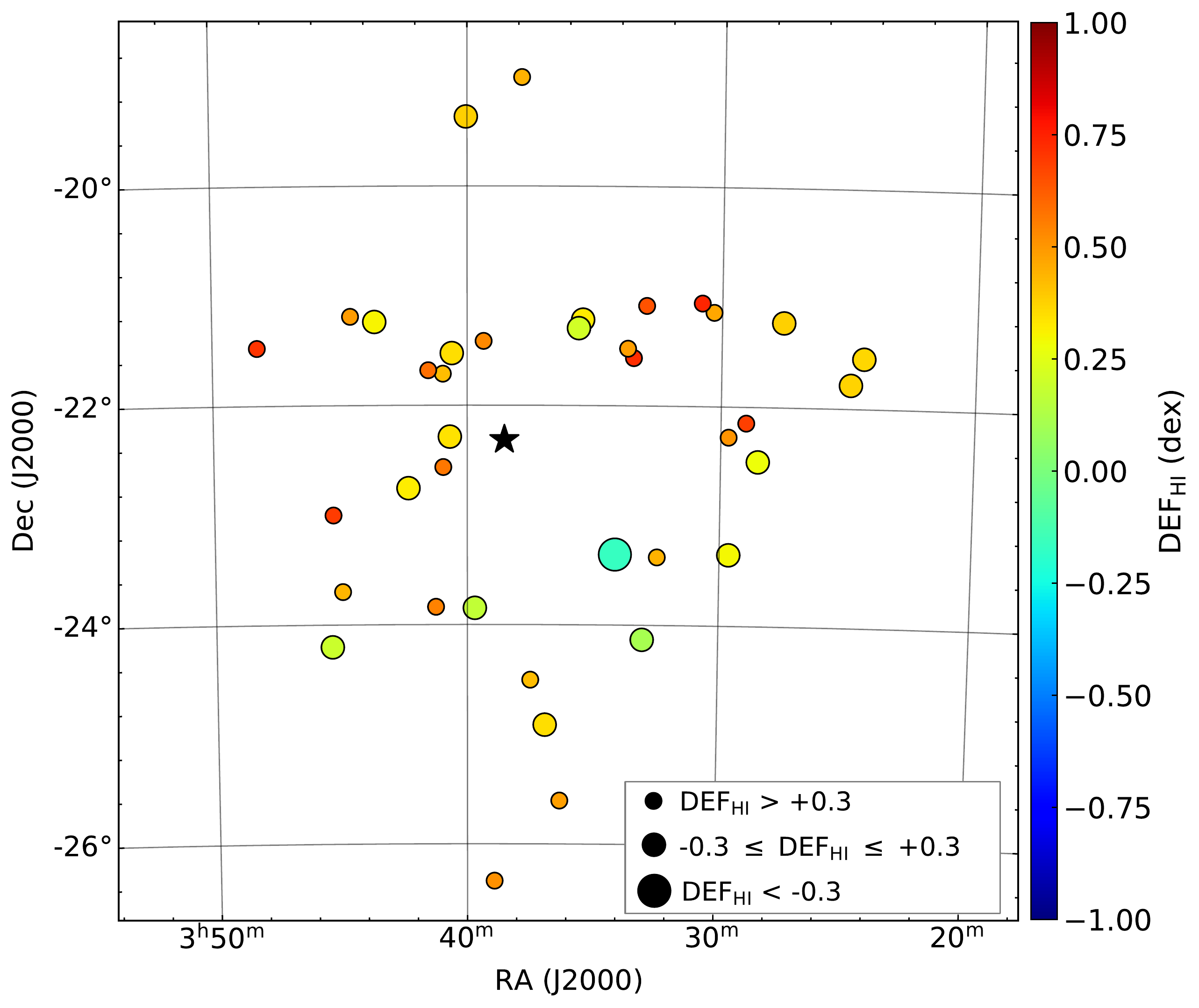}
  \caption{On-sky distribution of \HI\ deficiency parameters of galaxies in the Eridanus supergroup. Colour and size correpond to their DEF$_{\rm HI}$.
    The Eridanus group centre is marked with the star. 
    \label{onsky_DEF}}
\end{figure*}

The on-sky distribution of DEF$_{\rm HI}$
of galaxies in the Eridanus supergroup is shown in Figure~\ref{onsky_DEF}, where 
the majority of \HI\ deficient galaxies are in close proximity with other \HI\ detected galaxies. 
We note that the empirical scaling relation has a large instrinsic scatter with stellar luminosity alone as
a prediction of \HI\ mass. 
A more physically motivated derivation of the \HI\ deficiency parameter has been explored using 
the stellar mass and angular momentum (see \citealp{O16, Li20}). This method is adopted and the
interpretation of \HI\ deficiency of galaxies is further discussed in \citet{CM21}.

Overall, the Eridanus group consists of more \HI\ deficient galaxies than other Local Volume galaxy groups (see e.g. \citealp{Reynolds19}).  
We also find that the \HI\ deficiency parameter does not show any correlation as a function of projected distance from
the Eridanus group centre. This is consistent with the study of \HI\ galaxies in 16 loose groups, where \HI\ deficient galaxies
are not necessarily found at the centres of groups but within $\sim$1~Mpc from the group centre \citep{Kilborn09}.  
The non-correlation suggests that the pre-processing in the Eridanus group is yet to reach a stable stage of evolution.
This is somewhat expected for a dynamically young system that is in the process of merging.  

\begin{sidewaystable*}
  \centering
    \scriptsize
  \begin{minipage}{210mm}
  \caption{Photometry, morphology and derived parameters.}\label{catalogue2}
  \begin{tabular}{lllcccccccccc}
    \hline
ID	&	Designation	&       Other ID        	       &		$E(B-V)$	&	$g$ 	&	$r$ 	&	$M_{\rm r}$	&	log $L_{\rm r}$	&	log $M_{*}$	&	log $M_{\rm HI}$	&	DEF$_{\rm HI}$	&	Morphology	&	Note	\\
	&			&		&		(mag)	&	(mag)	&	(mag)	&	(mag)	&  (\lsun)	& (\msun)		& (\msun)		&		&		&		\\
(1)	&	(2)		&	(3)	&	(4)	&	(5)	&	(6)	&	(7)	&	(8)	&	(9)	&	(10)	&	(11)	&	(12)	&	(13)		\\
\hline
\multicolumn{13}{c}{Eridanus Group} \\
\hline
1	&	WALLABY~J032831$-$222957	& ESO 481$-$G028	&   0.0284$\pm$0.0005  	&	16.144	&	15.782	&	$-$15.829	&	8.176	&	8.05$\pm$0.27     &	8.12$\pm$0.22	&	0.06	&	\ldots	&	edge-on	\\
2	&	WALLABY~J032900$-$220851	& ESO 548$-$G025	&   0.0324$\pm$0.0012  	&	14.652	&	14.066	&	$-$17.545	&	8.862	&	8.98$\pm$0.25     &	7.82$\pm$0.24	&	0.87	&	(R')SB(s)a                          	&	peculiar	\\
3	&	WALLABY~J032937$-$232103	& ESO 481$-$G030	&   0.0243$\pm$0.0017  	&	15.457	&	15.071	&	$-$16.540	&	8.460	&	8.36$\pm$0.26     &	8.30$\pm$0.21	&	0.09	&	Sc                                      	&	\ldots	\\
4	&	WALLABY~J032941$-$221642	& NGC 1347	        &   0.0355$\pm$0.0011  	&	13.594	&	13.194	&	$-$18.417	&	9.211	&	9.13$\pm$0.23     &	8.45$\pm$0.21	&	0.51	&	SB(s)c? 	&	peculiar, pair	\\
5	&	WALLABY~J033019$-$210832	& LEDA 832131	        &   0.0345$\pm$0.0006  	&	16.400	&	15.960	&	$-$15.652	&	8.105	&	8.07$\pm$0.28     &	7.69$\pm$0.26	&	0.43	&	\ldots	&	Irr dwarf?	\\
6	&	WALLABY~J033047$-$210333	& ESO 548$-$G029	&   0.0369$\pm$0.0007  	&	13.848	&	13.265	&	$-$18.346	&	9.182	&	9.30$\pm$0.24     &	7.95$\pm$0.24	&	0.98	&	SB?                                     	&	\ldots	\\
7	&	WALLABY~J033228$-$232245	& ESO 482$-$G003	&   0.0299$\pm$0.0009  	&	15.845	&	15.541	&	$-$16.070	&	8.272	&	8.09$\pm$0.27     &	7.87$\pm$0.23	&	0.39	&	Sc                                      	&	\ldots	\\
8	&	WALLABY~J033257$-$210513	& ESO 548$-$G034	&   0.0294$\pm$0.0006  	&	14.067	&	13.507	&	$-$18.104	&	9.086	&	9.18$\pm$0.24     &	8.07$\pm$0.23	&	0.79	&	SB?                                     	&	LSB	\\
9	&	WALLABY~J033302$-$240756	& ESO 482$-$G005	&   0.0261$\pm$0.0008  	&	14.994	&	14.620	&	$-$16.991	&	8.640	&	8.53$\pm$0.25     &	8.81$\pm$0.20	&	$-$0.29	&	SB(s)dm?                         	&	edge-on	\\
10	&       WALLABY~J033326$-$234246	& IC 1952$^\dagger$	&   0.0265$\pm$0.0005  	&	13.024	&	\ldots	&	\ldots    	&	\ldots	&    \ldots$\pm$\ldots   &	8.72$\pm$0.20	&	\ldots	&	SB(s)bc?                                	&	\ldots	\\
11	&	WALLABY~J033327$-$213352	& ESO 548$-$G036	&   0.0307$\pm$0.0003  	&	13.862	&	13.189	&	$-$18.422	&	9.213	&	9.43$\pm$0.23     &	8.01$\pm$0.23	&	0.95	&	Sc                                      	&	\ldots	\\
12	&	WALLABY~J033341$-$212844	& IC 1953	        &   0.0303$\pm$0.0001  	&	12.128	&	11.610	&	$-$20.001	&	9.845	&	9.89$\pm$0.22     &	8.97$\pm$0.20	&	0.46	&	SB(rs)d                                 	&	\ldots	\\
13	&	WALLABY~J033408$-$232125& GALEXASC J033408.06$-$232130.1&   0.0254$\pm$0.0004  	&	19.428	&       19.497	&	$-$12.114	&	6.690	&	6.09$\pm$0.36     &	7.89$\pm$0.23	&	$-$0.83	&	\ldots	&	LSB dwarf	\\
14	&	WALLABY~J033501$-$245556	& NGC 1367/NGC1371	&   0.0267$\pm$0.0021  	&	\ldots	&	\ldots	&	\ldots	        &	\ldots	&    \ldots$\pm$\ldots   &	9.76$\pm$0.20	&	\ldots	&	SAB(rs)a                                	&	\ldots	\\
15	&	WALLABY~J033527$-$211302	& ESO 548$-$G049	&   0.0267$\pm$0.0006  	&	15.171	&	14.843	&	$-$16.768	&	8.551	&	8.39$\pm$0.26     &	8.31$\pm$0.21	&	0.15	&	S?                                      	&	\ldots	\\
16	&	WALLABY~J033537$-$211742	& IC 1962	        &   0.0279$\pm$0.0005  	&	14.420	&	14.007	&	$-$17.604	&	8.886	&	8.82$\pm$0.25     &	8.79$\pm$0.20	&	$-$0.08	&	SB(s)dm                                 	&	\ldots	\\
17	&	WALLABY~J033617$-$253615	& ESO 482$-$G011	&   0.0152$\pm$0.0008  	&	14.599	&	14.063	&	$-$17.548	&	8.863	&	8.93$\pm$0.25     &	8.23$\pm$0.22	&	0.46	&	Sb                                      	&	LSB	\\
18	&	WALLABY~J033653$-$245445	& ESO 482$-$G013	&   0.0175$\pm$0.0009  	&	14.690	&	14.364	&	$-$17.247	&	8.743	&	8.58$\pm$0.25     &	8.41$\pm$0.21	&	0.20	&	Sb?                                     	&	\ldots	\\
19	&	WALLABY~J033723$-$235753	& \ldots	        &   0.0247$\pm$0.0012  	&	\ldots	&	\ldots	&         \ldots	&	\ldots	&    \ldots$\pm$\ldots   &	8.06$\pm$0.22	&	\ldots	&	\ldots	&	\HI\ cloud	\\
20	&	WALLABY~J033728$-$243010	& NGC 1385	        &   0.0201$\pm$0.0002  	&	11.218	&	10.785	&	$-$20.826	&	10.174	&	10.13$\pm$0.21     &	9.34$\pm$0.20	&	0.34	&	SB(s)cd                                 	&	\ldots	\\
21	&	WALLABY~J033854$-$262013	& NGC 1398	        &   0.0134$\pm$0.0002  	&	10.387	&	9.609	&	$-$22.002	&	10.645	&	10.98$\pm$0.20     &	9.50$\pm$0.20	&	0.53	&	(R')SB(r)ab                             	&	\ldots	\\
22	&	WALLABY~J033911$-$222322	& \ldots	        &   0.0252$\pm$0.0015  	&	\ldots	&	\ldots	&	\ldots          &	\ldots	&    \ldots$\pm$\ldots   &	8.29$\pm$0.21	&	\ldots	&	\ldots	&	\HI\ cloud	\\
23	&	WALLABY~J033921$-$212450	& LEDA 13460	        &   0.0273$\pm$0.0005  	&	14.971	&	14.480	&	$-$17.131	&	8.696	&	8.71$\pm$0.25     &	8.02$\pm$0.22	&	0.55	&	S                                       	&	\ldots	\\
24	&	WALLABY~J033941$-$235054	& ESO 482$-$G027	&   0.0211$\pm$0.0004  	&	16.308	&	15.934	&	$-$15.677	&	8.115	&	8.00$\pm$0.28     &	8.30$\pm$0.22	&	$-$0.17	&	Irr                                     	&	\ldots	\\
25	&	WALLABY~J034036$-$213129	& ESO 548$-$G069	&   0.0250$\pm$0.0009  	&	15.431	&	14.971	&	$-$16.640	&	8.500	&	8.48$\pm$0.26     &	8.22$\pm$0.21	&	0.20	&	Irr                                     	&	\ldots	\\
26	&	WALLABY~J034040$-$221711	& ESO 548$-$G070	&   0.0263$\pm$0.0013  	&	15.349	&	14.814	&	$-$16.797	&	8.563	&	8.63$\pm$0.26     &	8.29$\pm$0.21	&	0.18	&	SBd?                      	&	edge-on	\\
27	&	WALLABY~J034056$-$223350	& NGC 1415	        &   0.0241$\pm$0.0003  	&	11.933	&	11.156	&	$-$20.455	&	10.026	&      10.36$\pm$0.21     &	8.93$\pm$0.20	&	0.63	&	(R)SAB0/a(s)                            	&	\ldots	\\
28	&	WALLABY~J034057$-$214245	& NGC 1414	        &   0.0233$\pm$0.0005  	&	14.435	&	13.957	&	$-$17.654	&	8.906	&	8.91$\pm$0.24     &	8.40$\pm$0.21	&	0.33	&	SB(s)bc?                      	&	edge-on, pair	\\
29	&	WALLABY~J034114$-$235017	& ESO 482$-$G035	&   0.0214$\pm$0.0011  	&	13.270	&	12.700	&	$-$18.911	&	9.408	&	9.51$\pm$0.23     &	8.52$\pm$0.21	&	0.58	&	SB(rs)ab                                	&	\ldots	\\
30	&	WALLABY~J034131$-$214051	& NGC 1422	        &   0.0248$\pm$0.0010  	&	13.758	&	13.095	&	$-$18.516	&	9.250	&	9.46$\pm$0.23     &	8.33$\pm$0.21	&	0.66	&	SBab 	&	peculiar, edge-on, pair	\\
31	&	WALLABY~J034219$-$224520	& ESO 482$-$G036	&   0.0199$\pm$0.0006  	&	15.164	&	14.627	&	$-$16.984	&	8.637	&	8.70$\pm$0.25     &	8.40$\pm$0.21	&	0.13	&	SAB(s)m                                 	&	LSB	\\
32	&	WALLABY~J034337$-$211418	& ESO 549$-$G006	&   0.0400$\pm$0.0018  	&	15.014	&	14.483	&	$-$17.129	&	8.695	&	8.76$\pm$0.25     &	8.47$\pm$0.21	&	0.10	&	IB(s)m                                  	&	\ldots	\\
33	&	WALLABY~J034434$-$211123	& LEDA 135119	        &   0.0544$\pm$0.0029  	&	15.832	&	15.365	&	$-$16.246	&	8.342	&	8.33$\pm$0.27     &	7.82$\pm$0.24	&	0.48	&	S                                       	&	\ldots	\\
34	&	WALLABY~J034456$-$234158	& LEDA 13743$^{*}$	&   0.0194$\pm$0.0009  	&	15.781	&	15.333	&       $-$16.278	&	8.355	&	8.33$\pm$0.27     &	7.94$\pm$0.25	&	0.37	&	\ldots	&	pair	\\
35	&	WALLABY~J034517$-$230001	& NGC 1438	        &   0.0197$\pm$0.0010  	&	12.786	&	12.091	&	$-$19.520	&	9.652	&	9.89$\pm$0.22     &	8.40$\pm$0.21	&	0.89	&	SB0/a?(r)                               	&	\ldots	\\
36	&	WALLABY~J034522$-$241208	& LEDA 792493	        &   0.0142$\pm$0.0005  	&	17.268	&	16.816	&	$-$14.795	&	7.762	&	7.74$\pm$0.29     &	7.98$\pm$0.23	&	$-$0.12	&	\ldots	&	LSB dwarf	\\
37	&	WALLABY~J034814$-$212824	& ESO 549$-$G018	&   0.0603$\pm$0.0015  	&	13.088	&	12.397	&	$-$19.215	&	9.530	&	9.77$\pm$0.22     &	8.29$\pm$0.22	&	0.91	&	SAB(rs)c                                	&	\ldots	\\
\hline                                                                                                                                                                        
  \end{tabular}                                                                                                                                                                   
  {\it Note.} This table is available in its entirety as Supporting Information with the electronic version of the paper. A portion is shown here for guidance regarding its form and content.
  $^{\dagger}$: $r$-band image has been contaminated with artefact. 
  $*$: Unresolved \HI\ source. Other possible optical ID is 6dFGS~J034456.8-234200. 
  Cols (1)--(3): Identification number, designation and other identification. 
  Col (4): $E(B-V)$ is the Galactic dust extinction.   
  Cols (5)--(6): Intrinsic $g$ and $r$-band photometry derived from the DR8 Legacy Survey images (see Section~\ref{smass}). 
  Cols (7)--(8): $M_{\rm r}$ and $L_{\rm r}$ are the $r$-band absolute magnitude and luminosity of the source (see Section~\ref{smass}). 
  Col (9): $\log M_{*}$/\msun\ is the derived stellar mass in logarithm scale and error is derived from the error propagation equation with the
  assumption of 10$\%$ uncertainty for the distance (see Section~\ref{smass}).
  Col (10): $\log M_{\rm HI}$/\msun\ is the derived \HI\ mass using the corrected $S_{\rm int}$ in logarithm scale and error is derived
  from the error propagation equation with the
  assumption of 10$\%$ uncertainty for the distance (see Section~\ref{himass_def}). 
  Col (11): DEF$_{\rm HI}$ is the \HI\ deficiency parameter (see Section~\ref{himass_def}).
  Cols (12)--(13): Morphology classification from NED and note for the source from this study. LSB: low surface brightness galaxy.  
\end{minipage}
\end{sidewaystable*}

\section{Atomic Gas Fraction--Stellar Mass Scaling Relation} \label{scaling}

\begin{figure*}
  \includegraphics[width=\columnwidth]{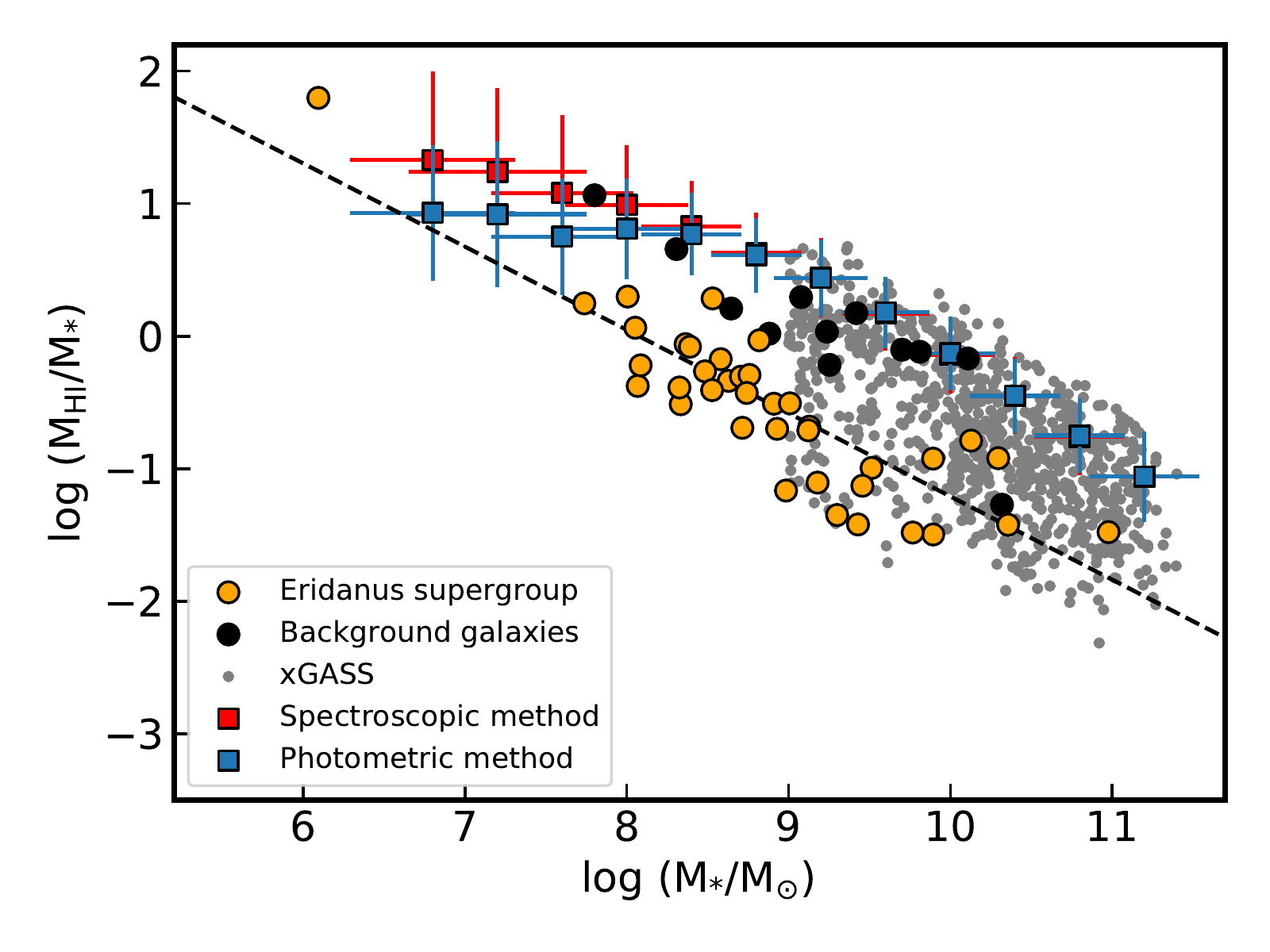}
  \includegraphics[width=\columnwidth]{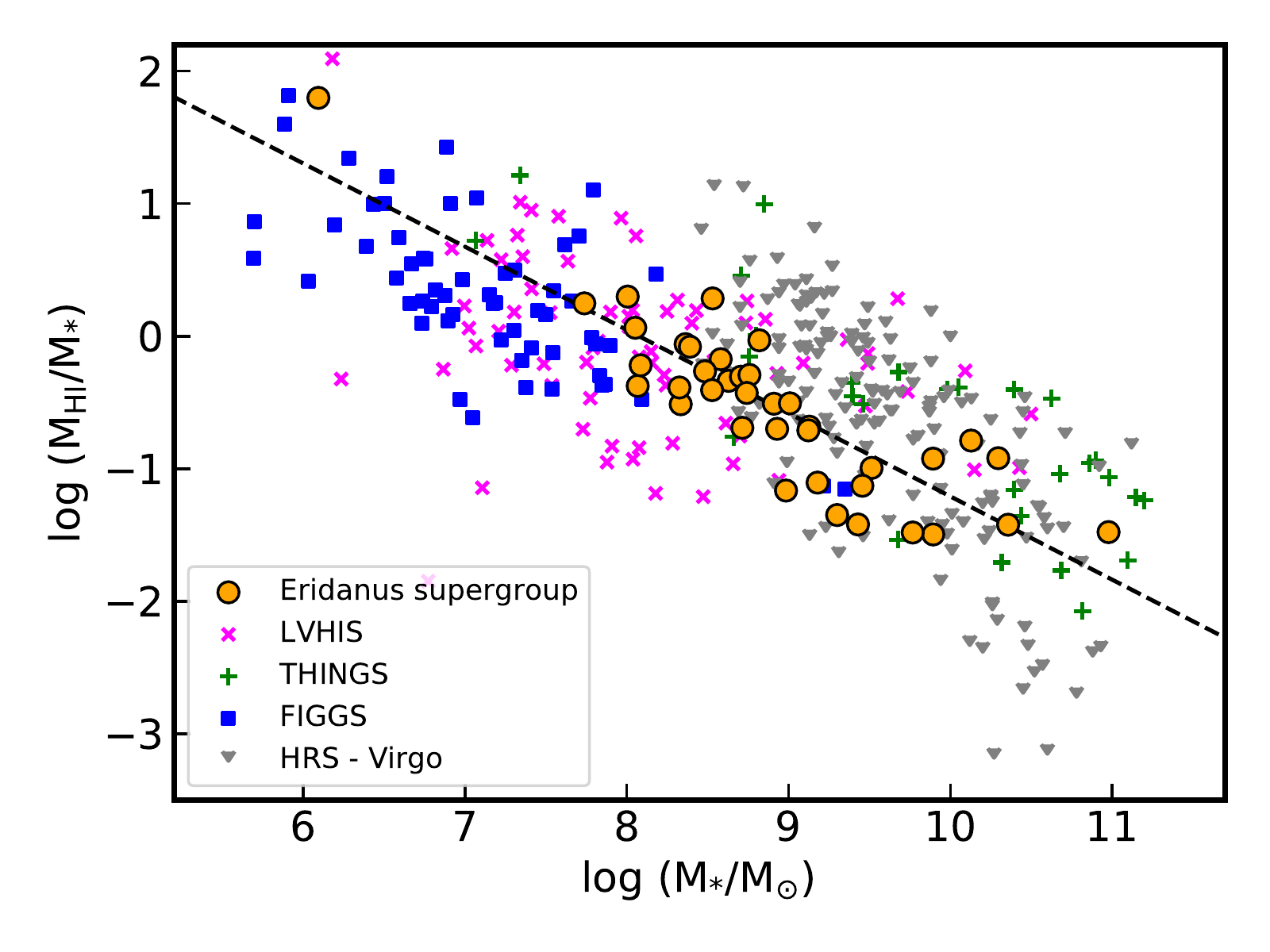}
  \includegraphics[width=\columnwidth]{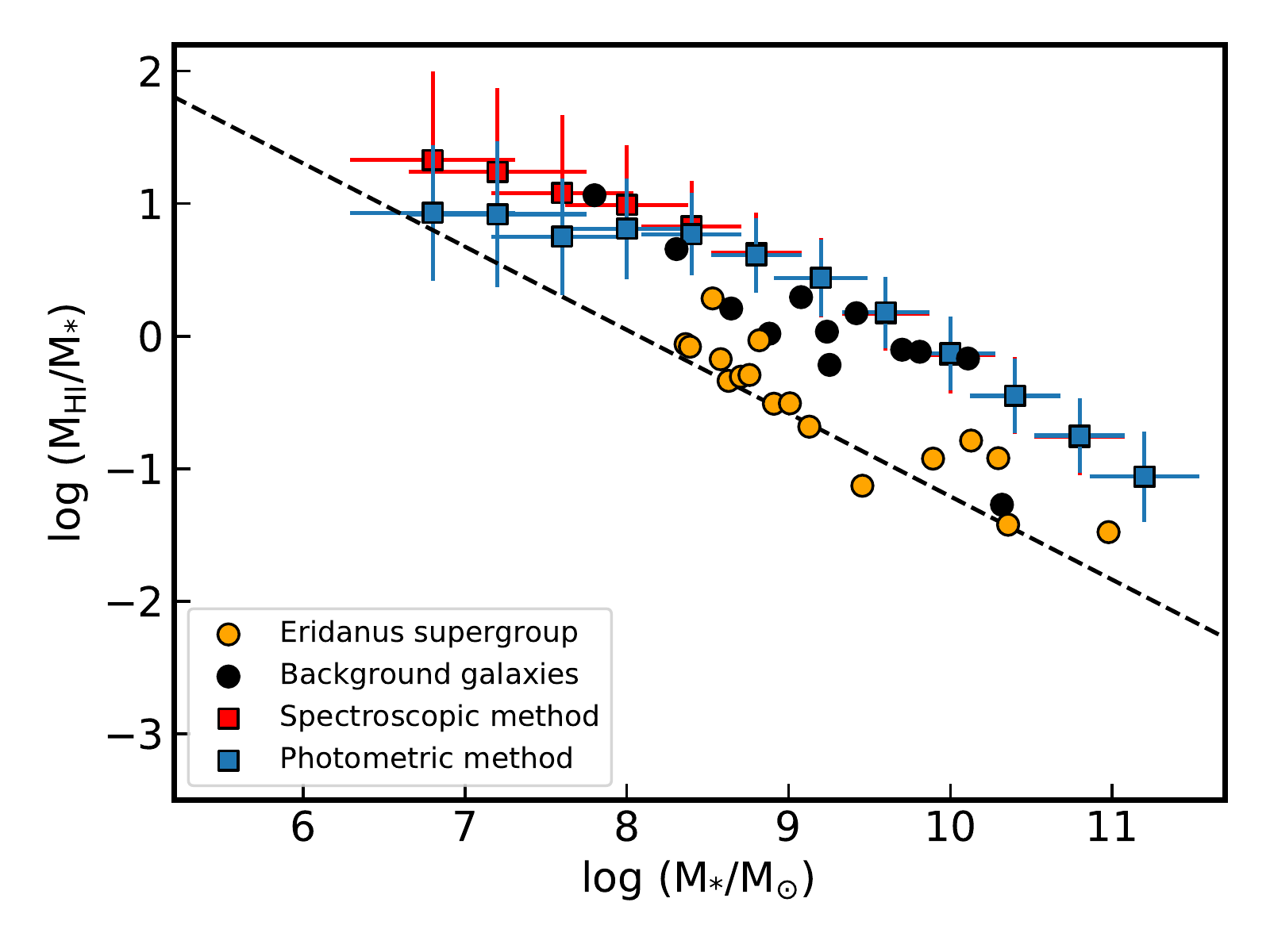}
  \includegraphics[width=\columnwidth]{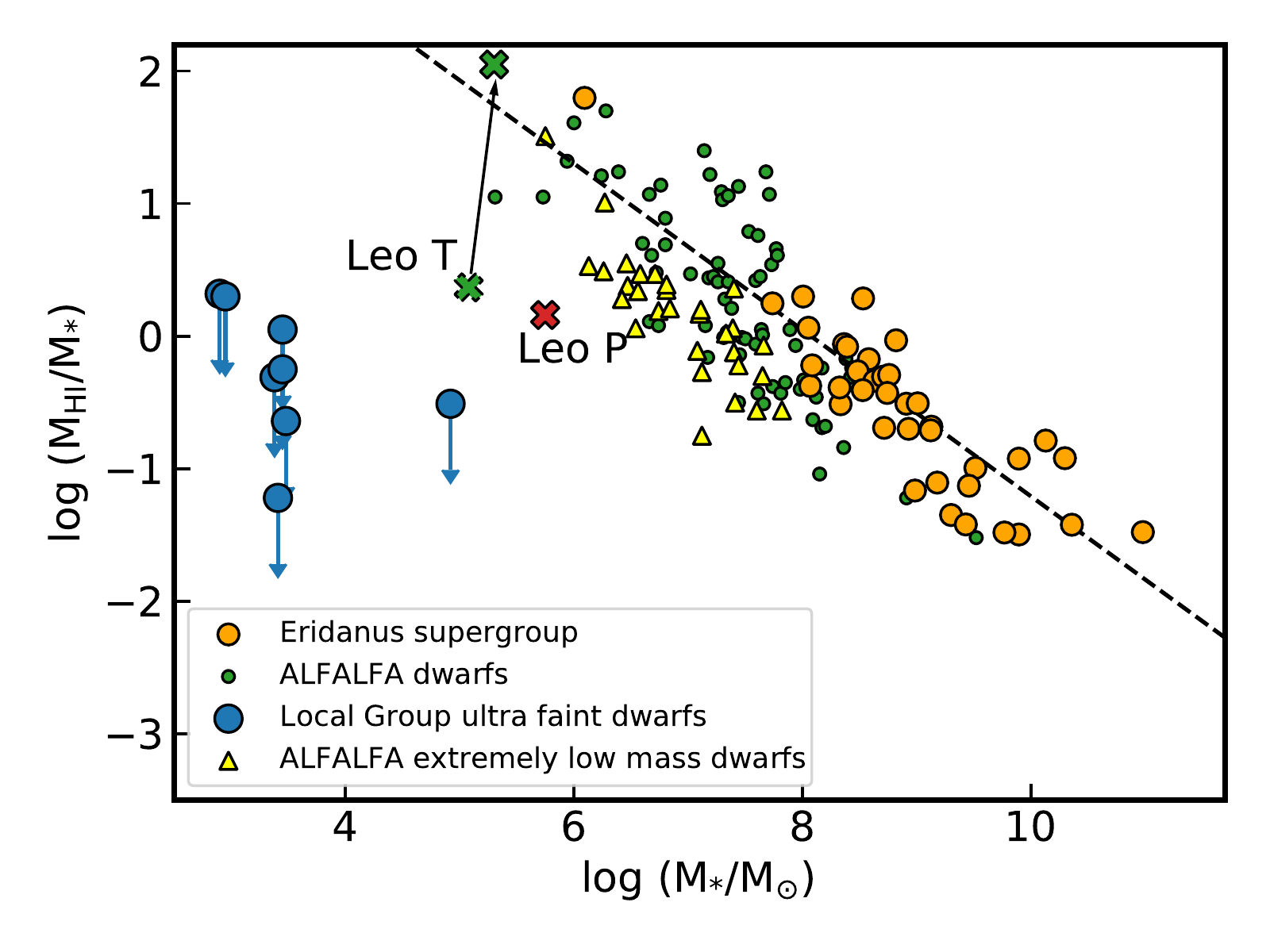}
  \caption{Atomic gas fraction scaling relation in logarithm scale ($\log$~($M_{\rm HI}/M_{*}$) versus $\log~M_{*}$).
    {\it Top left}: Eridanus supergroup and background galaxies in this study are represented by
    orange and black dots, respectively. 
    xGASS \HI\ detected sample is in grey dots. The empirical relations derived from 
    ALFALFA.40, SDSS spectroscopic (blue square) and photometric (red square) are overlaid on to the plot.
    {\it Top right}: LVHIS (magenta crosses), THINGS (green pluses), HRS - Virgo cluster members only (grey triangles) and FIGGS (blue squares)
    samples. The dashed line is used as a guide for the trend. Galaxies in the Eridanus supergroup (orange dots) are
    also overplotted and scaled to Chabrier IMF. Other studies used $H_{\rm 0}$ = 70 \kms\ Mpc$^{-1}$, which the resulting \HI\ and stellar masses are
    still within errors of our derived $M_{\rm HI}/M_{*}$ and $M_{*}$.
    {\it Bottom left}: Same as top left panel except with Eridanus supergroup galaxies that are detected at an assumed distance of 52~Mpc. 
    {\it Bottom right}: Adding dwarf galaxies from ALFALFA.40 sub-sample (green dots), dwarf satellites within the Milky Way (blue circles), extremely
    low mass dwarfs from ALFALFA (yellow triangles), Leo T (green cross) and Leo P (red cross) for comparison.
    The shift of the Leo T data point is due to the use of a different $M_{\rm HI}$. 
    \label{gasfrac}} 
\end{figure*}

The atomic gas fraction ($M_{\rm HI}/M_{*}$) versus stellar mass scaling relation is
a useful tool in investigating the influence of different environments
on galaxy properties and \HI\ content. 
In Figure~\ref{gasfrac}, we show the atomic gas fraction as a function of stellar mass.
The top left shows xGASS \HI\ detected sample (grey circles; \citealp{C18}), ALFALFA.40 samples (blue and red squares; \citealp{Maddox15})
and samples from this study (orange and black circles). 
xGASS is a stellar mass selected \HI\ survey and its sample is
limited to 9 < $\log$ ($M_{*}$/\msun) < 11.5 and 0.01 < $z$ < 0.05.
The xGASS sample shows a clear trend of linear relation of decreasing $\log M_{\rm HI}/M_{*}$ with increasing $\log M_{*}$.
The blue and red squares represent the atomic gas fraction (median log~$M_{\rm HI}$--log~$M_{*}$) of ALFALFA.40 samples
with stellar masses 
derived from SDSS spectra and photometry, respectively (see tabulated values in Table 1 of \citealp{Maddox15}).
These empirical relations show that the gas fraction follows the same trend as in the study of xGASS 
at higher stellar mass (> $10^{9}$~\msun) but flatten out at lower stellar mass indicating relatively higher gas content in
this regime. Deviation at the lower stellar mass end is seen between the two SDSS samples. 
The transition to this plateau presumably traces a real change in galaxy population as galaxies with
high gas content ($M_{\rm HI}$/$M_{*}\sim 100$)
would have been detected by ALFALFA.40. 

Our data for the Eridanus supergroup and background galaxies are
represented in orange and black circles, respectively.
We find that galaxies in the Eridanus supergroup tend to have a
lower atomic gas fraction for a given stellar mass than our background or xGASS or Maddox's galaxies.
In top right panel of Figure~\ref{gasfrac}, we show the trend being consistent with Local Volume \HI\ Survey (LVHIS; \citealp{Koribalski18}),
Virgo cluster only Herschel Reference Survey (HRS; \citealp{Boselli14}), The \HI\ Nearby Galaxy Survey (THINGS; \citealp{Walter08})
and Faint Irregular Galaxy Survey (FIGGS; \citealp{Begum08})
samples presented in \citet{Wang17}\footnote{Chabrier IMF and $H_{\rm 0}$ = 70 \kms\ Mpc$^{-1}$ were adopted.} (hereafter W17). 
The dashed line is a robust median linear fit of the data in W17, $\log$~($M_{\rm HI}/M_{\rm *}) = -0.63\times\log~M_{\rm *}$+5.07,
and it is used as a guide for the trend only. 
As stated in W17, such lower atomic gas fraction
trend is a result of the selection effect and is not caused by the use of interferometer versus single-dish data. 
We also rule out the use of different colour-stellar mass relations and errors of stellar/\HI\ masses as the cause of this
trend for galaxies in the Eridanus supergroup.
The shift in trend toward lower atomic gas fraction is not caused by the closer vs further distance
of the sample either, as can be supported by the following test. 
If the bias is due to the \HI\ detection limit, we expect the
data points of Eridanus supergroup galaxies to be shifted upward once they are placed in the same distance as background galaxies.
To achieve this, we assume a distance of 52~Mpc (comparable to the distance of our background galaxies)
and calculate their S/N of $S_{\rm int}$. 
We consider those with S/N of $S_{\rm int}$ $\geq$ 5.9 to be detectable at this assumed distance (see bottom left panel).
This value is the lowest detectable S/N of $S_{\rm int}$ of our galaxies sample (see Table~\ref{catalogue1}).   
We find that the lower gas fraction trend remains when comparing to our background galaxies.
This indicates that environmental effects might be at play for the galaxies in the Eridanus supergroup.   
It would be interesting to re-visit the \HI\ scaling
relation with the full WALLABY survey using group, void and cluster samples to probe the environmental effects in the future.

We also investigate the trend in the low mass regime by overplotting the
gas-rich Local Volume dwarf samples selected from ALFALFA.40 (\citealp{Huang12}; green circles),  
ultra-faint dwarf satellites of the Milky Way studied by \citet{Westmeier15} (blue circles), sample from
the Survey of \HI\ in Extremely Low-mass Dwarfs (SHIELD; \citealp{McQuinn21}) (yellow triangles), 
``almost dark'' galaxies, Leo T and Leo P \citep{Adams18, RW08, MCQuinn15} (green and red crosses) 
onto the scaling relation as
shown in the bottom right panel of Figure~\ref{gasfrac}. 
Sample of \citet{Huang12} is a subset of ALFALFA.40, which stellar masses etc of
\HI\ selected dwarf galaxies are re-derived via the spectral energy distribution (SED) fitting.
The selection eliminates the bias of ALFALFA.40 sample that tends to
under estimate the stellar masses of low mass galaxies with the SDSS reduction pipeline.
This bias contributes to a higher gas fraction and the flattening trend at the low
mass end as seen in \citet{Maddox15}. The SHIELD
dwarf galaxies sample ($M_{\rm HI} \lesssim 10^{7.2}$~\msun) is a volumetrically complete \HI\ selected
sample from ALFALFA. The overall gas fraction as a sample group is slightly lower than Huang's but
within the scatter. We note that the SHIELD sample mostly consists of isolated dwarf galaxies. 

As for the ultra-faint dwarf samples, the gas fraction being presented here are their upper limit.
It is known that dwarf satellites located within the Local Group
are \HI\ deficient \citep{GP09,P21}. The loss of gas in these satellites is hypothesised to be the result
of tidal interaction and ram-pressure stripping
during perigalactic passages. The plot shows that the gas fraction of the dwarfs in the Local Group does not follow
the general trend of the Local Volume ($M_{*} < 10^{6}$~\msun), although a future study is needed with a larger sample. 
It is also interesting to note that
both Leo T (adopt $M_{\rm HI}$ of Leo T in \citealp{RW08}) and Leo P appear to be outliers and are located closer to the ultra-faint dwarf samples on the guided 
\HI\ scaling relation. If we adopt the $M_{\rm HI}$ of Leo T in the latest
study of \citet{Adams18}, the data point falls onto the guided \HI\ scaling relation. This is the result of
a significant amount of faint \HI\ emission being recovered with deep \HI\ observations. 
It is unclear if this would be the case for Leo P. 

\section{Star Formation Rate} 

To derive the star formation rate (SFR), we
use the W4 band of the neoWISE resolution enhanced mosaics (HiRes; \citealp{Masci09}),
far-ultraviolet (FUV) and near-ultraviolet (NUV) images of
the Galaxy Evolution Explorer ($GALEX$; \citealp{Martin05})\footnote{GALEX images are retrieved from the
  Mikulski Archive Space Telescope (MAST).}. We  
follow the method as described in
W17 to perform the photometric measurements of these images for our galaxies.
Subsequently, we calculate the total SFR as follows:
\begin{equation}
\mathrm{SFR}~(\msun~\mathrm{yr^{-1}}) = \mathrm{SFR}_\mathrm{FUV/NUV} + \mathrm{SFR}_\mathrm{W4},  
\end{equation}
where the SFR of each band is calculated using the model-based luminosity-to-SFR calibrations in \citet{Calzetti13}.
We adopt a fixed $D_{\rm L}$ of 21~Mpc for the Eridanus, NGC~1407 and NGC~1332~groups and $D_{\rm L}$ 
for the background galaxies (see Table~\ref{catalogue1}). 
The FUV/NUV and W4 luminosities indicate the dust unattenuated and attenuated part of the SFR, respectively.
The FUV luminosities are preferred over the NUV in the estimates of SFR to minimize contamination from old stars,
but when the FUV images are unavailable we use the NUV luminosities instead.
We note that all galaxies in our sample have FUV detections if they have FUV observations. For galaxies
that are not detected in the W4 band (mostly low-mass galaxies), the dust attenuated part of the SFR is set to zero.
If there is no FUV/NUV detection, the SFR is considered to be the lower limit.

\begin{figure}
  \includegraphics[width=\columnwidth]{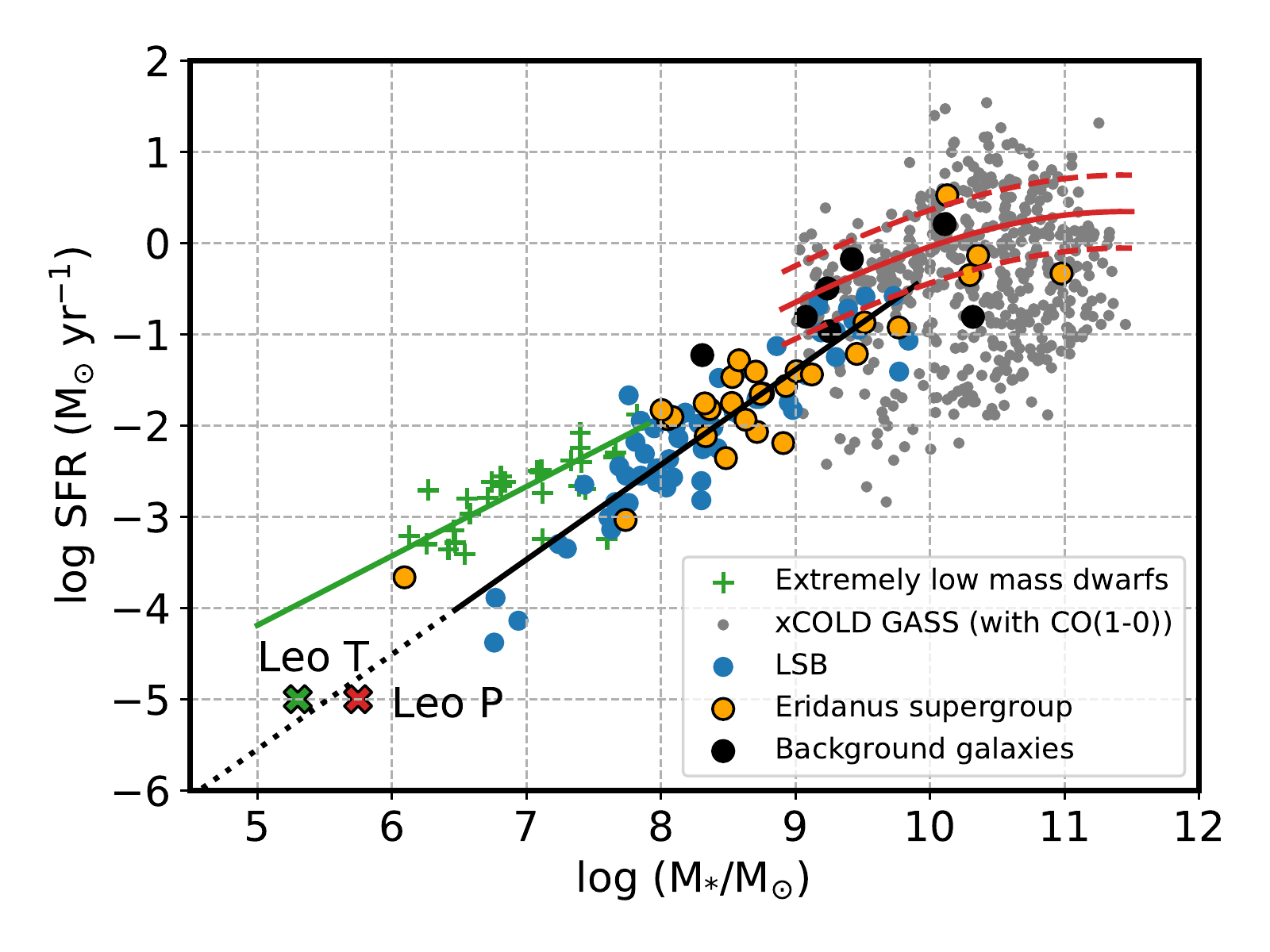}
  \caption{Star forming main sequence relation in logarithmic scale ($\log$ SFR vs $\log M_{*}/M_{\rm *}$).
    Eridanus supergroup and background galaxies in this study are represented by
    orange and black dots, respectively. xCOLD GASS (grey dots), LSB galaxies sample (blue dots), extremely low mass dwarfs (green pluses),
    Leo T (green cross) and Leo P (red cross) are also plotted for comparison. 
    The solid black and dashed lines represent the fit and extended
    relation from MG17. The solid and dashed red lines represent the fit and $\pm$0.4~dex scatter boundaries from S17. The green solid line
    represents the fit from MQ21. 
    \label{sfr}}
\end{figure}

We investigate if our galaxies
lie on the star-forming main-sequence (SFMS). The SFMS is a relation on the SFR-$M_{*}$ plane.
Galaxies that lie on this SFMS relation are 
actively forming stars, are quenched if they lie below and are starbursting if they lie above. 
In Figure~\ref{sfr}, 
we show the fitted SFR-$M_{*}$ relations derived from a sample of star forming 
galaxies in xCOLD GASS (\citealp{Saintonge17}; hereafter S17) (red line), LSB galaxies (\citealp{McGaugh17}; hereafter MG17)
(black line) and extremely low mass dwarfs (\citealp{McQuinn21}; hereafter MQ21) (green line).
The xCOLD GASS data with CO(1-0) (grey dots)
represent the SFMS relation derived for high
mass galaxies. The LSB galaxies data (blue dots) are at the low mass end.
The study of MG17 shows a steep slope for the SFMS using the low mass samples but a flatter slope with the high
mass samples, which is similar to the relation derived with the xCOLD GASS samples.
A shallower SFMS relation slope for a similar stellar mass range is seen in MQ21's study as compared to the MG17's study.
This shallower SFMS relation is consistent with the SFMS relation derived from spiral galaxies. However, 
with a lower birth rate parameter, MQ21 finds that a subset of the SHIELD sample galaxies would fall onto the steeper SMFS of MG17. 
This suggests that these subset of galaxies possess lower recent star formation that is similar to the star formation
properties of LSB galaxies. 

We also overplot Leo T (green cross), Leo P (red cross),
our galaxies in the Eridanus supergroup (orange dots) and background galaxies (black dots)
onto the SFR--$M_{*}$ plane for comparison. Our lower mass (< $10^{9}$~\msun)
Eridanus sample lies on the SFMS of MG17's study. Among the more massive galaxies (>$10^{9}$~\msun), three of them (ESO~482-G035,
NGC 1422 and ESO~548-G021) lie below the S17 SFMS relation but on MG17's. The first two are \HI\ deficient within the Eridanus group.
ESO~548-G021 is \HI\ normal within the NGC~1332 group. It is unclear if these galaxies are quenching due to them being located in the
overlapping region of these two SFMS relations. 
The background galaxies are in general following the SFMS relation of S17 with one (LEDA~798516) slightly above MG17's but it is still within
the scatter. LEDA~798516 is possibly interacting with surrounding galaxies and this could result in bursting of star formation.
Leo T and Leo P fall onto the extended relation of MG17's while the lowest stellar mass dwarf in our sample lies above it but closer
to MQ21's. 

\begin{table*}
  \centering
    \scriptsize
  \begin{minipage}{160mm}
  \caption{Star formation rates.}\label{sfr_cat}
  \begin{tabular}{lllcccccccc}
    \hline
ID	&	Designation	&	SFR$_{\rm FUV}$	&	$\sigma_{\rm FUV}$	&	SFR$_{\rm NUV}$	&	$\sigma_{\rm NUV}$	&	SFR$_{\rm W4}$	&	$\sigma_{\rm W4}$	& SFR$_{\rm total}$ & $\sigma_{\rm total}$	& $\log$ SFR	\\
        &			&	($\msun$~yr$^{-1}$)&	($\msun$~yr$^{-1}$)	&       ($\msun$~yr$^{-1}$)&	($\msun$~yr$^{-1}$)	&	($\msun$~yr$^{-1}$)&	($\msun$~yr$^{-1}$)	&	($\msun$~yr$^{-1}$)	&	($\msun$~yr$^{-1}$)& ($\msun$~yr$^{-1}$)		\\

(1) & (2) & (3) & (4) & (5) & (6) & (7) & (8) & (9) & (10) & (11) \\
\hline
\multicolumn{11}{c}{Eridanus Group} \\
\hline
1	&	WALLABY~J032831$-$222957	&	0.012	&	0.241	&	0.002	&	0.081	&	\ldots	&	\ldots	&    0.012	&	0.241	&	$-$1.93	\\         	   
2	&	WALLABY~J032900$-$220851	&	\ldots	&	\ldots	&	\ldots	&	\ldots	&	0.006	&	0.085	&    >0.006	&	0.085	&	>$-2.22$  \\          
3	&	WALLABY~J032937$-$232103	&	0.015	&	0.101	&	0.004	&	0.075	&	\ldots	&	\ldots	&    0.015	&	0.101	&	$-$1.82  	\\         
4	&	WALLABY~J032941$-$221642	&	\ldots	&	\ldots	&	\ldots	&	\ldots	&	0.050	&	0.070	&    >0.050	&	0.070	&	>$-1.30$  \\          
5	&	WALLABY~J033019$-$210832	&	\ldots	&	\ldots	&	\ldots	&	\ldots	&	0.007	&	0.098	&    >0.007	&	0.098	&	>$-2.15$  \\          
6	&	WALLABY~J033047$-$210333	&	\ldots	&	\ldots	&	\ldots	&	\ldots	&	0.046	&	0.070	&    >0.046	&	0.070	&	>$-1.34$  \\          
7	&	WALLABY~J033228$-$232245	&	0.012	&	0.276	&	0.002	&	0.075	&	\ldots	&	\ldots	&    0.012	&	0.276	&	$-$1.91  	\\         
8	&	WALLABY~J033257$-$210513	&	\ldots	&	\ldots	&	\ldots	&	\ldots	&	0.008	&	0.082	&    >0.008	&	0.082	&	>$-2.10$  \\          
9	&	WALLABY~J033302$-$240756	&	0.034	&	0.090	&	0.006	&	0.067	&	\ldots	&	\ldots	&    0.034	&	0.090	&	$-$1.47  	\\         
10	&	WALLABY~J033326$-$234246	&	0.057	&	0.077	&	0.014	&	0.067	&	0.127	&	0.069	&    0.184	&	0.017	&	$-$0.74  	\\         
11	&	WALLABY~J033327$-$213352	&	\ldots	&	\ldots	&	\ldots	&	\ldots	&	0.270	&	0.069	&    >0.270	&	0.069	&	>$-0.57$  \\          
12	&	WALLABY~J033341$-$212844	&	\ldots	&	\ldots	&	\ldots	&	\ldots	&	1.263	&	0.069	&    >1.263	&	0.069	&	>0.101    \\          
13	&	WALLABY~J033408$-$232125	&	\ldots	&	\ldots	&	0.0002	&	0.4131	&	\ldots	&	\ldots	&    0.0002	&	0.4131	&	$-$3.66  	\\    
14	&	WALLABY~J033501$-$245556	&	0.308	&	0.065	&	0.063	&	0.065	&	0.040	&	0.070	&    0.348	&	0.023	&	$-$0.46  	\\         
15	&	WALLABY~J033527$-$211302	&	\ldots	&	\ldots	&	\ldots	&	\ldots	&	0.006	&	0.102	&    >0.006	&	0.102	&	>$-2.22$  \\          
16	&	WALLABY~J033537$-$211742	&	\ldots	&	\ldots	&	\ldots	&	\ldots	&	0.004	&	0.093	&    >0.004	&	0.093	&	>$-2.40$  \\          
17	&	WALLABY~J033617$-$253615	&	0.020	&	0.077	&	0.004	&	0.067	&	0.007	&	0.077	&    0.027	&	0.002	&	$-$1.57  	\\         
18	&	WALLABY~J033653$-$245445	&	0.047	&	0.065	&	0.010	&	0.065	&	0.004	&	0.105	&    0.052	&	0.003	&	$-$1.29  	\\         
19	&	WALLABY~J033723$-$235753$^{*}$	&	\ldots	&	\ldots	&	\ldots	&	\ldots	&	\ldots	&	\ldots	&    \ldots	&	\ldots	&	\ldots    \\          
20	&	WALLABY~J033728$-$243010	&	0.824	&	0.065	&	0.199	&	0.065	&	2.507	&	0.069	&    3.331	&	0.278	&	0.52      \\               
21	&	WALLABY~J033854$-$262013	&	0.463	&	0.065	&	0.113	&	0.065	&	\ldots	&	\ldots	&    0.463	&	0.065	&	$-$0.33  	\\         
22	&	WALLABY~J033911$-$222322$^{*}$	&	\ldots	&	\ldots	&	\ldots	&	\ldots	&	\ldots	&	\ldots	&    \ldots	&	\ldots	&	\ldots    \\          
23	&	WALLABY~J033921$-$212450	&	\ldots	&	\ldots	&	0.004	&	0.076	&	0.005	&	0.108	&    0.008	&	0.001	&	$-$2.07  	\\         
24	&	WALLABY~J033941$-$235054	&	0.015	&	0.097	&	0.003	&	0.071	&	\ldots	&	\ldots	&    0.015	&	0.097	&	$-$1.83  	\\         
25	&	WALLABY~J034036$-$213129	&	\ldots	&	\ldots	&	0.004	&	0.066	&	\ldots	&	\ldots	&    0.004	&	0.066	&	$-$2.36  	\\         
26	&	WALLABY~J034040$-$221711	&	0.012	&	0.068	&	0.003	&	0.066	&	\ldots	&	\ldots	&    0.012	&	0.068	&	$-$1.94  	\\         
27	&	WALLABY~J034056$-$223350	&	0.048	&	0.077	&	0.017	&	0.072	&	0.681	&	0.069	&    0.728	&	0.073	&	$-$0.14  	\\         
28	&	WALLABY~J034057$-$214245	&	\ldots	&	\ldots	&	0.006	&	0.065	&	\ldots	&	\ldots	&    0.006	&	0.065	&	$-$2.19  	\\         
29	&	WALLABY~J034114$-$235017	&	0.079	&	0.073	&	0.017	&	0.066	&	0.055	&	0.070	&    0.134	&	0.011	&	$-$0.87  	\\         
30	&	WALLABY~J034131$-$214051	&	0.013	&	0.073	&	0.005	&	0.066	&	0.048	&	0.070	&    0.061	&	0.006	&	$-$1.22  	\\         
31	&	WALLABY~J034219$-$224520	&	0.039	&	0.088	&	0.007	&	0.067	&	\ldots	&	\ldots	&    0.039	&	0.088	&	$-$1.41  	\\         
32	&	WALLABY~J034337$-$211418	&	0.022	&	0.079	&	0.005	&	0.066	&	\ldots	&	\ldots	&    0.022	&	0.079	&	$-$1.65  	\\         
33	&	WALLABY~J034434$-$211123	&	0.008	&	0.110	&	0.002	&	0.073	&	\ldots	&	\ldots	&    0.008	&	0.110	&	$-$2.12  	\\         
34	&	WALLABY~J034456$-$234158	&	\ldots	&	\ldots	&	0.005	&	0.072	&	0.013	&	0.080	&    0.017	&	0.002	&	$-$1.76  	\\         
35	&	WALLABY~J034517$-$230001	&	\ldots	&	\ldots	&	\ldots	&	\ldots	&	0.052	&	0.070	&    >0.052	&	0.070	&	>$-1.28$ 	\\    
36	&	WALLABY~J034522$-$241208	&	\ldots	&	\ldots	&	0.001	&	1.138	&	\ldots	&	\ldots	&    0.001	&	1.138	&	$-$3.04	\\                 
37	&	WALLABY~J034814$-$212824	&	0.058	&	0.069	&	0.013	&	0.066	&	0.060	&	0.070	&    0.119	&	0.009	&	$-$0.93	\\                 
        \hline                                                                                                                                                                        
  \end{tabular}                                                                                                                                                                   
  {\it Note.} This table is available in its entirety as Supporting Information with the electronic version of the paper. A portion is shown here for guidance regarding its form and content. $^{*}$: \HI\ clouds.
  Cols (1)--(2): Identification and designation. 
  Cols (3)--(4): Star formation rate and its error in the $GALEX$ FUV. 
  Cols (5)--(6): Star formation rate and its error in the $GALEX$ NUV. 
  Cols (7)--(8): Star formation rate and its error in the neoWISE W4 band. 
  Cols (9)--(10): Total star formation rate and its error. 
  Col (11): Total star formation rate in the logarithmic scale.
  \end{minipage}
\end{table*}

\section{Environment of the Eridanus Supergroup}

\begin{figure*}
  \includegraphics[scale=0.2]{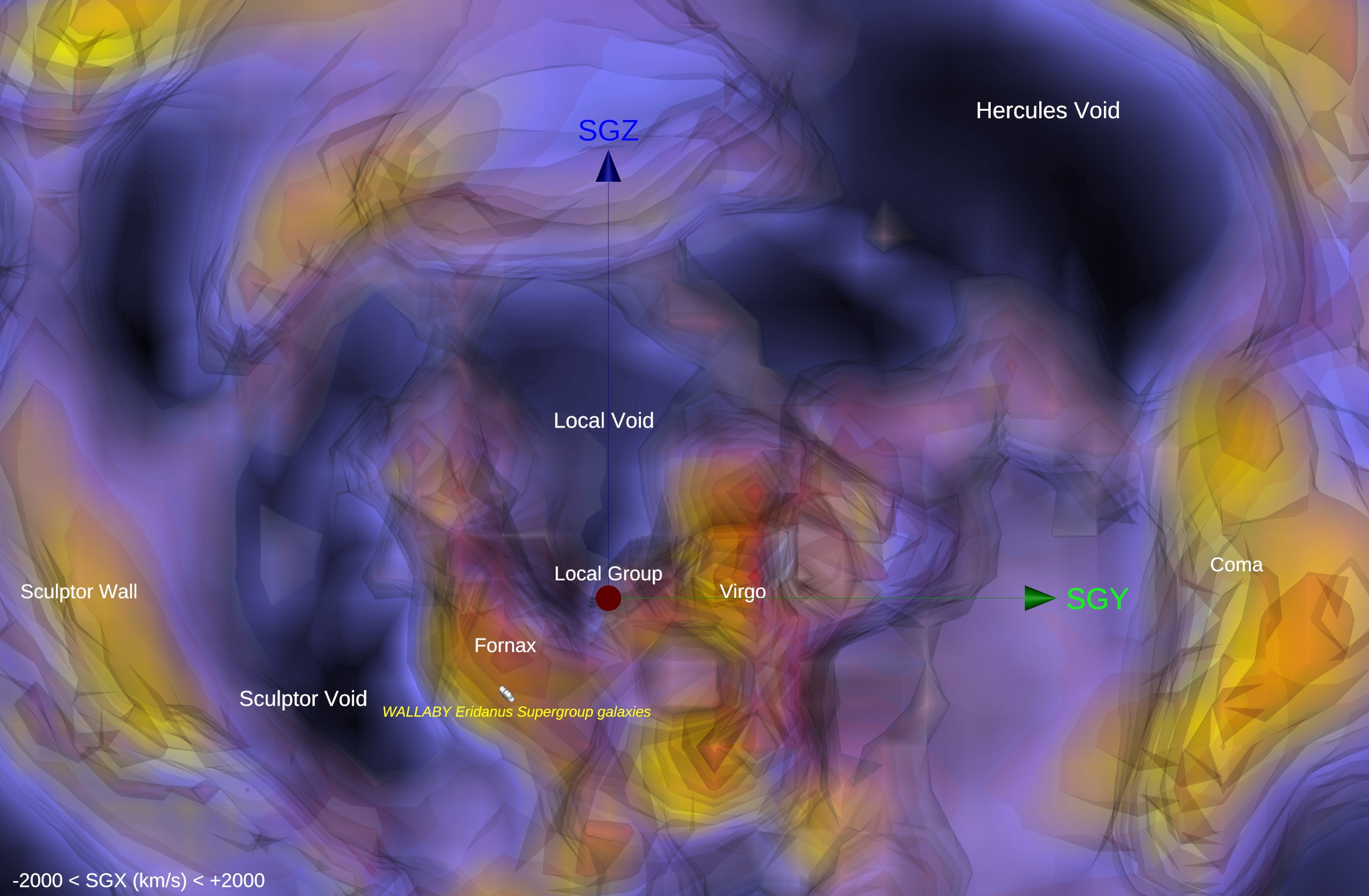}
  \caption{
    A slice of 3-dimensional model contains planes within the SGX=$-$2000~\kms\ and SGX=+2000~\kms\ is shown 
    face-on from the positive SGX direction.
    The positions of the Eridanus supergroup galaxies are plotted against a map of
    the density contrast $\delta$ reconstructed from the Cosmicflows-3 Catalog of peculiar velocities \citep{Graziani19}.
    This map is obtained with the ray-casting volume rendering technique,
    resulting in a smooth representation of the $\delta$ field
    ranging from most underdense (deep blue color) to most overdense (yellow),
    combined with a series of semi-transparent iso-surface polygons
    resulting in a sharp materialization of the surfaces ranging from $\delta=0$ (grey surface)
    to highly overdense ($\delta=2.8$ in red). The grey surface delineates the
    frontier between underdense and overdense patches of the local Universe.
    Scale and orientation are provided by the 5000~\kms\ long arrows
    emanating from our position and associated with the cardinal axes of the supergalactic coordinate system.
    An interactive 3-dimensional model is available at \url{https://sketchfab.com/3d-models/wallaby-eridanus-supergroup-vs-cf3-density-2d92e8a9f4b74f4293d9fabb9a6e73b3}. 
  }
  \label{environment}
\end{figure*}

The cosmic web is referred as a network of filaments with
clustered galaxies and gases stretched across the Universe and separated by
voids \citep{DP17}. It is seen in simulations and maps derived from galaxy redshift surveys. 
The so called the cosmic flow models are 
represented by the observed distribution of galaxies, peculiar velocities, and the underlying density field (see 
\citealp{Tully16, DP17} for detail on Cosmic V-web). 
In Figure~\ref{environment}, 
we show a 3-dimensional slice of the Universe centred at supergalactic position SGX=0 and
the environment
that the Eridanus supergroup resides in. 
The model is computed 
using the overdensity field, $\Delta$, by the means of four
iso-surface levels, namely 0.5, 1.0, 1.5 and 2.0. 
We find that all 43 galaxies are located near the knot of
Fornax. If we adopt a fixed $D_{\rm L}$ of 21~Mpc for all galaxies,
we also find that all of them fall in the same $\Delta$ = 1.428 grid. 
The eigenvalues $\lambda_{\rm 1}$, $\lambda_{\rm 2}$ and $\lambda_{\rm 3}$ of the shear tensor are
0.621, 0.090 and 0.030, respectively. The positive eigenvalues suggest that they are residing
in a knot cell of the V-web. However, the two lower eigenvalues ($\lambda_{\rm 2}$, $\lambda_{\rm 3}$)
indicate that some are edging toward the Sculptor Void. 
Overall, the model suggests that Eridanus supergroup is
currently merging/falling onto the Fornax V-web knot.
In future work, We will re-examine
these findings with the derived individual distances from the
Tully-Fisher relation paper.

\section{Summary and Conclusions}

We present the first WALLABY pre-pilot results from (almost) all the ASKAP antennas. 
These observations mimic the observation strategy
for the WALLABY pilot survey. We assess the data quality by using a set of statistical metrics. 
We also perform source finding using \sofia\ and obtain a catalogue of 55 \HI\ sources within $\sim$37~MHz, 
including two \HI\ clouds without stellar counterparts. In conjunction with deep optical images from the DR8 DESI Legacy Imaging Survey,
we recover new gas-rich low surface brightness galaxies and dwarfs.
We compare the integrated fluxes of WALLABY with  
the HIPASS and Parkes basketweave Eridanus survey.
This reveals a $20\%$ deficit of integrated flux for the pre-pilot observations, which
appears to arise from an accumulation of flux density scale errors, missing extended flux, low S/N
ratio, and the use of inaccurate primary beam models for the edge beams. All of these are better addressed in the WALLABY
pilot survey data. In the meantime, a uniform 20$\%$ correction is applied to the Eridanus
fluxes for statistical purposes. 
Based on the galaxy group catalogues of
G93, T15 and B06, we re-define membership of the Eridanus supergroup for our analysis. The mean recession velocity
of the Eridanus group is $\sim$1500~\kms. 

We also perform $g$ and $r$-band photometry measurements using \profound\ and derive
the corresponding stellar mass based on the $M/L$-colour relation in \citet{Bell03}. By comparing with
a more recently derived $M/L$-colour relation of \citet{Zibetti09}, we find that the stellar masses
are generally consistent except for one low mass galaxy.
We also calculate the \HI\ masses and the \HI\ deficiency parameter. 
The \HI\ masses of Eridanus group and background galaxies range from $10^{7.5-9.8}$ \msun\ and $10^{9-10.1}$ \msun, respectively.
There are 20 galaxies in the Eridanus group considered to be \HI\ deficient (DEF$_{\rm HI}$ > 0.3~dex)
and most of them are near other \HI\ detected sources. All galaxies show signs of
disturbance in their \HI\ morphology. Both tidal interaction and ram-pressure stripping mechanisms 
contribute gas loss in the Eridanus group \citep{CM21}. 
There is no correlation between the \HI\ deficiency
parameter and the projected distance from the Eridanus group centre.

Two massive previously known \HI\ clouds have been detected (see \citealp{Wong21}).
Smaller \HI\ clouds are also detected as part of a single source detection. 
The most prominent tidal debris field is seen in NGC~1359 (an interacting pair) and NGC~1385. This debris
has no optical counterpart. Some galaxies also show evidence of extra-planar gas extending out of the disk.

Comparing the gas fraction scaling relation with xGASS and the study of \citet{Maddox15}, we find that the gas fraction of
background galaxies follows the general trend of decreasing $M_{\rm HI}/M_{*}$ with increasing $M_{*}$.
However,
the gas fraction of galaxies in the Eridanus supergroup is lower for a given $M_{*}$. To further investigate if the
lower gas fraction trend is real in the Eridanus supergroup,
we run a \HI\ detection limit test. We find that a lower gas fraction trend remains for the galaxies in the Eridanus supergroup. 
We also compare the galaxies in the Eridanus supergroup with the gas fraction
of dwarfs in the Local Group. The trend of $M_{\rm HI}/M_{*}$ versus $M_{*}$ among
the Local Group dwarf population is not clear but an indication of
various \HI\ scaling relations are needed for different environment density.

To investigate if galaxies in our study are actively forming stars, we 
compare them with the star-forming main-sequence determined by S17, MG17 and MQ21. 
Overall, our galaxies in the Eridanus supergroup and background galaxies are actively forming stars.
We rule out the possibility of a gas accreting event given that they generally follow the
gas fraction scaling relation in the high mass regime.

\section*{Acknowledgements}

This research was supported by the Australian Research Council Centre of Excellence for All Sky Astrophysics in 3 Dimensions (ASTRO 3D),
through project number CE170100013. PK is partially supported by the BMBF project 05A17PC2 for D-MeerKAT.
LVM acknowledges financial support from the grants AYA2015-65973-C3-1-R and RTI2018-096228-B-C31 (MINECO/FEDER, UE),
as well as from the State Agency for Research of the Spanish MCIU through the
”Center of Excellence Severo Ochoa” award to the Instituto de Astrofisica de Andalucia (SEV-2017-0709).
The Australian SKA Pathfinder is part of the Australia Telescope National Facility which is funded by
the Australian Government with support from the National Collaborative Research Infrastructure Strategy and Industry Endowment Fund.
ASKAP uses the resources of the Pawsey Supercomputing Centre with funding
provided by the Australian Government under the National Computational Merit Allocation Scheme (project JA3)
We acknowledge the Wajarri Yamatji as the traditional owners of the Murchison Radio Observatory (MRO) site and thank
the operational staff onsite. 
This research has made use of images of the Legacy Surveys. The Legacy Surveys consist of three individual and
  complementary projects: the Dark Energy Camera Legacy Survey (DECaLS; Proposal ID \#2014B-0404; PIs: David Schlegel and Arjun Dey),
  the Beijing-Arizona Sky Survey (BASS; NOAO Prop. ID \#2015A-0801; PIs: Zhou Xu and Xiaohui Fan), and the Mayall z-band Legacy Survey
  (MzLS; Prop. ID \#2016A-0453; PI: Arjun Dey). DECaLS, BASS and MzLS together include data obtained, respectively, at the Blanco telescope,
  Cerro Tololo Inter-American Observatory, NSF’s NOIRLab; the Bok telescope, Steward Observatory,
  University of Arizona; and the Mayall telescope, Kitt Peak National Observatory, NOIRLab.
  The Legacy Surveys project is honored to be permitted to conduct astronomical research on Iolkam Du’ag (Kitt Peak),
  a mountain with particular significance to the Tohono O’odham Nation. 
BQF thanks A. Robotham and L. Davies for assisting the use ProFound and VISTAview cutout service.


\section*{Data availability}
The data underlying this article are available in the
article and in its online supplementary material. The processed
ASKAP data can be retrieved via CSIRO ASKAP Science Data Archive (CASDA)
with a given scheduling block identification number.
The DOI for the Eridanus data is \url{https://dx.doi.org/10.25919/0yc5-f769}. 
The data products
from \sofia\ is currently available within the WALLABY team and will be release to the public
at a later date.




\bibliographystyle{mnras}
\bibliography{ref} 


\section*{Supporting information}

Additional Supporting Information can be found in the online version
of this article.

Figure~\ref{combine}: Integrated \HI\ column density maps of
individual sources overlaid onto the DR8 DESI Legacy Imaging Survey $g$-band stacked images. Velocity field maps and spectrum of individual sources.

Table~\ref{catalogue1}: Source catalogue and derived parameters.

Table~\ref{catalogue2}: Photometry, morphology and derived parameters. 

Table~\ref{sfr_cat}: Star formation rates. 




\appendix

\section{\HI\ and Optical Morphologies} \label{morp}


In this section, we examine the \HI\ and optical morphologies of  
galaxies.
We refer the reader to Figures~\ref{onsky} and \ref{combine} for \HI\, optical morphologies and ID number of galaxies. 
All of the galaxies in the Eridanus supergroup show signs of disturbance in \HI\ morphology. 
The \HI\ detection of NGC~1347 (\#4) includes its interacting galaxy, PGC~816443.
An \HI\ extension is visible at $N_{\rm HI}$ $\sim5\times10^{20}$ cm$^{-2}$.
LEDA~832131 (\#5) does not have a morphological classification in NED. We identify it as an irregular dwarf with multiple star-forming
regions visible in the optical image. It is located to the south-west of both ESO~548$-$G029 (\#6)
and ESO~548$-$G034 (\#8), which are also distorted in \HI\ morphology. This indicates that the three are likely interacting with each other.
ESO~548$-$G036 (\#11) is located south-west of IC~1953 (\#12). There is strong evidence that
these two galaxies are interacting with
an \HI\ extension seen in IC~1953 towards ESO~548$-$G036. An \HI\ cloud is also detected north of ESO~548$-$G036.
NGC~1385 (\#20) is a face-on spiral galaxy that is located 0.6\degr\
south of WALLABY~J033723$-$235753. Its \HI\ morphology is very distorted and there is a southern tidal debris field.
There is no optical counterpart for the \HI\ debris. 
ESO~482$-$G035 (\#29) is a face-on SBab galaxy with a distorted and elongated \HI\ morphology, which
is in contrast to the non-fully developed \HI\ disk as shown in \citet{OD05b}. 
There is also a known LSB dwarf, F482$-$018 \citep{ME99}, south of ESO~482$-$G035, which does not have an \HI\ detection.
Given the elongation of the \HI, this dwarf could potentially be a satellite of ESO~482$-$G035.
GALEXASC~J033408.06$-$232130.1 (\#13; DEF$_{\rm HI}$ =$-$0.83) is a LSB dwarf and we find a \HI\ cloud is seen on the eastern side of this galaxy.
ESO~482$-$G005 (\#9; DEF$_{\rm HI}$ =$-$0.29) is an edge-on galaxy. Both galaxies show signatures of disturbance.

NGC~1422 (\#30) shows a truncated \HI\ disk that is smaller than the stellar disk.
LEDA~135119 (\#33) is classified as a spiral galaxy in NED but no distinctive spiral arms are seen in our optical image. 
Its \HI\ morphology is asymmetric and an \HI\ cloud north of the galaxy is detected. It is located at the edge of the observed field,
and hence, we cannot rule out the possibility that the galaxy is interacting
with another nearby galaxy. 
Both NGC~1398 (\#21) and NGC~1415 (\#27) are large bright galaxies with $B$-band optical isophotal diameter
measured at 25 mag arcsec$^{-2}$ of 425\arcsec\ and 208\arcsec, respectively. The \HI\ traces various star-forming sites and
is extended beyond the optical disk.
NGC~1398 (SBab galaxy) also has a ring like \HI\ that resembles a similar feature as seen in NGC~1533 \citep{RW03}.
Modeling of NGC~1533 suggests that the ring like \HI\ is formed 
as a result of an unequal-mass merging event between gas-rich LSB galaxies and host SB galaxy's disk \citep{Bekki04}.
NGC~1398 is located at the edge of the observed field and furthest from the Eridanus group centre. 
NGC~1438 (\#35) is a SB0/a galaxy that has a peculiar \HI\ morphology. There are two tidal tails south of the galaxy
and no optical counterpart has been identified along the tails.  

Other interesting \HI\ morphologies within the Eridanus group include ESO~482$-$G013 (\#18) and ESO~482$-$G027 (\#24). 
There are \HI\ extensions seen above and below the \HI\ disk of ESO~482$-$G013 (\#18).
Some edge-on galaxies also show evidence of extraplanar gas with plumes of gas extended out of the disk (e.g. IC~1952, ESO~482$-$G013 and ESO~482$-$G011). 
The latter is an irregular galaxy and its \HI\ morphology resembles a head-tail structure with the tail pointing away
from the Eridanus group centre. It is also located 0.53\degr\ and 1.46\degr\ angular distance from the two massive \HI\ clouds,
WALLABY~J033723$-$235753 (\#19) and WALLABY~J033911$-$222322 (\#22). We refer the reader to \citet{Wong21} for the discussion
of the origin of these two \HI\ clouds.

Among the three detections in the NGC~1407 group, a large and spectacular \HI\ tidal debris field is seen around the NGC~1359, an interacting galaxy pair (\#41).
Both ESO~548$-$G065 (\#43) and NGC~1390 (\#42) show disruption of the \HI\ outer disk.
Within the NGC~1332 group, NGC~1325 (\#38) is a SAbc galaxy
that forms a pair along with NGC~1325A. There is no \HI\ detection of NGC~1325A in our study but there is in W15.
Evidence of interaction is seen with
gas extending from the north corner of LSB galaxy, ESO~548-G011 (\#39), towards NGC~1325. 

The \HI\ morphology of all background galaxies also shows signs of disturbance.
ESO~549$-$G023 (\#50) is a SBa galaxy and is also classified as an emission line
galaxy. It has an extended \HI\ ``tail'' toward the south-east and some small \HI\ debris to the north-east of the galaxy.
LEDA~798516 (\#48) is a LSB galaxy and is possibly interacting with a few galaxies south of it, including LEDA~798377 (unresolved).
Another \HI\ detection is also seen at the north-east side of LEDA~798516.
The location coincides with an edge-on galaxy (GALEXASC J033838.83$-$233810.9), which is not identified as a separate source by \sofia. 
%

\bsp	
\label{lastpage}
\end{document}